\documentclass[12pt] {amsart}
\usepackage{amsfonts,amssymb,amsmath,epsfig,graphics,graphicx,rotating}
\usepackage{a4wide}
\usepackage[english]{babel}
\usepackage{setspace}
\usepackage[margin=1in]{geometry}

\newcommand{\halmos}{\rule{1ex}{1.4ex}}
\makeatletter \@addtoreset{equation}{section} \makeatother

\newtheorem{ittheorem}{Theorem}
\newtheorem{itlemma}{Lemma}
\newtheorem{itproposition}{Proposition}
\newtheorem{itcorollary}{Corollary}
\newtheorem{itdefinition}{Definition}
\newtheorem{itremark}{Remark}
\newtheorem{itexamples}{Examples}

\newenvironment{theorem}{\addtocounter{equation}{1}
\begin{ittheorem}}{\end{ittheorem}}
\newenvironment{lemma}{\addtocounter{equation}{1}
\begin{itlemma}}{\end{itlemma}}
\newenvironment{proposition}{\addtocounter{equation}{1}
\begin{itproposition}}{\end{itproposition}}
\newenvironment{corollary}{\addtocounter{equation}{1}
\begin{itcorollary}}{\end{itcorollary}}
\newenvironment{definition}{\addtocounter{equation}{1}
\begin{itdefinition}}{\end{itdefinition}}
\newenvironment{remark}{\addtocounter{equation}{1}
\begin{itremark}}{\end{itremark}}

\newenvironment{examples}{\addtocounter{equation}{1}
\begin{itexamples}}{\end{itexamples}}

\newenvironment{proofs}{\noindent {\em Proof}.\,\,\,}
{\hspace*{\fill}$\halmos$\smallskip}

\newenvironment{proofsm}{\noindent}
{\hspace*{\fill}$\halmos$\smallskip}

\newcommand{\beq}{\begin{eqnarray}}
\newcommand{\eeq}{\end{eqnarray}}

\newcommand{\beqq}{\begin{eqnarray*}}
\newcommand{\eeqq}{\end{eqnarray*}}

\newcommand{\be}{\begin{equation}}
\newcommand{\ee}{\end{equation}}

\newcommand{\bl}{\begin{lemma}}
\newcommand{\el}{\end{lemma}}

\newcommand{\br}{\begin{remark}}
\newcommand{\er}{\end{remark}}

\newcommand{\bex}{\begin{examples}}
\newcommand{\eex}{\end{examples}}

\newcommand{\bt}{\begin{theorem}}
\newcommand{\et}{\end{theorem}}

\newcommand{\bd}{\begin{definition}}
\newcommand{\ed}{\end{definition}}

\newcommand{\bp}{\begin{proposition}}
\newcommand{\ep}{\end{proposition}}

\newcommand{\bc}{\begin{corollary}}
\newcommand{\ec}{\end{corollary}}

\newcommand{\bpr}{\begin{proofs}}
\newcommand{\epr}{\end{proofs}}

\newcommand{\bprm}{\begin{proofsm}}
\newcommand{\eprm}{\end{proofsm}}

\newcommand{\bi}{\begin{itemize}}
\newcommand{\ei}{\end{itemize}}

\newcommand{\ben}{\begin{enumerate}}
\newcommand{\een}{\end{enumerate}}

\newcommand{\Z}{\mathbb Z}
\newcommand{\R}{\mathbb R}
\newcommand{\N}{\mathbb N}
\newcommand{\C}{\mathbb C}

\newcommand{\E}{\mathbb E}
\newcommand{\F}{\mathbb F}

\newcommand{\Pf}{\operatorname{Pf}}

\begin{document}

\title[Pfaffian representations]{
Graph theory and  Pfaffian representations 
\\of Ising partition function.}
\author{Thierry Gobron}
\address{ LPTM, UMR 8089, CNRS -
Universit\'e de Cergy-Pontoise. 2, avenue Adolphe Chauvin, Pontoise
\\95031 Cergy-Pontoise cedex, France}

\date{}
\begin{abstract}
      A well known theorem due to Kasteleyn  states that the partition function of an Ising model on an arbitrary planar graph
can be represented as the Pfaffian of a skew-symmetric 
matrix associated to the graph. This results both 
embodies the free fermionic nature of any planar Ising model and eventually gives an effective way of computing its partition 
functions in closed form.
 An extension of this result to non planar models expresses the partition function as a sum of 
Pfaffians which number is related to the genus of the oriented surface on which the graph can be embedded.

      In graph theory, McLane's theorem (1937) gives a characterization of planarity as a property of the cycle space of a graph, 
and recently, Diestel et al. (2009)  extended this approach to embeddings in arbitrary surfaces. 

       Here we show that McLane's approach naturally leads to Kasteleyn's results: McLane characterization of planar graphs
is just what is needed to turn an Ising partition function into a Pfaffian.

Using this approach, we prove that the Ising partition function on an arbitrary non planar graph can be written as \emph{the real part }of the Pfaffian of a single matrix with coefficients taken in a multicomplex algebra $\C_{\tilde g}$, where $\tilde g$ is the non-orientable  genus, or crosscap number, of the embedding surface.

Known representations as sums of Pfaffians follow from this result. In particular, Kasteleyn's result which involves $4^g$ matrices with real coefficients,
$g$ orientable genus, is also recovered through some algebraic reduction. 

\end{abstract}
\maketitle
\setcounter{equation}{0}
\section {Introduction}
\label{sec:1}

A few years after Onsager's solution of the Ising model on a square lattice \cite{O}, an alternative, combinatorial method
has been elaborated through the pioneering works of Kac and Ward \cite{KW}, Potts and Ward \cite{PW} and Hurst and Green \cite{HG}.
In these works, the evaluation of Ising partition function on a rectangular array was reduced to a combinatorial enumeration of perfect matchings
(dimer coverings in the physics litterature), leading to an expression in terms of the Pfaffian of a related skew-symmetric matrix.
This method acquired a deeper signification about fifty years ago, when Fisher \cite{F2} and Kasteleyn \cite{K1,K2,K3} showed that 
it applies equally to an arbitrary planar graph.

Numerous attempts have been made to generalize these results to non-planar graphs.  Kasteleyn's landmark result  \cite{K3} states that the partition function 
can be written as a sum of Pfaffians, whose number grows exponentially with the orientable genus of the graph.  This statement has been given a better mathematical basis only rather recently \cite{GL, CR} and also extended to non orientable surfaces \cite{Te}. Finding a sensible lower bound on this number is still an open problem.

 Enumeration of dimer coverings and evaluation of Ising partition function are clearly two deeply connected problems, but it should be stressed that they differ somewhat on this point:

In the dimer problem, one counts the number of dimer coverings on a given graph, or, in other words, the number of subgraphs 
for which the incidence number is exactly one on every vertex. For bipartite graphs,
it reduces to the old permanent-determinant problem \cite{Po} and there are well known complexity issues \cite{V,T}. However the graphs for which the dimer covering enumeration can be turned into a Pfaffian are not necessarily planar \cite{Tu, LP}. Accordingly, some exact results 
on nonplanar graphs have been obtained in the context of statistical physics \cite{LW1,LW2}.

In the Ising case, one computes a weighted sum over all closed curves of a given graph, that is over all subgraphs with an even incidence number on each vertex. In contradistinction with the dimer problem, the enumeration of these subgraphs on an arbitrary connected graph 
$G=(V,E)$ is trivial and equal to $2^{\beta_1}$ where $\beta_1$ is the first Betti number, $\beta_1(G)=|E|-|V|+1$. The complexity 
in evaluating the partition function comes from the introduction of a weight function on the edge set and there seems to be no exception to Kasteleyn's planarity rule.

In the present paper, we consider again Kasteleyn's combinatorial approach,  giving another glimpse to its  graph theoretical foundations. 

Our starting point consists in considering the Ising model on an arbitrary graph and derive a set of nonlinear algebraic equations in the entries of an associated skew-symmetric matrix, so that any consistent solution to these equations would lead to a representation of the Ising partition function as a Pfaffian. 
We thus let aside Kasteleyn's edge orientation method, valid when matrix entries are chosen in $\{-1,0,1\}$ and look for a solution with coefficients in $\R$ or $\C$. 
Rather surprisingly, the existence (or not) of such a solution derives straightforwardly from a classical planarity criterion  \cite{ML}. 
\bt\label{thm:ml}  (MacLane 1937): 
A Graph is planar if and only if its cycle space admits a basis in which every edge appears at most
twice.
\et
\smallskip\noindent

This criterion allows us to prove in our setting that planarity is a necessary and sufficient condition for representing the Ising partition function as a single Pfaffian. 

Recently, a generalization of McLane's criterion to non planar graphs has been considered \cite{BD},
from which we draw some new results on non planar graphs. Our main result is the following:
the Ising partition function on a non planar graph can be written as the real part of the Pfaffian of a single matrix, which coefficients are chosen in a multicomplex algebra $\C_{\tilde g}$, where $\tilde g$ is the non-orientable  genus, or crosscap number,  of the embedding surface.

Old and new representations as sums of Pfaffians with real or complex coefficients follow from this result as corollaries. Basically, such sums involve $2^{\tilde g}$ terms rather than $4^g$ (g orientable genus).
For orientably simple graphs, $\tilde g = 2 g +1$, we recover Kasteleyn's results through some further algebraic reduction. For non orientably simple graphs, $\tilde g < 2 g +1$ and our result does improve Kasteleyn's one. However the number of Pfaffians remains exponential in $\tilde g$.

In Section $2$, we define our setting and present some preliminary results that we use later in this work;  when using terminology and concepts from graph theory, we try to follow standard textbooks such as Harary's \cite{H} or Diestel's \cite{D}, to which we refer for a thorough exposition. 
The Pfaffian reduction formula is known since a long time but we found rather few quotations \cite{GH}. It is related, if not equivalent, 
to the more well-known  Pl\"ucker-Grassmann relations between determinants. The connection between minor graphs and Pfaffian
 reduction formula is essential in the present approach as it allows to a substantial reduction of the algebraic problem at hand.
 In Section $3$, we state our main results. In Section $4$, we give the proofs. In Section $5$, we introduce a Grassman
 representation \cite{B, S} which allows us to state the Ising representation problem in terms of a system of algebraic equations. For completeness, we also provide a proof of the Pfaffian reduction formula.

\medskip\noindent
\section {Preliminaries}
\label{sec:2}

The problem we want to address is the representation of the Ising partition function on an  arbitrary finite graph as the Pfaffian of some related matrix.
Hereafter, we recall the first steps in this approach, fix some notations 
and give precise definitions that will be used to state the results of next section.
We conclude this section with some preliminary results.

$\bullet$ Ising Partition Function.

Let $G=(V,E)$ be a finite simple graph. To each vertex $v\in V$, we attach a variable 
$\sigma_v\in\{-1,1\}$. Given a collection of real valued interaction terms, $\underline J = \{J_{e}\ge 0,  e\in E\}$, we define the inhomogeneous Ising Hamiltonian on the space of configurations $\Omega =  \{-1,1\}^{|V|}$, as
\be
\mathcal H(\underline \sigma ) = - \sum_{\{x,y\}\in E} J_{x,y} \sigma_x \sigma_y
\ee

and the associated Ising partition function $Z_{\text{Ising}}^{\beta}(G,\underline J)$ as
\be 
Z_{\text{Ising}}^{\beta}(G,\underline J) = \sum_{\underline \sigma\in \Omega} \exp\{ - \beta \mathcal H (\underline \sigma )\}
\ee

Note that for clarity of the exposition we consider here only simple graphs so that (un-oriented) edges can be identified with (un-ordered) pairs of vertices. Without loss of generality, the graph we will consider  in the sequels are 2-connected simple graphs without loops. In more general cases, the partition function either factorizes or can be trivially rewritten so that one falls back into the previous class of graphs. In the same line of thought, we do not introduce explicitely boundary terms or external fields, which can be considered through a modification of the underlying graph and/or
a particular choice of interaction strength on some edges.
Finally, we also reduce to ferromagnetic interactions ($J_{.}\ge 0$), since we want to keep with the most standard concept of weight functions, but dealing with ``signed weight functions''
would work as well.

The well known high temperature expansion of the Ising partition function leads to an expression of 
$Z_{\text{Ising}}^{\beta}(G,\underline J)$ as a sum over a class of subgraphs of $G$. 

\beq\label{Z:HTE}
Z_{\text{Ising}}^{\beta}(G,\underline J) =  2^{|V|} \Bigl(\prod_{e\in E} \cosh(\beta J_e)\Bigr) \sum_{C\in \mathcal C(G)} \prod_{e\in C} \tanh(\beta J_e)
\eeq
where $ \mathcal C(G)$ is the set of closed curves on $G$, that is the set of all edge subsets $C\subset E$ such that each vertex $v\in V$ is incident with an even number of edges in $C$.

Note that if one introduce  the operation of symmetric difference  $\triangle$:
\be
A \triangle B = \bigl( A\setminus B\bigr) \cap \bigl(B\setminus A\bigr) \quad \hbox{for all sets } A,B
\ee
then $(\mathcal C(G), \triangle)$ is the cycle space of $G$ and has the structure of a vector space
over the field $\F_2 = \Z/{2 \Z}$ .  Under this structure, the set of all cycles on $G$ 
(more precisely their edge sets) is a generating family for $\mathcal C(G)$\cite{D}.

Here we leave aside the physical meaning of \eqref{Z:HTE} and restrict our interest to the representation of the sum in the right hand side. In the sequels, we consider the following closely related quantity, which we still call a partition function for obvious reasons.
\bd\label{def:pfg}
Let $G = (V,E)$ be a finite, non-oriented graph  and $w : E \rightarrow \R ^+$ a weight function defined on the edge set.

We define the {\sl partition function of
  $G$ } with weight function $w$, as
\be\label{def:ZG}
Z_G(w) = \sum_{C \in {\mathcal C}(G)} w(C)
\ee
where the weight of any closed curve $C \subset E$ is defined as the product of its edge weights:
\be
w(C)=
\begin{cases}
1 & \hbox{ if } C=\emptyset\cr
\displaystyle{\prod_{e\in C} w(e)} & otherwise
 \end{cases}
 \ee
 \ed
The partition function $Z_G(w)$ is clearly related to the Ising partition function on the same graph. Choosing the weight function as,
\be
w(e) = \tanh(\frac{\beta J_e}{2})\qquad \forall e\in E
\ee
we have the correspondence
\beq
Z_{\text{Ising}}^{\beta}(G,\underline J) =  2^{|V|} \Bigl( \prod_{e\in E} \cosh(\beta J_e)\Bigr) Z_G(w) 
\eeq

$\bullet$ {\bf Dart graphs and perfect matchings}.

The next step consists in mapping the expression of the (Ising) partition function, onto a weighted dimer problem on an auxiliary graph. Such graphs have been originally called ``terminal graphs'' \cite{K3},
but are named hereafter {\sl dart graphs} in order to emphasize their relation to other classes of derived graphs, such as the  ``line graphs'' \cite{HN}. The choice of this new name stems from the fact that definitions of both classes are identical,   just changing edges (or lines)  by ``half-edges'' or darts. 
 
\begin{definition}
Let $G=(V,E)$ a graph. Its dart graph $\mathcal D(G) = (V_{\mathcal D}, E_{\mathcal D})$ is the 
simple unoriented graph which vertex set $V_{\mathcal D}$  identifies with the set of darts on $G$:
\be
V_{\mathcal D}=\Bigl\{ (v,e) \in V\times E \hbox{ such that $v$ is incident with $e$ }\bigr\}
\ee
and such that there is an edge beween two vertices if and only if they have exactly one common element,
\be
E_{\mathcal D} =\Bigl\{ \bigl\{(v,e),(v',e')\bigr\} \in V_{\mathcal D}\times V_{\mathcal D}\hbox{ such that either }
v=v' \hbox{  or } e=e' \bigr\}
\ee
\end{definition}

A dart graph has necessarily  an even number of vertices and not every graph is the dart graph of some other. 

The edge set of a dart graph $\mathcal D(G)$ splits in two disjoint subsets,
\beq
E_{\mathcal D} &=& E_{\mathcal D}^V \cup E_{\mathcal D}^E,\\
E_{\mathcal D}^V &=&\Bigl\{ \bigl\{(v,e),(v',e')\bigr\} \in V_{\mathcal D}\times V_{\mathcal D}\hbox{ such that }
v=v' \hbox{  and } e\not=e' \bigr\}\\
E_{\mathcal D}^E &=&\Bigl\{ \bigl\{(v,e),(v',e')\bigr\} \in V_{\mathcal D}\times V_{\mathcal D}\hbox{ such that  }
v\not=v' \hbox{  and } e=e' \bigr\}
\eeq
If $G$ has no isolated point, the graph $(V_{\mathcal D},E_{\mathcal D}^V )$ has $|V|$ connected components, each one being a complete graph. 
Moreover, there is a one to one correspondence which associates every edge $e\in E$ with $((v,e),(v',e))\in E_{\mathcal D}^E$ so that both sets can be identified
\be\label{equiv}
E_{\mathcal D}^E\equiv E
\ee

As noted by Kasteleyn \cite{K3}, interest in dart graphs lies in the close connection between the set of closed curves on a graph $G$, and the perfect matchings on $\mathcal D(G)$.
\begin{definition}
Let $G=(V,E)$ be a graph. A perfect matching (or equivalently a dimer configuration ) on $G$ is a subset $E'\subset E$, such that each vertex $v\in V$ is incident with exactly one edge in $E'$. 
\end{definition}

For an arbitrary graph, the existence (or not) of a perfect matching can be proven using Tutte's characterization theorem\cite{Tu}. On a dart graph, it is a straightforward property:
\begin{proposition}\label{prop:rdc}
Let $G=(V,E)$ be a graph. If it is a dart graph, then  it admits at least one perfect matching.
\end{proposition}

Under the identification \eqref{equiv}, the connection between closed curves on a graph $G$ and perfect matchings on its dart graph $\mathcal D(G)$ can be expressed as follows:
\begin{proposition}\label{prop:pm-cc}
Let $M_1$, $M_2$ be any pair of perfect matchings on $\mathcal D(G)$, the set $C = (M_1\triangle M_2)\cap E_{\mathcal D}^E$  is a closed curve on $G$.
\end{proposition}

Denote by $\mathcal M(G)$ the set of perfect matchings on $\mathcal D(G)$. Then, given a fixed perfect matching
$M_0\in \mathcal M(G)$, one can construct a mapping $\Phi_{M_0} :   \mathcal M(G) \rightarrow \mathcal C(G)$, which associates any perfect matching on $\mathcal D(G)$ to a closed curve on $G$, as
\be\label{corresp}
\Phi_{M_0} (M) = (M_0\triangle M)\cap E_{\mathcal D}^E 
\ee

$\Phi_{M_0}$ is 
clearly surjective,
but generally not one-to-one. This induces an equivalence relation on $\mathcal M(G)$ as
\be
M_1\sim M_2 \Longleftrightarrow \Phi_{M_0} (M_1) = \Phi_{M_0} (M_2)\quad \forall M_1, M_2 \in \mathcal M
\ee

On an arbitrary connected graph $G=(V,E)$, the number of closed curves is $2^{\beta_1}$ where $$\beta_1=|E|-|V|+1$$ is the first Betti number of $G$. This is obviously  the same as counting the number of equivalence classes in $\mathcal M(G)$. These equivalence classes have not necessarily the same 
number of elements and enumeration of perfect matchings on a dart graph remains non trivial in the general case.  A particular exception are the $3$-regular graphs, on which the mapping $\Phi_{M_0}$  is one-to-one \cite{F}. 
 
 Using this correspondence, the notion of Pfaffian representation can be given a precise meaning.

$\bullet$ {\bf incidence matrices, weights and Pfaffian representations}.\hfill\break\noindent
We introduce a generalized notion of incidence matrix
 with complex valued coefficients.
 \begin{definition}\label{def:im}
 Let $G=(V,E)$ a connected graph and $\mathcal D(G)$ its dart graph. 
 $A^G \in M_{|V_{\mathcal D}(G)|}(\C)$ is an incidence matrix on $\mathcal D(G)$ if $A^G$ is an antisymmetric matrix  which entries are in one-to-one correspondence with the 
 elements of $V_{\mathcal D}(G)$, so that
 \be\label{def:im:1}
  \{d_1,d_2\}\not\in E_{\mathcal D}(G)\Longrightarrow A^G_{d_1,d_2}  = 0 
 \ee
  \end{definition}

 Note that the coefficients of the matrix need not be taken in $\{-1,0,1\}$, but are arbitrary complex numbers.
 We reserve the name of weighted incidence matrix to the following.
 \begin{definition}
  Let $G=(V,E)$ a connected graph, $M_0$ a fixed perfect matching on $\mathcal D(G)$ and $A^G$ an incidence matrix defined as above. 
  Given a weight function $w$ on $E\equiv E_{\mathcal D}^E(G)$ with value in $\R^{+,*}$, 
  the matrix $A^{G,M_0}(w) \in M_{|V_{\mathcal D}(G)|}(\C)$ with coefficients
  \be\label{def:wm}
  A^{G,M_0}_{d_1,d_2} (w)= 
  \begin{cases}
  w^{-1}(\{d_1,d_2\})  A^G _{d_1,d_2} & \hbox{ if } \{d_1,d_2\} \in E_{\mathcal D}^E(G) \cap M_0\cr
  w(\{d_1,d_2\})  A^G _{d_1,d_2} & \hbox{ if } \{d_1,d_2\} \in E_{\mathcal D}^E(G)  \setminus M_0\cr
  A^G _{d_1,d_2} & \hbox{ otherwise.}
  \end{cases}
  \ee
  is the weighted incidence matrix associated to $(G,M_0, A^G,w)$.
 \end{definition}

We now can state a precise definition of a Pfaffian representation:

\begin{definition}\label{def:pfaffrep}
Let $G=(V,E)$ a connected simple graph. The partition function on $G$ admits a Pfaffian representation if there exists
a perfect matching $M_0$ on $\mathcal D(G)$, an incidence matrix $A^G$ and a constant $\Lambda\not = 0$
such that for all weight functions $w:  E \rightarrow \R^+$, one has
\be\label{def:pfaffrep:1}
Z_G(w) = \frac{1}{\Lambda} \;w( M_0\cap E_{\mathcal D}^E(G) ) \;\Pf\bigl(A^{G,M_0}(w)\bigr)
\ee
\end{definition}

The main motivation for such a definition is the fact that the Pfaffian of a weighted incidence matrix $A^{G,M_0}(w)$, can be always written as a sum over all closed curves on $G$, so that the identity \eqref{def:pfaffrep:1} derives from identification of the coefficients of two similar expansions.
Starting from the definition of a Pfaffian and setting $|V_{\mathcal D}(G)| = 2 n$, $n\in \N$, we have

\be\label{pfaffAexp}
Pf({ A^{G,M_0}(w)}) = {1\over 2^n\,n! } \sum_\sigma (-1)^{|\sigma|}
 A^{G,M_0}_{\sigma(d_1),\sigma(d_2)}(w)
  \cdots A^{G,M_0}_{\sigma(d_{2 n -1}),\sigma(d_{2 n})} (w)
\ee
where the summation runs over the set $S_{V_{\mathcal D} (G)}$  of all permutations on $V_{\mathcal D} (G)$. Taking into account the above definition of an incidence matrix, each non zero term can be associated to a perfect matching on $\mathcal D (G)$ and, collecting all edge weights,  the expansion can be written as
 
\beq\label{pfaffAexp1}
&&Pf({ A^{G,M_0}(w)}) = \sum_{M\in\mathcal M(G)}{1\over 2^n\, n ! } \sum_{\sigma\in \Pi(M)} (-1)^{|\sigma|}
 A^{G}_{\sigma(d_1),\sigma(d_2)}
  \cdots A^{G}_{\sigma(d_{2 n -1}),\sigma(d_{2 n})} \nonumber\\
  &&\phantom{Pf({ A^{G,M_0}(w)}) = \sum_{M\in\mathcal M(G)}{1\over 2^n} {1\over ({n\over 2})! } \sum_{\sigma\in \Pi(M)} }
  \times
  w(M\cap (E_{\mathcal D}^E(G)\setminus M_0)) w^{-1}(M\cap E_{\mathcal D}^E(G)\cap M_0)) 
\eeq
where $ \Pi(M)$ is the set of $2^n\, n! $ permutations of $S_{V_{\mathcal D} (G)}$  contributing to the same perfect matching $M$,
\be
 \Pi(M) =\bigl\{ \sigma \in S_{V_{\mathcal D} (G)}, \{\{\sigma(d_1),\sigma(d_2)\}, \cdots ,
 \{\sigma(d_{2 n -1}),\sigma(d_{2 n})\}\} = M \bigr\}
 \ee
As a consequence of the antisymmetry of $A^G$, the $2^n\, n! $ terms of the summation on $\Pi(M)$ are equal for every $M\in \mathcal M(G)$.
Now the weights appearing in the right hand side of \eqref{pfaffAexp1} can be rewritten as
\beq
&&w(M\cap (E_{\mathcal D}^E(G)\setminus M_0))\; w^{-1}(M\cap E_{\mathcal D}^E(G)\cap M_0)) \nonumber\\
&&\qquad\qquad\qquad\qquad= w^{-1}(M_0\cap E_{\mathcal D}^E(G)) \; w( E_{\mathcal D}^E(G)\cap (M\triangle M_0))
\eeq

and the expression \eqref{pfaffAexp1} can be turned into a weighted sum over the closed curves in $\mathcal C(G)$,
\be\label{pfaffAexp2}
Pf({ A^{G,M_0}(w)}) = w^{-1}(M_0\cap E_{\mathcal D}^E(G)) \sum_{C\in \mathcal C(G)} w(C) F_{A^G}(M_0, C)
\ee

where for all $C\in \mathcal C(G)$ and any reference perfect matching $M_0$, $F_{A^G}(M_0, C)$
is the sum over all perfect matchings mapped to  $C$ by $\Phi_{M_0}$,
\be\label{FA}
F_{A^G}(M_0, C) = 
\sum_{M\in \Phi_{M_0}^{-1}(C)}{1\over 2^n\, n! } \sum_{\sigma\in \Pi(M)} (-1)^{|\sigma|}
 A^{G}_{\sigma(d_1),\sigma(d_2)}
  \cdots A^{G}_{\sigma(d_{2 n -1}),\sigma(d_{2 n})} 
\ee

Using this expression and the vector space structure of $ \mathcal C(G)$, we have the following characterization, 

\bl \label{lem:fa}
The partition function of $G$ admits a Pfaffian representation if and only if there exists an incidence matrix $A^G$,
a reference dimer configuration $M_0$ and 
a cycle basis ${\mathcal B}_G$ on $(\mathcal C(G), \triangle)$ such that $\forall C\in \mathcal C(G)$, $F_{A^G}(M_0,C)\not=0$ and
\be\label{syseqr}
F_{A^G}(M_0, C) = F_{A^G}(M_0, C\triangle \gamma) \qquad \forall C\in  {\mathcal C}(G),\;
 \forall \gamma \in  {\mathcal B}_G
\ee
\el

\begin{remark}
A change of reference perfect matching  just induces a shift in the mapping for any two reference mappings $M_0, M_1 \in \mathcal M(G)$, one has
\be\label{mapping}
\Phi_{M_1} (M) =   \Phi_{M_0} (M)  \,\triangle\, \Phi_{M_1} (M_0)  \quad \forall M \in \mathcal M(G) 
\ee
 The set of equations  \eqref{syseqr}
 remains globally invariant under such a shifh and the existence (or not) of a Pfaffian representation \eqref{def:pfaffrep:1} is independent on the choice of the reference perfect matching. 
\end{remark}

\begin{remark}
When all vertices of $G$ have an even degree, the reference perfect matching $M_0$ can be chosen so that $E_{\mathcal D}^E(G)\cap M_0 =\emptyset$, the dependence on the weight function in definitions \eqref{def:wm} and \eqref{def:pfaffrep:1} simplifies and more classical expressions are recovered. In the general case, such a reference perfect matching does not exist and Proposition \ref{prop:rdc} suggests to take instead  $E_{\mathcal D}^E(G)$ as a reference perfect matching.
See also reference \cite{K3} for an alternative approach in the later case.
 \end{remark}

We end this section with a relation between the notion of minor graph 
and the Pfaffian reduction formula.

 $\bullet$ {\bf Minor graphs and Pfaffian reduction formula}.
 
 Informally a graph $G_1$ is a minor of  another graph $G_2$ (noted  $G_1\preccurlyeq G_2$ ) if it can be obtained from it by repeating in some arbitrary order the two following elementary transformations: (a) deleting one edge; (b) contracting one edge, i.e. identifying its two end-vertices  into a single vertex and deleting any loop or parallel edges that may arise. We refer to  \cite{D} for a more formal definition. 

\begin{figure}[htbp]
\begin{center}
\includegraphics[scale=.5,  angle = - 90, trim=  4cm 0cm 4cm 0cm, clip=true ]{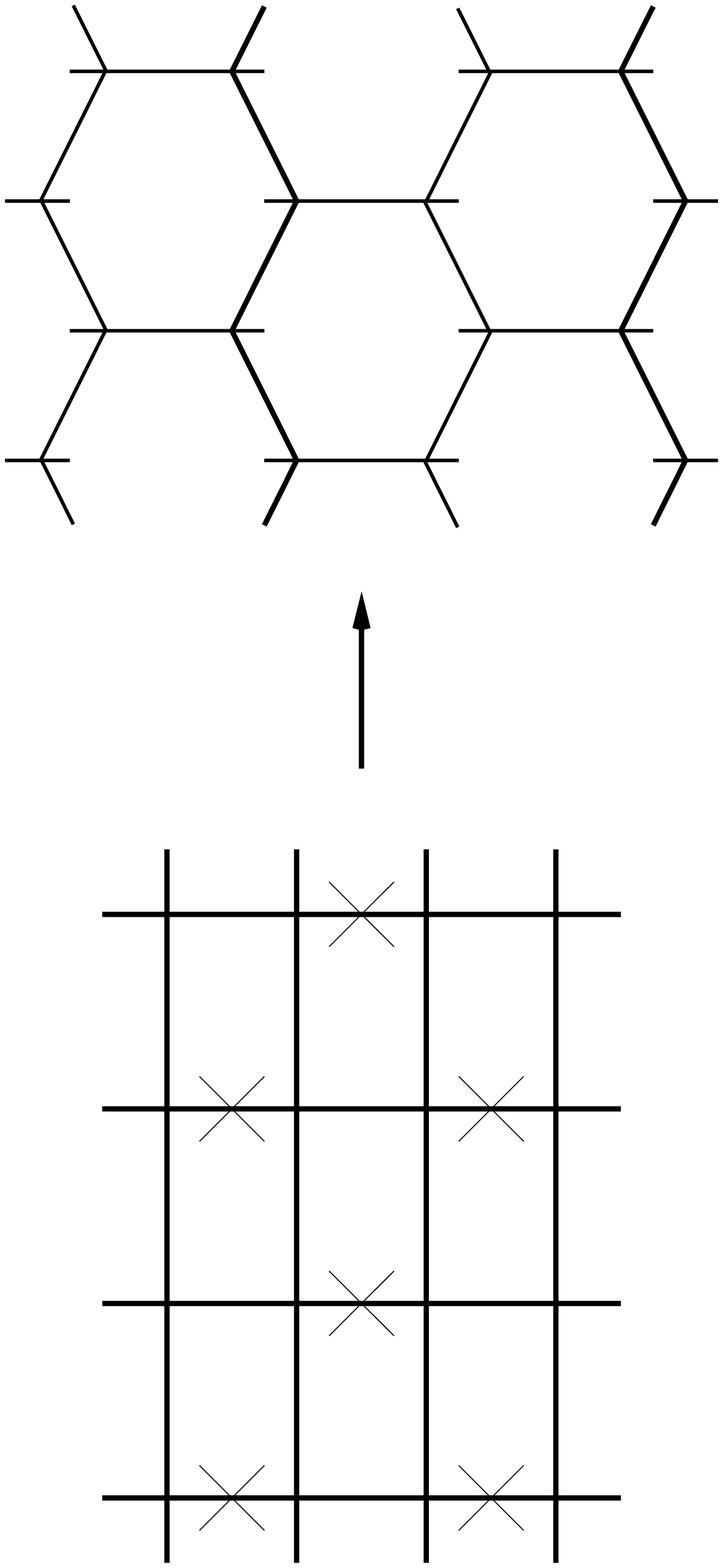}
\end{center}
\caption{Example of minor graph obtained by deleting a subset of edges:
an hexagonal lattice is formed out of a square lattice.
}
\label{figure:1}
\end{figure}

\begin{figure}[htbp]
\begin{center}
\includegraphics[scale=.5,  angle = - 90, trim=  4cm 0cm 4cm 0cm, clip=true]{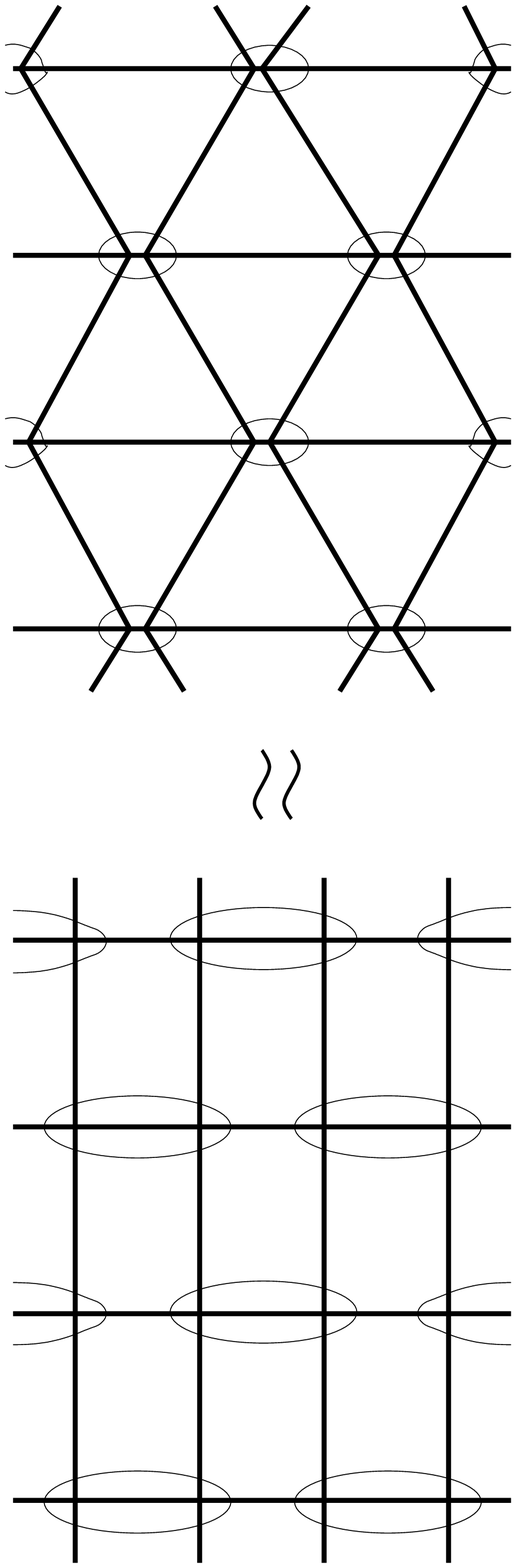}
\end{center}
\caption{Example of minor graph obtained by contracting a subset of edges:
a triangular lattice is formed out of a square lattice.}
\label{figure:1t}
\end{figure}

In the next two figures, two simple examples of these transformations are shown, which have some relevance in the present context: starting from a planar square lattice, one may form an hexagonal lattice by deleting one edge per site (Figure \ref{figure:1}),  while the contraction
of the same set of edges leads to the triangular lattice (Figure \ref{figure:1t}).

On the other side, the Pfaffian reduction formula \cite{GH} is a well known property of Pfaffians and  can be stated as follows. 

Let $A = \bigl( A_{ij}\bigr)_{i,j \in \{1,\dots,2 n\}}$ a matrix of order $2n$ and $K$ a subset of the set of indices $\{1,\dots,2 n\}$. We denote 
by $A_K$ the sub-matrix of $A$ obtained by deleting all
rows and columns not indexed by an element of $K$. 

\be\label{submatrix}
A_K = \bigl( A_{ij}\bigr)_{i,j \in K}
\ee
Similarly, we denote by $A^{\bar K}$ the antisymmetric matrix indexed by the complementary set of 
indices $\bar K = \{1,\dots, 2 n\} \setminus K$,
which elements are:
\be\label{supmatrix}
\bigl( A^{\bar K} \bigr)_{ij} = - \bigl( A^{\bar K} \bigr)_{ji} = \Pf\bigl( A_{(K \cup \{i,j\})} \bigr)\qquad \hbox{for all } i <  j \in \bar K
\ee
We stress that in definitions \eqref{submatrix}--\eqref{supmatrix}, the order on any subset of indices is the one induced by the order on $\{1,\dots,2 n\}$. Pfaffian reduction formula reads:
\bp\label{hiera}
Let $A$ an antisymmetric matrix of order $2n$ and $K$ a subset of indices 
of order $2p$, $0 < p < n$ such that $\Pf ( A_K )\not= 0$.  We have:
\be\label{eq-hiera}
\Pf(A) =\bigl[ \Pf ( A_K ) \bigr]^{-(n-p-1)} \Pf (A^{\bar K})
\ee
\ep

The connection between minor graphs and Pfaffian reduction formula is as follows:
\bp\label{prop:mi_pf}
Suppose that  $G_1=(V_1,E_1)$ and $G_2=(V_2,E_2)$ are two simple graphs such that $G_1\preccurlyeq G_2$. Consider a transformation which reduces $G_2$ to $G_1$ and denote by $E_2^d$ (respectively $E_2^c$ ) the set of deleted (respectively contracted) edges in this transformation.
Let $A^{G_2}$ be an incidence matrix (in the sense of \ref{def:im} ) on $\mathcal D(G_2)$ such that 
\be\label{prop:mi_pf:1}
A^{G_2}_{d_1,d_2}=0 \hbox{ if either $d_1$ or $d_2$ is incident with an edge in } E_2^d
\ee

Let $K$ be the set of vertices in $\mathcal D(G_2)$ incident with an edge in $E_2^d\cup E_2^c$. 
Then the set $\bar K = V_{\mathcal D}(G_2) \setminus K$ can be identified with $V_{\mathcal D}(G_1)$,
and the matrix  $\bigl[A^{G_2}\bigr]^{\bar K}$ defined as in \eqref{supmatrix} is an incidence matrix on $\mathcal D(G_1)$.
\ep
In the above proposition, edge contraction (respectively edge deletion) is given a counterpart as a transformation of the incidence matrix
associated to the dart graph. Contracting one edge means simply integrating out (using the Pfaffian reduction formula) the 
corresponding matrix entries.  Deleting one edge may have various representations depending on the choice of the reference 
dimer configuration. Here we take $M_0= E_{\mathcal D}^E (G)$, which is  consistent with the general case (proposition \ref{prop:rdc}),  and edge deletion corresponds to
factorizing out the associated edge weight. Note in particular that by Proposition \ref{hiera}, both Pfaffians are proportional in some functional sense.
\br\label{rem:fisher}
The minor relation is a partial order on the set of graphs, and in particular the genus of a graph is always larger or equal than the genus of any of its minors. This property does not transfer to the associated dart graphs and $\mathcal D(G_1)$ may have a larger genus than  $\mathcal D(G_2)$ even if  $G_1\preccurlyeq G_2$. 
\er

\medskip\noindent
\section {Main results.}
\label{sec:3}

In this section, we state our main results on the Pfaffian representations of  partition functions on a graph $G$, starting from definitions  \ref{def:pfg} and \ref{def:pfaffrep}.  

Our first result is the well known theorem due to Kasteleyn \cite{K3}.

\bt\label{thm:planar}
The partition function on a graph $G$ admits a Pfaffian representation if and only if $G$ is planar.
\et

The proof presented here is independent on the orientability criterion used by Kasteleyn in his original proof. Instead, we emphasize  a connection between this representation problem and McLane's planarity criterion, Theorem \ref{thm:ml}, which is interesting in its own right. Incidentally, we prove also that in contradistinction with the dimer problem, there is no ``Pfaffian'' non-planar graph for the Ising model. 

Furthermore, the approach presented here allows for some new representations of the partition function when $G$ is a non-planar graph.  For all positive integer $n$, let $\C_n$ be the multicomplex algebra of order $n$ \cite{BP} and  $\operatorname{Re}$ the linear operator $\C_n\rightarrow \R$ which associates  to any element its real part. Our main result is the following:
\bt\label{thm:mc}
Let $G=(V,E)$ be a graph of nonorientable genus $\tilde g$. There exist  an incidence matrix $A^G$ on $\mathcal D(G)$ (in the sense of \ref{def:im} ) 
with coefficients in $\C_{\tilde g}$, a constant $\Lambda\in \C_{\tilde g}$  and a reference perfect matching $M_0$  such that for all edge weight functions on $G$, the partition function on $G$ can be written as
\be\label{ZNP}
 Z_G(w) =  \;w( M_0\cap E^E_{\mathcal D}(G)) \;\operatorname{Re}\bigl[\Lambda\;\Pf\bigl(A^{G,M_0}(w)\bigr) \bigr]
\ee
\et

Note that the above expression depend on the non orientable genus \cite{MT}, that is the minimal number of crosscaps a surface should have to embed  $G$ without edge crossings. 

In the present work,  we limit ourselves to the simplest application of Equation \eqref{ZNP}, that is the derivation of an expression of the partition function of a nonplanar graph  as a sum of Pfaffians. The first one is in terms of the non orientable genus and involves matrices with complex coefficients:
\bc\label{cor:sum_no}
Let $G=(V,E)$ be a graph of nonorientable genus $\tilde g$. There exist
a family of  incidence matrices $(A_{j}^{G})_{j\in\{1,\cdots,2^{\tilde g}\}}$ on $\mathcal D(G)$
(in the sense of \ref{def:im} ) 
with coefficients in $\C$, constants $(\Lambda_j)_{j\in\{1,\cdots,2^{\tilde g}\}}$  in $\C$  and a reference perfect matching $M_0$  
 such that for all edge weight functions on $G$, the partition function on $G$ can be written as
\be\label{cor:sum_noe}
 Z_G(w) =  \;w( M_0\cap E^E_{\mathcal D}(G)) \operatorname{Re}\bigl[\sum_{j=1}^{2^{\tilde g}} \Lambda_j\;\Pf\bigl(A_{j}^{G,M_0}(w)\bigr) \bigr]
\ee
\ec
A similar expression has been derived by Tessler \cite{Te}, building on Kasteleyn's approach.
The nonorientable genus $\tilde g$ of a graph $G$ is related to its orientable genus $g$ through the inequality \cite{St}
\be\label{nog_og}
\tilde g \le 2 g +1
\ee
When the graph is non orientably simple (e.g. projective grids),  the non-orientable genus can be much smaller than this bound, so that the expansion \ref{cor:sum_noe}
is a real improvement over an expansion in terms of the orientable genus.   

A proof of inequality \eqref{nog_og} comes from the fact that
starting from any orientable embedding in a surface of genus $g$, one can construct a
nonorientable embedding of nonorientable genus $2 g +1$.
 We give an explicit construction on a surface with $2 g +1$ crosscaps, 
 showing that every edge passes through them an even number of times. As a consequence, both entries of the incidence matrix and constant that appear in Theorem \ref{thm:mc}
 can be taken in the subalgebra of $\C_{2 g+1}$ generated by even products of generators.
This leads to an expansion in terms of $2^{2 g}$ Pfaffians of matrices with real coefficients, equivalent to the one stated first by Kasteleyn.

\bc\label{cor:sum_o}
Let $G=(V,E)$ be a graph of orientable genus $g$. There exist
a family of  incidence matrices $(A_{j}^{G})_{j\in\{1,\cdots,4^{ g}\}}$ on $\mathcal D(G)$
(in the sense of \ref{def:im} ) 
with coefficients in $\R$, constants $(\Lambda_j)_{j\in\{1,\cdots,4^{ g}\}}$  in $\R$  and a reference perfect matching $M_0$  
 such that for all edge weight functions on $G$, the partition function on $G$ can be written as
\be\label{cor:sum_oe}
 Z_G(w) =  \;w( M_0\cap E^E_{\mathcal D}(G)) \sum_{j=1}^{4^{g}} \Lambda_j\;\Pf\bigl(A_{j}^{G,M_0}(w)\bigr) 
\ee
\ec

In the rest of the section, we give a sketch of the proof of Theorem \ref{thm:planar}, explain the connection with McLane's criterion and prepare for the derivation of Theorem \ref{thm:mc}. 
The first step is built on Proposition \ref{prop:mi_pf} and is the following

\bt\label{thm:mi_pf}
Let  $G_1$ and $G_2$ be two simple graphs such that $G_1\preccurlyeq G_2$ and $A^{G_2}$ an incidence matrix on $\mathcal D(G_2)$.
There exists an incidence matrix $A^{G_1}$ on $\mathcal D(G_1)$,  a constant $\Lambda_{1,2}$ and two reference perfect matchings, $M_1$ on  $\mathcal D(G_1)$ and $M_2$ on  $\mathcal D(G_2)$,
such that for every closed curve $C$ in $\mathcal C(G_1)$,
\be 
F_{A^{G_1}} (M_1, C) = \Lambda_{1,2} \; F_{A^{G_2}} (M_2, \Pi(C)) 
\ee
where $\Pi : \mathcal C(G_1) \rightarrow \mathcal C(G_2)$ is the canonical mapping induced by the transformation from $G_2$ to its minor $G_1$.

In particular, if the partition function on $G_2$ admits a Pfaffian representation 
then so does the partition function on $G_1$.

\et
Theorem \ref{thm:mi_pf} allows here to reduce the class of graphs one has to consider to prove both necessity and sufficiency of the planarity condition.
Incidentally, it also clarifies a point raised by Fisher 
long ago\cite{F}, related to the fact that a planar graph has a non planar dart graph if some of its vertices are incident with more than three edges. In such a case, the related dimer problem is no longer solvable
and two approaches have been proposed:  In Kasteleyn approach \cite{K3}, one keeps the original graph,
and has to argue that the multiplicities encountered are exactly compensated by some sign rule; In Fisher's approach, one deals with a larger, $3-$regular graph so that the dart graph is also planar, and reduces 
later to the original graph. A consequence of Theorem \ref{thm:mi_pf} is that both approaches are not only always simultaneously possible, but that Fisher's planar dimer construction implies Kasteleyn's sign rule.

In the course of the proof of Theorem \ref{thm:planar}, we use Theorem \ref{thm:mi_pf} twice.
On one hand, the proof that planarity is a necessary condition is reduced to an application of  
Kuratowski's planarity criterion \cite{K}, starting from the following constatation:

\bp\label{prop:K5K33}
The partition functions on the complete bipartite graph $K_{3,3}$ and on the complete graph $K_5$  have no Pfaffian representation.
\ep

On the other hand, Theorem \ref{thm:mi_pf}  and the following lemma allows to arbitrarily restrict the proof of sufficiency to $4$-regular graphs,
\bl\label{lem:4r}
Given a $2$-connected simple  graph $G$ embedded in some surface $\mathcal S$, there exists a $2$-connected $4$-regular simple graph $\tilde G$ embeddable in the same surface  such that $G\preccurlyeq \tilde G$.
\el 
Note that  $4$-regular graphs  are just chosen for convenience.
We are left with the following 
\bp\label{prop:planar4r}
Let $G=(V,E)$ be a $4$-regular, $2$-connected simple graph with no loops.
If $G$ is planar, then its partition function admits a Pfaffian representation.
\ep

Proof of Proposition \ref{prop:planar4r} consists in two parts. First, reduction to $4$-regular graphs helps us to write Lemma \ref{lem:fa} as a set of algebraic equations in the incidence matrix entries, depending on 
the choice of the cycle basis. We prove next that McLane criterion ( Theorem \ref{thm:ml} ) is just what is needed to pick  a cycle basis so that these algebraic equations do have a consistent solution. First, we introduce some notations.

Let $G=(V,E)$ be as in Proposition \ref{prop:planar4r}.
The vertex set of its dart graph $\mathcal D(G)$ is
 \be
 V_{\mathcal D} (G) =\bigl\{ (v,e), v\in V, e\in E(v)\bigr\}
 \ee
 where $E(v)\subset E$ is the set of four edges at vertex $v\in V$.
 We suppose that the sets $V$ and $E(v)$ for all $v\in V$ have been ordered in some arbitrary way and for all $v\in V$ we denote by $e_v^k$ the $k^{th}$ element of $E(v)$. On  $V_{\mathcal D} (G)$, we consider the induced lexicographic order. 
 \be
 (v,e) \le (v',e') \Longleftrightarrow 
 \begin{cases}
 v <v'\cr
 \hbox{or}\cr
 v= v' \hbox{ and } e\le e'
 \end{cases}
 \forall ((v,e), (v',e'))\in V_{\mathcal D} (G)\times V_{\mathcal D} (G)
\ee
 
 Since there is an even number of edge at each vertex, there exists a perfect matching $M_0$ on $\mathcal D(G)$  such that
$M_0\cap E = \emptyset$.  Actually there are $3^{|V|}$ such matchings and we take arbitrarily the following  one as our  reference perfect matching,
\be\label{def:RPM}
M_0 =\bigcup_{v\in V} \bigl\{ \bigl\{(v,e_v^1),(v,e_v^2)\bigr\},\bigl\{(v,e_v^3),(v,e_v^4)\bigr\}\bigr\}
 \ee
 We construct an incidence matrix $A^G$ of order $|V_{\mathcal D}(G)|= 4 |V|$.

\be\label{def:A}
{\bf A^G} = \bigl(A^G_{(v,e),(w,f)}\bigr)_{(e,f)\in E(v)\times E(w)
\atop (v,w) \in V\times V\hfill }
\ee
By Definition \ref{def:im}, entries of $A^G$ are zero except those corresponding to an edge in $\mathcal D(G)$.
If the edge is in $E_{\mathcal D}^E(G)$ (respectively in $E_{\mathcal D}^V(G)$), $e=f$ (respectively $v=w$), the coefficent $A^G_{(v,e),(w,f)}$ is termed an ``edge entry'' (respectively a ``site entry'').

Recalling that $A^G$ is an antisymmetric matrix, we rename all its edge and site entries as follows.
For every edge $e$ in $E$ with endvertices $v$ and $w$ such that $v<w$, we write
\be\label{6a}
b_e =
A^G_{(v,e),(w,e)}
\ee

while for every vertex $v\in V$ and every permutation $\sigma$ on $E(v)$,
we define

\beq\label{6b}
&&S_v(\sigma)=A^G_{(v,\sigma(e_v^{\scriptstyle{1}})),(v,\sigma(e_v^{\scriptstyle{2}}))}
\qquad
\bar S_v(\sigma)=A^G_{(v,\sigma(e_v^{\scriptstyle{3}})),(v,\sigma(e_v^{\scriptstyle{4}}))} \nonumber\\
&& T_v(\sigma)=A^G_{(v,\sigma(e_v^{\scriptstyle{1}})),(v,\sigma(e_v^{\scriptstyle{3}}))}
\qquad
\bar T_v(\sigma)=A^G_{(v,\sigma(e_v^{\scriptstyle{2}})),(v,\sigma(e_v^{\scriptstyle{4}}))}\\
&&
U_v(\sigma)=A^G_{(v,\sigma(e_v^{\scriptstyle{1}})),(v,\sigma(e_v^{\scriptstyle{4}}))}
\qquad
\bar U_v(\sigma)=A^G_{(v,\sigma(e_v^{\scriptstyle{2}})),(v,\sigma(e_v^{\scriptstyle{3}}))} \nonumber
\eeq

When $\sigma$ is the identity, we simplify the notations further as
\beq\label{def:se}
&&s_v = S_v(\underline 1) \qquad 
t_v = T_v(\underline 1) \qquad 
u_v = U_v(\underline 1)\nonumber\\
&& \bar s_v = \bar S_v(\underline 1) \qquad 
\bar t_v =  \bar T_v(\underline 1) \qquad 
 \bar u_v = \bar U_v(\underline 1)
\eeq

Now let $\gamma=(V_\gamma, E_\gamma)$ be a cycle of length $r_\gamma$ on $G$. We consider a cyclic order on both its edge and vertex sets, 
\beq
 V_{\gamma} &=&(v^{{\gamma}}_i)_{i\in \{1,\cdots,r_\gamma\}}
\label{vgamma}\\
\label{egamma}
E_{\gamma} &=& (e^{{\gamma}}_i)_{i\in \{1,\cdots,r_\gamma\}}
\eeq
where the indices are defined modulo $r_\gamma$ so that $v^{{\gamma}}_i$  is incident with $e^{{\gamma}}_{i-1}$and $e^{{\gamma}}_{i}$ for all $i \in \{1,\cdots,r_\gamma\}$. 

We want to relate both the order on $V_{\mathcal D} (G)$ and the order along $\gamma$ and thus introduce $\gamma$-related notations.

For all $i$ in $\{1,\cdots,r_\gamma\}$, we rename the edge entry in $\bf A^G$ associated to $e^\gamma_i$ as,
\be\label{def:eeg}
B^\gamma_i =A_{(v^\gamma_i,e^\gamma_i),(v^\gamma_{i+1},e^\gamma_i)}
\ee

Note that $B^\gamma_i$ is equal to either  $+ b_{e^\gamma_i}$ or $- b_{e^\gamma_i}$ according to whether $\gamma$ goes through edge $e^\gamma_i$ compatibly or not with the order on $V_{\mathcal D}(G)$. In order to rename the site entries along $\gamma$, we consider the following

\smallskip\noindent
\bl\label{lem:perm}
Let $\gamma$ a cycle of length $r_\gamma$ ordered as above. For every $i \in \{1,\cdots, 
r_\gamma\}$, there exists a unique permutation $\sigma^{{\gamma}}_i$
on $E(v_i^\gamma)$ with positive signature such that,
\beq
e^{\gamma}_{i-1} &=& \sigma^{\gamma}_i (e_{v^{\gamma}_i}^1)\\
e^{\gamma}_{i} &= & \sigma^{\gamma}_i (e_{v^{\gamma}_i}^4)
\eeq
\el
\smallskip
The effect of the $\sigma^{\gamma}_i$'s is to reorder the edges on each site visited by the cycle, so that it enters a site by the first edge and leave by the forth. When considering a given cycle $\gamma$, we also simplify notations \eqref{6b} as
\beq\label{def:seg}
&&S_i^\gamma = S_{v_i^{\gamma}}(\sigma^{\gamma}_i) \qquad 
 T_i^\gamma = T_{v_i^{\gamma}}(\sigma^{\gamma}_i)\qquad 
 U_i^\gamma = U_{v_i^{\gamma}}(\sigma^{\gamma}_i)\nonumber\\
&&\bar S_i^\gamma = \bar S_{v_i^{\gamma}}(\sigma^{\gamma}_i) \qquad 
 \bar T_i^\gamma =  \bar T_{v_i^{\gamma}}(\sigma^{\gamma}_i) \qquad 
  \bar U_i^\gamma = \bar U_{v_i^{\gamma}}(\sigma^{\gamma}_i)
\eeq
The relations between notations \eqref{def:se} and \eqref{def:seg} depend on the actual value of  $\sigma^{\gamma}_i$ among twelve possible realizations and are listed in Table \ref{table:1}.

\begin{table}[htb] 
\begin{center}
\hskip.0truecm\vbox{\offinterlineskip
\hrule
\halign{\vrule#&\strut\enskip\hfil#\hfil&\enskip\hfil#\hfil&\enskip\hfil#\hfil&\enskip\hfil#\hfil\enskip&&\vrule#&\strut\quad\hfil#\quad\cr
height4pt&\omit&\omit&\omit&\omit& &\omit& &\omit& &\omit& &\omit& &\omit& &\omit&\cr
&$\sigma_i^\gamma(e_v^1)$&$\sigma_i^\gamma(e_v^2)$&$\sigma_i^\gamma(e_v^3)$&$\sigma_i^\gamma(e_v^4)$& &$S_i^\gamma$& &$\bar S_i^\gamma$& &$T_i^\gamma$& &$\bar T_i^\gamma$& &$U_i^\gamma$& &$\bar U_i^\gamma$& \cr
height4pt&\omit&\omit&\omit&\omit& &\omit& &\omit& &\omit& &\omit& &\omit& &\omit&\cr
\noalign{\hrule}
height4pt&\omit&\omit&\omit&\omit& &\omit& &\omit& &\omit& &\omit& &\omit& &\omit&\cr
&$e_v^1$&$e_v^2$&$e_v^3$&$e_v^4$& &          $s_v$& &  $\bar s_v$& &         $t_v$& & $\bar t_v$& &          $u_v$& &  $\bar u_v$& \cr
height4pt&\omit&\omit&\omit&\omit& &\omit& &\omit& &\omit& &\omit& &\omit& &\omit&\cr
&$e_v^2$&$e_v^1$&$e_v^4$&$e_v^3$& &        $-s_v$& &$-\bar s_v$& &$-\bar t_v$& &        $-t_v$& & $\bar u_v$& &          $u_v$& \cr
height4pt&\omit&\omit&\omit&\omit& &\omit& &\omit& &\omit& &\omit& &\omit& &\omit&\cr
&$e_v^3$&$e_v^4$&$e_v^1$&$e_v^2$& & $\bar s_v$& &          $s_v$& &        $-t_v$& &$-\bar t_v$& &$-\bar u_v$& &        $-u_v$& \cr
height4pt&\omit&\omit&\omit&\omit& &\omit& &\omit& &\omit& &\omit& &\omit& &\omit&\cr
&$e_v^4$&$e_v^3$&$e_v^2$&$e_v^1$& &$-\bar s_v$& &       $-s_v$& &  $\bar t_v$& &         $t_v$& &        $-u_v$& &$-\bar u_v$& \cr
height4pt&\omit&\omit&\omit&\omit& &\omit& &\omit& &\omit& &\omit& &\omit& &\omit&\cr
\noalign{\hrule}
height4pt&\omit&\omit&\omit&\omit& &\omit& &\omit& &\omit& &\omit& &\omit& &\omit&\cr
&$e_v^1$&$e_v^3$&$e_v^4$&$e_v^2$& &$t_v$& &  $\bar t_v$& &  $u_v$& & $\bar u_v$& &  $ s_v$& & $\bar s_v$& \cr
height4pt&\omit&\omit&\omit&\omit& &\omit& &\omit& &\omit& &\omit& &\omit& &\omit&\cr
&$e_v^3$&$e_v^1$&$e_v^2$&$e_v^4$& &$-t_v$& &$-\bar t_v$& & $-\bar u_v$& &$-u_v$& &$\bar s_v$& &  $s_v$& \cr
height4pt&\omit&\omit&\omit&\omit& &\omit& &\omit& &\omit& &\omit& &\omit& &\omit&\cr
&$e_v^4$&$e_v^2$&$e_v^1$&$e_v^3$& & $\bar t_v$& &$t_v$& &$-u_v$& &$-\bar u_v$& &  $-\bar s_v$& &$-s_v$& \cr
height4pt&\omit&\omit&\omit&\omit& &\omit& &\omit& &\omit& &\omit& &\omit& &\omit&\cr
&$e_v^2$&$e_v^4$&$e_v^3$&$e_v^1$& &$-\bar t_v$& &$-t_v$& &$\bar u_v$& &  $u_v$& &$-s_v$& &$-\bar s_v$& \cr
height4pt&\omit&\omit&\omit&\omit& &\omit& &\omit& &\omit& &\omit& &\omit& &\omit&\cr
\noalign{\hrule}
height4pt&\omit&\omit&\omit&\omit& &\omit& &\omit& &\omit& &\omit& &\omit& &\omit&\cr
&$e_v^1$&$e_v^4$&$e_v^2$&$e_v^3$& &  $u_v$& &$\bar u_v$& &$s_v$& & $\bar s_v$& & $t_v$& & $\bar t_v$& \cr
height4pt&\omit&\omit&\omit&\omit& &\omit& &\omit& &\omit& &\omit& &\omit& &\omit&\cr
&$e_v^4$&$e_v^1$&$e_v^3$&$e_v^2$& &$-u_v$& &$-\bar u_v$& &$-\bar s_v$& &$-s_v$& &$\bar t_v$& &  $t_v$& \cr
height4pt&\omit&\omit&\omit&\omit& &\omit& &\omit& &\omit& &\omit& &\omit& &\omit&\cr
&$e_v^2$&$e_v^3$&$e_v^1$&$e_v^4$& &$\bar u_v$& & $u_v$& &$-s_v$& &$-\bar s_v$& &$-\bar t_v$& &$-t_v$& \cr
height4pt&\omit&\omit&\omit&\omit& &\omit& &\omit& &\omit& &\omit& &\omit& &\omit&\cr
&$e_v^3$&$e_v^2$&$e_v^4$&$e_v^1$& &$-\bar u_v$& &$-u_v$& &$\bar s_v$& &$s_v$& &$-t_v$& &$-\bar t_v$& \cr
height4pt&\omit&\omit&\omit&\omit& &\omit& &\omit& &\omit& &\omit& &\omit& &\omit&\cr
}
\hrule}
\medskip
 \caption{Correspondence between the 6 site entries in the upper triangular part of the incidence matrix for a given vertex $v$ (Equation \eqref{def:se}), and the new names (Equation \eqref{def:seg}) after reordering of $E(v)=\{e_v^1,e_v^2,e_v^3,e_v^4\} $ according to a cycle $\gamma$ passing through $v$
 ($=v_i^\gamma$ for some index $i$).
 $\sigma_i^\gamma$ is the even permutation on $E(v)$ such that  $\gamma$ enters $v$ through edge $\sigma_i^\gamma(e_v^1)$, and leaves it through $\sigma_i^\gamma(e_v^4)$ (Lemma \ref{lem:perm}).}
 \label{table:1}
\end{center}
\end{table}  

Using these notations, we can state the following proposition:
\smallskip\noindent
\bp\label{prop:cycle_eq}
Let $\gamma$ a cycle on $G$ of length $r_\gamma$ and notations as above. 
If the entries of the incidence matrix $\bf A^G$ associated to elements of $\gamma$ have non zero values and verify the following set of equations:
\beq\label{seq}
&& S^\gamma_i {\bar S}^\gamma_i +
T^\gamma_i {\bar T}^\gamma_i = 0\qquad \forall i \in \{1,\cdots,r_\gamma\}\\
\label{eeq}
&& \bigl(B^\gamma_i\bigr)^2 = - {{U}^\gamma_i {\bar T}^\gamma_i \over S^\gamma_i } \; {{U}^\gamma_{i+1} T^\gamma_{i+1} \over {\bar S}^\gamma_{i+1} } \qquad 
\forall i \in \{1,\cdots,r_\gamma\}\\
\label{ceq}
&&\prod_{i=1}^{r_\gamma} B^\gamma_i = - \prod_{i=1}^{r_\gamma} {U}^\gamma_i
\eeq
Then, for any perfect matching $M_0$  such that $M_0\cap E = \emptyset$, the contribution of every closed curve in ${\mathcal C}(G)$  is invariant under addition of $\gamma$,
\be\label{cycle_eq0}
F_{A^G}(M_0,C) = F_{A^G}(M_0, C \triangle \gamma) \qquad\forall C\in {\mathcal C}(G).
\ee
\ep
\smallskip\noindent
Proposition \ref{prop:cycle_eq} will be proven in Section \ref{sec:5}, where we introduce a representation in terms of a Grassmann algebra. Equations \eqref{seq}--\eqref{ceq} can be obtained by considering all local configurations around $\gamma$ with an even number of edges at each site.  Some of these configurations may not correspond to an actual element $C\in {\mathcal C}(G)$, for instance  when a cut along $\gamma$ splits the graph $G$ into two pieces, so that some of these equations are possibly not
necessary. The set of equations \eqref{seq}--\eqref{ceq} always admits a nowhere zero solution, when considering a single cycle. However, compatibility of these conditions for an arbitrary  collection of cycles is not granted, even for planar graphs.  

Here, MacLane's planarity criterion comes into play as it provides us with a cycle basis $\mathcal B_0$ with specific properties. In particular, elements of such a basis cannot form a cluster (a double cover of a proper subset of $E(v)$, at some vertex $v\in V$ ) \cite{BD}. 
Then the absence of clusters with $3$ edges implies compatibility of equations \eqref{seq} for all $\gamma\in \mathcal B_0$ and all $v\in V$.
Similarly, supposing that site entries have been chosen so that equations \eqref{seq} hold for all $\gamma\in \mathcal B_0$, the absence of clusters with $2$ edges implies that edge entries can be chosen so that equations \eqref{eeq} also hold for all $\gamma\in \mathcal B_0$.

The remaining equations \eqref{ceq} are strongly reminding of Kasteleyn's orientation prescription. Note that  equations \eqref{seq}--\eqref{eeq} are independent on the sign of the edge entries $B_i^\gamma$,
and that once they are solved, one has
\be
\label{ceq2}
\prod_{i=1}^r \bigl(B^\gamma_i\bigr)^2 = \prod_{i=1}^r \bigl({U}^\gamma_i\bigr)^2
\ee
so that equation \eqref{ceq} just amounts to choose independently the sign of the product of edge entries along $\gamma$. Existence of a solution to the equations \eqref{ceq} for all $\gamma\in \mathcal B_0$ just derives from independence of the elements in that basis.

Proposition \ref{prop:cycle_eq} implies that conditions of  Lemma \ref{lem:fa}
are verified for any planar $4$-regular graph, and Proposition \ref{prop:planar4r} follows.

We now generalize these results to non planar graphs.

By Theorem \ref{thm:planar}, we already know that a non planar 
graph $G$  cannot have a cycle basis such that Equations \eqref{seq}--\eqref{ceq} can be solved simultaneously for all its elements. The strategy we adopt here consists in finding the largest free family of cycles on which 
these equations may hold simultaneously and looking at what happens when completing it to a cycle basis.

Following the generalization to non planar graphs of McLane's Theorem \cite{BD}, we say that a family of cycles on a graph is ``sparse'' if it contains no cluster, where a cluster is a subfamily of cycles
covering twice a proper subset of $E(v)$, for some vertex $v$. Just as in the planar case, Equations \eqref{seq}--\eqref{ceq}  can be proven to be simultaneously solvable for all cycles forming a sparse family. 
 
Intuitively, a maximal sparse family of cycles has to be related to the embedding properties of the graph, but this relation is rather intricate: in a non planar 2-cell embedding, face boundaries are closed walks but not necessarily cycles and another definition for sparseness is required \cite{BD}. The following lemma together with Theorem \ref{thm:mi_pf} allows us to bypass this point: 
\bl\label{faces=cycles}
Let $G_1$ be a $2$-connected simple graph embedded in some surface $\mathcal S$. There exists a $4$-regular $2$-connected graph $G_2 \succcurlyeq G_1$ 
embeddable in the same surface $\mathcal S$ so that all its face boundaries are cycles.
\el
If the genus of the embedding in $\mathcal S$ is minimal for $G_1$, it is also minimal for $G_2$ since $G_2 \succcurlyeq G_1$. 
When considering such an embedding for $G_2$ in $\mathcal S$,
the family of face boundaries is a collection of cycles and a sparse family of smallest codimension in $\mathcal C(G_2)$. Theorem \ref{thm:mi_pf} allows then to transfer related results back to $G_1 \preccurlyeq G_2$, even if $G_1$  has no strong embedding of its own genus.

Hence, we consider a $4$-regular graph $G=(V,E)$ with a strong 2-cell embedding in some surface $\mathcal S$.  We denote by $\mathcal F_0(G)$
the set of cycles which are face boundaries in that embedding and by  $\mathcal F_0^*(G)$ a maximal free subfamily, obtained by dropping out one element of $\mathcal F_0(G)$.  

Clearly $\mathcal F_0^*(G)$ 
is a sparse familly and as in the planar case,
this property leads to the proof that the algebraic equations 
 \eqref{seq}--\eqref{eeq} can be simultaneously solved for all $\gamma\in \mathcal F_0^*(G)$. 
 
A non orientable embedding can be described by an embedding scheme, that is a pair $(\Pi,\lambda)$ where $\Pi=(\pi_v)_{v\in V}$ defines a cyclic ordering of edges at every vertex, and $\lambda : E \rightarrow \{-1,1\}$ is a signature on the edge set \cite{MT}. Note that the correspondence between embeddings and  embedding schemes is not one to one, and for a graph with $|V|$ vertices there can be as much as $2^{|V|}$  embedding schemes describing the same embedding. 
We prove the following Lemma.
\bl\label{nonorientable=complex}
Let $G$ a $4$-regular graph with a strong 2-cell embedding in some surface $\mathcal S$, defined up to homeomorphism by the embedding scheme $(\Pi,\lambda)$. There exists an incidence matrix $A^G$
with coefficients
\beq\label{nonorientable=complex0}
A^G_{(v,e),(w,f)}\in 
\begin{cases}
 i \R &\hbox{  if }  e=f \hbox{ and }\lambda(e)=-1\cr
 \R & \hbox{ otherwise.}\cr
\end{cases}
\hbox{  for all } (v,e),(w,f) \in V_{ \mathcal D}(G)
\eeq
and a reference perfect matching $M_0$ as in \eqref{def:RPM}  such that 
\beq\label{cycle_eq0c}
F_{A^G}(M_0,C) = F_{A^G}(M_0, C \triangle \gamma) \qquad\forall C\in {\mathcal C}(G), \forall \gamma\in\mathcal F_0^*(G) 
\eeq
In particular the coefficients of  $A^G$ can be chosen in $\R$ if and only if the embedding is orientable.
\el

When the incidence matrix is chosen as in  Lemma \ref{nonorientable=complex}, each closed curve $C$ has a weight in the Pfaffian expansion which depends only on its homology class on $\mathcal S$. Proof of Theorem \ref{thm:mc} uses  the fact that for a sufficiently large class of graphs, these weights can be easily computed. However, the coefficients have to be chosen in a suitable multicomplex algebra $\C_n$ and Lemma \ref{nonorientable=complex} stated in this larger algebraic context.

The more classical expansions \eqref{cor:sum_noe} and \eqref{cor:sum_oe} are then obtained directly 
by an algebraic reduction from $ \C_n$ to $\C$ and $\R$, respectively.
This suggests that expression \eqref{ZNP} does contain more information than classical expansions,
and we believe that it will prove useful to get new results on non planar Ising model.
\section {Proofs}
\label{sec:4}
In this section, we present proofs of all results, with the exception 
of Propositions  \ref{prop:cycle_eq} and \ref{hiera}  which are considered in the next section.

\noindent 
{\bf Proof of Proposition \ref{prop:rdc}.}\par 

 If $G=(V,E)$ is the dart graph of some other graph, says $G=\mathcal D(G')$ and $G'=(V',E')$,  its edge set
is the union of two distinct parts, $E= E_{\mathcal D}^{V'} \cup E_{\mathcal D}^{E'}$. Its subgraph
$(V,E_{\mathcal D}^{E'})$ forms a perfect matching, since by definition, each vertex in $V$ is
a dart on $G'$ and, as such, is incident with exactly one edge of $E_{\mathcal D}^{E'} $.
\hfill\halmos

\noindent 
{\bf Proof of Proposition \ref{prop:pm-cc}.}\par 
Let $M$ be a perfect matching on $\mathcal D(G)$. For each vertex $v\in V$, denote by $E(v)$
the set of edges at $v$, and by $V_{\mathcal D}(v)$ the set of darts at $v$
\be
V_{\mathcal D}(v)= \Bigl\{ (v,e), e\in E(v)\bigr\}
\ee
An edge of $M$ is either in $E_{\mathcal D}^V$ and have thus its two endpoints in the same cluster
$V_{\mathcal D}(v)$ for some $v \in V$, or is in $E_{\mathcal D}^E$ and hits two distinct clusters.
Since $M$ is a perfect matching on $\mathcal D(G)$, it hits each of its vertices exactly once.
Thus for all $v\in V$, the number of edges in $M\cap E_{\mathcal D}^E$ which hit $V_{\mathcal D}(v)$
has the same parity as $|V_{\mathcal D}(v)|$. This number, modulo $2$ is thus independent on the matching. In particular, if $M_1$, $M_2$ are two perfect matchings on $\mathcal D(G)$,we have
 \be
 C = (M_1\triangle M_2)\cap E_{\mathcal D}^E = 
(M_1\cap E_{\mathcal D}^E)\triangle(M_2\cap E_{\mathcal D}^E)
\ee
and for all $v\in V$, $C$ hits $V_{\mathcal D}(v)$ an even number of times.
\hfill\halmos

\bigskip\noindent 
{\bf Proof of Lemma \ref{lem:fa}}\par 

Substituting both Expressions \eqref{def:ZG} and  \eqref{pfaffAexp2} in the definition \eqref{def:pfaffrep:1} and owing to the algebraic independence of the weights, it is clear that the partition function on $G$ admits a Pfaffian representation in the sense of definition \ref{def:pfaffrep} if and only if there exists an incidence matrix $A^G$ a reference perfect matching $M_0$ and a constant $\Lambda\not = 0$ such that
\be\label{FA=K}
F_{A^G}(M_0, C) = \Lambda  \qquad \forall C\in \mathcal C(G)
\ee

The actual value of constant $\Lambda$ is irrelevant , so we only need to verify that for some reference perfect matching,
 $F_{A^G}(M_0, \cdot) $ is non zero and is equal on any two closed curves in $ {\mathcal C}(G)$,
\be\label{syseq}
F_{A^G}(M_0, C)  =F_{A^G}(M_0, C') \qquad \hbox{ for all } C, C' \in  {\mathcal C}(G)
\ee
Now $ ({\mathcal C}(G),\triangle)$ is a vector space and admits a cycle basis, say  ${\mathcal B}_G$.
Therefore the above set of equations can be reduced to an invariance property under addition of any cycle in that basis
\be\label{syseqr3}
F_{A^G}(M_0, C)  =F_{A^G}(M_0, C \triangle \gamma) \qquad \hbox{ for all } C \in  {\mathcal C}(G)
 \hbox{ and all }g \in  {\mathcal B}_G
\ee
Equations \eqref{syseqr3} are clearly a subset of Equations \eqref{syseq}; by chain rule, it is easy to show that
they generate all of them, so that both sets are equivalent.  
\hfill\halmos

\bigskip\noindent 
{\bf Proof of Proposition \ref{prop:mi_pf}.}\par

Consider two graphs $G_1=(V_1,E_1)$ and $G_2=(V_2,E_2)$ such that $G_1\preccurlyeq G_2$. Consider a given transformation which
send $G_2$ on $G_1$, and denote by $E_2^c$ (respectively $E_2^d$ ) the set of contracted (respectively deleted) edges of $E_2$
in that transformation. By construction, each connected component of $(V_2,E_2^c)$ is a tree  $T_v$, which is mapped under contraction to a given vertex $v$ in $G_1$. Furthermore, the edges in $E_2\setminus (E_2^c\cup E_2^d)$ are in 
one to one correspondence with those of $E_1$, which implies that there is also a one to one correspondence between
the set of darts $ V_{\mathcal D}(G_1)$ and the subset of  $V_{\mathcal D}(G_2)$ defined as
\be
\overline K = \bigl\{(v,e)\in V_{\mathcal D}(G_2), e\in E_2\setminus (E_2^c \cup E_2^d)\}
\ee

For all $d\in V_{\mathcal D}(G_1)$, we denote by $\tilde d$ its image in $\overline K$ and assume that the order on $V_{\mathcal D}(G_1)$ is 
induced from the order on  $V_{\mathcal D}(G_2)$ through  this correspondence.
 
Let $A^{G_2}$ be an incidence matrix on $\mathcal D(G_2)$ such that $A^{G_2}_{d_1,d_2}=0$ if either $d_1$ or $d_2$ is incident with an edge in $E_2^d$;  Then, by  \eqref{supmatrix}, the matrix $[A^{G_2}]^{\overline K}$ is an antisymmetric matrix of order $|\overline K| = |V_{\mathcal D}(G_1)|$, and  for every pair of darts $(d_1,d_2)$ in $V_{\mathcal D}(G_1)$ with $d_1 < d_2$, its entries are
\be\label{supm2}
A^{\overline K}_{\tilde d_1,\tilde d_2} = \Pf \bigl(A_{(K\cup \{\tilde d_1, \tilde d_2\}}\bigr)
\ee
where 
\be\label{K=EcEd}
K = \bigl\{(v,e)\in V_{\mathcal D}(G_2), e\in E_2^c \cup E_2^d\}
\ee
In order to prove that $A^{\overline K}$ is an incidence matrix on $\mathcal D(G_1)$, we have to check that 
Equation \eqref{def:im:1} holds for all pairs $(d_1,d_2)\in V_{\mathcal D}(G_1)$. Let us consider a pair of darts $(d_1,d_2)$ such that $A^{\overline K}_{\tilde d_1,\tilde d_2}\not=0$. Then in the expansion of the right hand side of \eqref{supm2}, there is at least one non zero term which necessarily contains as a factor the term $A_{\tilde d_1, \tilde d_2}$, or a product of terms 
$A_{\tilde d_1, \tilde g_1} A_{\tilde g_1', \tilde g_2} \cdots A_{\tilde g_k', \tilde d_2}$
for some $k>0$ where $( \tilde g_i,\tilde g_i')$, $1\le i\le k$,  are pairs of darts associated to the same edge in $E_2^c$ (by condition \eqref{prop:mi_pf:1}). 

In the first case, $A_{\tilde d_1, \tilde d_2}\not = 0$
implies $(\tilde d_1, \tilde d_2)\in E_{\mathcal D} (G_2)$ since $A$ is an antisymmetric incidence matrix. Hence  $\tilde d_1$ and $\tilde d_2$ contain the same edge in $E_2\setminus (E_2^c \cap E_2^d)$, which implies that $d_1$ and $d_2$ also contain the same edge in $E_1$.

In the second case, $A_{\tilde d_1, \tilde g_1}\not= 0 $, $A_{\tilde g_1', \tilde g_2}\not= 0$, $\cdots$,  $A_{\tilde g_k', \tilde d_2}\not= 0$ imply that each pair of darts, $\tilde d_1$ and $\tilde g_1$, $\tilde g_1'$ and $ \tilde g_2$, $\cdots$,  $\tilde g_k'$ and $\tilde d_2$, share the same vertex, respectively. Since $(\tilde g_i, \tilde g_i')\in E_2^c$ for all $1\le i\le k$, these vertices are in the same connected components of $(V_2,E_2^c)$
and thus shrink to the same vertex of $G_1$ under contraction. Thus $d_1$ and $d_2$  contain the same vertex in $V_1$.

In both cases, $(d_1,d_2)\in E_{\mathcal D}(G_1)$. $A^{\overline K}$ is thus an incidence matrix on $\mathcal D(G_1)$.
\hfill\halmos

\noindent 
{\bf Proof of Theorem \ref{thm:planar}.}\par

Let $G$ be a planar, $2$-connected simple graph. By Lemma  \ref{lem:4r}, there exists a planar, $2$-connected,  $4$-regular graph $\tilde G \succcurlyeq \tilde G$, which partition function admits a
Pfaffian representation by Proposition  \ref{prop:planar4r}. By Theorem \ref{thm:mi_pf}, this property extends to all its minors and in particular
the partition function on $G$ admits a Pfaffian representation.

Let $G$ be a non planar graph. Kuratowski's planarity criterion \cite{K}, one of its minors is homeomorphic to the complete graph $K_5$ or the complete bipartite graph $K_{3,3}$. By Proposition  \ref{prop:K5K33}, the partition function on either of these two graphs has no Pfaffian representation and  by Theorem \ref{thm:mi_pf}, there is also no such Pfaffian representation for the partition function on
$G$.
\hfill\halmos

\bigskip\noindent 
{\bf Proof of Theorem \ref{thm:mc}.}\par

Let $G=(V,E)$ be a graph of nonorientable genus $\tilde g$. We first suppose that $G$ is $4$-regular and construct 
another $4$-regular graph $\tilde G=(\tilde V,\tilde E)$ with $\tilde G \preccurlyeq G$ in the following way: we consider a surface $S$
with $\tilde g$ crosscaps and decompose it  into $\tilde g +1$ domains
$D_0$, $D_1$,$\cdots$,$D_{\tilde g}$ so that $D_0$ is homeomorphic to a sphere with $\tilde g$ holes and each of the
$D_k$, $k>0$ is homeomorphic to a Moebius strip. We draw $G$ on $S$ so that all its vertices are in $D_0$.  On every domain 
 $D_k$, $k>0$ , we draw two nested non intersecting simple curves around  the crosscap. We call $\tilde G$ the $4$-regular graph 
 which representation on $S$ results from the superposition of the representation of $G$ and the $2 \tilde g$ closed curves, adding
 each intersection point to the vertex set , and associating each line segment to an edge.
 
 Finally, we possibly use Lemma \ref{faces=cycles} and modify $\tilde G$  accordingly so that the all face boundaries of $\tilde G$
 on $S$ are cycles. For every $k\in \{1,\cdots, \tilde g\}$, we call  $\tilde G_k=(\tilde V_k,\tilde E_k)$ the subgraph  of $\tilde G$ with vertex set the subset of $\tilde V$ represented in $D_k$ and edge set the set of edges in $\tilde E$ with both endvertices in $\tilde V_k$.

 Note that in this embedding every edge of $\tilde G$ goes through at most one crosscap.
  We thus consider an embedding scheme $(\Pi,\lambda)$ which describes the embedding of $\tilde G$ on $S$ so that $\lambda(e)= -1$
 for every edge $e$  going through a crosscap and $\lambda(e)= +1$ otherwise.  By Lemma \ref{nonorientable=complex}, there exists an incidence matrix $A^{\tilde G}$ with coefficients as in \eqref{nonorientable=complex0},
 such that equations \eqref{seq}--\eqref{ceq} hold for all cycles $\gamma\in \mathcal F_0^*(\tilde G)$. 
 
 We introduce a multicomplex algebra $\C_{\tilde g}$ with generators $i_1$, $\cdots$, $i_{\tilde g}$ such that
 $i_\alpha^2= -1$ for all $\alpha \in \{1,\cdots, \tilde g\}$ and $i_\alpha i_\beta= i_\beta i_\alpha$  for all $\alpha, \beta\in \{1,\cdots, \tilde g\}$.
 We construct a new matrix $\tilde A^{\tilde G}$ with coefficients in $\C_{\tilde g}$, so that for every pair of darts $(v,e)$, $(w,f)$ in $V_{\mathcal D} (\tilde G)$
 \beq
 \tilde A^{\tilde G}_{(v,e),(w,f)} =
 \begin{cases}
 i_k \Im \bigl(A^{\tilde G}_{(v,e),(w,f)}\bigr) &\hbox{ if } e=f \in \tilde G_k  \hbox { and } \lambda(e) = -1 \cr
 A^{\tilde G}_{(v,e),(w,f)} & \hbox{otherwise.}
 \end{cases}
 \eeq
 where $\Im(\cdot)$ denotes the imaginary part.
By construction,  every  cycle in $\mathcal F_0^*(\tilde G)$ has its support in one of the
subgraphs $ \tilde G_k$, or has all edges with $+1$ signature.

Thus for every cycle $\gamma\in \mathcal F_0^*(\tilde G)$, the equations \eqref{seq}--\eqref{ceq} 
 contain at most one generator $i_k$, so that they hold for coefficients of  $ \tilde A^{\tilde G}$ 
 as soon as they hold for those of $ A^{\tilde G}$. Thus  $\tilde A^{\tilde G}$ is also an incidence matrix (with coefficients in $\C_{\tilde g}$ )
 such that equations \eqref{seq}--\eqref{ceq} hold for all cycles $\gamma\in \mathcal F_0^*(\tilde G)$. 
 
 The set of closed curves on graph $\tilde G$ has $2^{\tilde g}$ homology classes and any two elements in the same homology class are given the 
 same contribution in the Pfaffian expansion of $ \tilde A^{\tilde G}$. We now construct a particular element in each homology class which  contribution can be computed. 
 
 Let us denote by $\mathcal C_R(\tilde G)  \subset \mathcal C(\tilde G)$ the set of closed curves with support in $\cup_{1\le k\le \tilde g}  \tilde E_k$.
Given a reference perfect matching $M_0$ as in  \eqref{def:RPM}, we also denote by  $\mathcal M_R(\tilde G)$ the set of perfect matchings 
in $\mathcal M (\tilde G)$ which have their image in $\mathcal C_R(\tilde G) $.
\be
\mathcal M_R(\tilde G) = \{ M\in \mathcal M (\tilde G), \Phi_{M_0} (M) \in \mathcal C_R(\tilde G)\}
\ee
We also define the following subsets of darts
\beq
\mathcal D_0 = \bigl\{(v,e)\in V_{\mathcal D}(\tilde G), v\in \tilde V\setminus \cup_{1\le k\le \tilde g}\tilde V_k \bigr\}
\eeq
and for all $k\in \{1,\cdots, \tilde k\}$
\beq
\mathcal D_k = \bigl\{(v,e)\in V_{\mathcal D}(\tilde G), v\in \tilde V_k \bigr\}
\eeq
Obviously the chosen reference perfect matching has its support in $\cup_{0\le k\le \tilde g} \mathcal D_k \times \mathcal D_k $
and thus also every perfect matching in $\mathcal M_R(\tilde G)$. Now the sets of dimers $ \mathcal D_k \times \mathcal D_k$, $0\le k\le \tilde g$
have disjoint support, so that $\mathcal M_R(\tilde G)$ is a direct product of sets of perfect matchings on each component. 

Now define
  \beq
  f_0&= & F_{\tilde A^{\tilde G}} (M_0,\gamma_0) \hbox{ for some } \gamma_0 \in \mathcal F_0^*(\tilde G)\nonumber\cr
  f_k&= & F_{\tilde A^{\tilde G}} (M_0,\gamma_1) \hbox{ for some } \gamma_1^k \in \mathcal C(\tilde G_k)\setminus \mathcal F_0^*(\tilde G)
\eeq
Note that the value of $f_0$ is independent of the choice of $ \gamma_0 \in \mathcal F_0^*(\tilde G)$, 
and $\mathcal C(\tilde G_k)\setminus \mathcal F_0^*(\tilde G)$ is not empty ( $\tilde G_k$ is not planar) and the value of 
$f_1$ is independent on the choice of $\gamma_1^k$ in that set. 

 The subgraph $\tilde G_k$ contains $K_5$ as a minor (which can be formed from the two closed lines around the crosscap and two 
 edges of $G$ passing through the crosscap and their $8$ crossing points). Thus by Theorem \ref{thm:mi_pf}, there exists an incidence matrix on $K_5$ which Pfaffian expansion has weights proportional to $f_0$ and $f_k$.   In particular they verify Equation \eqref{K5-obstruction} which is homogeneous. Thus weights $f_0$ and $f_k$ verify the same equation, namely,
 \beq
 f_k f_0^3 = - f_k^3 f_0
 \eeq
 Since $\gamma_1^k$ passes through the crosscap an odd number of times, $f_k$ is necessarily proportional to $I_k$,
 which implies then
 \beq
 f_k= \pm  i_k f_0
 \eeq
For conveniency, we possibly change the sign of the coefficients of the matrix $ \tilde A^{\tilde G}$ which are proportional to $i_k$
(Indeed, for every cycle $\gamma\in \mathcal F_0^*(\tilde G)$,  equations \eqref{seq}--\eqref{ceq} keep unchanged in that transformation),
so that we can fix those signs to be
  \beq
 f_k= i_k f_0
 \eeq
 Now consider a closed curve $C\in \mathcal C_R(\tilde G)$.
 Denote by $C_k$ the closed curve in $\in \mathcal C(\tilde G)$ with support in $\tilde G_k$ which coincide with $C$
 on $\tilde G_k$ and by $C_0$ the null curve.
 By construction, we have
 \beq
 F_{\tilde A^{\tilde G}}(M_0,C) \bigl(F_{\tilde A^{\tilde G}}(M_0,C_0)\bigr)^{\tilde g-1}= \prod_{k=1}^{\tilde g} F_{\tilde A^{\tilde G}}(M_0,C_k)
 \eeq 
 and thus
 \beq
  F_{\tilde A^{\tilde G}}(M_0,C) =  f_0  \prod_{k=1}^{\tilde g}  (i_k)^{\epsilon_k}
  \eeq
  where 
  \beq
   \epsilon_k=
   \begin{cases}
   + 1 &\hbox{if } C|_{\tilde G_k} \hbox { has nonorientable genus} 1\cr
   0& \hbox{otherwise}
   \end{cases}
   \eeq
   where $C|_{\tilde G_k}$ is the restriction of $C$ on $\tilde G_k$. 
   
   Setting
   \beq\label{Lambda}
   \Lambda = \frac{1}{f_0} \prod_{k=1}^{\tilde g} (1-i_k)
   \eeq
  We get, for every curve $C\in \mathcal C(\tilde G)$, 
  \beq
  \;\operatorname{Re}\bigl[  \Lambda F_{\tilde A^{\tilde G}}(M_0,C) \bigr] =1
  \eeq
  The Theorem is proven.

\hfill\halmos

\bigskip\noindent 
{\bf Proof of Corollary \ref{cor:sum_no}.}\par 
Let $G=(V,E)$ be a graph of nonorientable genus $\tilde g$. By   Theorem \ref{thm:mc}, there is an incidence matrix $A^G$ on $\mathcal D(G)$ 
with coefficients in $\C_{\tilde g}$, a constant $\Lambda\in \C_{\tilde g}$  and a reference perfect matching $M_0$ 
such that

\be\label{ZNP1}
 Z_G(w) =  \;w( M_0\cap E^E_{\mathcal D}(G)) \;\operatorname{Re}\bigl[\Lambda\;\Pf\bigl(A^{G,M_0}(w)\bigr) \bigr]
\ee

Now there are $2^{\tilde g}$  distincts algebra homomorphisms $\tilde H_k : \C_{\tilde g}  \rightarrow \C$, $k\in \{1,\cdots,2^{\tilde g}\}$
such that for all $j\in \{1,\cdots, \tilde g\}$, $\tilde H_k (i_j) \in \{-i, + i\}$, and we have for every element  $w\in \C_{\mathcal g}$,
\beq
\operatorname{Re}(w) =\frac{1}{2^{\tilde g}}\; \sum_{k=1}^{2^{\tilde g}} \tilde H_k(w)
\eeq
Here we set
\beq
\Lambda_k = \frac{1}{2^{\tilde g}}\;  \tilde H_k(\Lambda)
\eeq
and for all $(d_1,d_2)\in V_{\mathcal D}(G)\times V_{\mathcal D}(G)$
\beq
\bigl(A_k^G\bigr)_{d_1,d_2} =  \tilde H_k\bigl(A^G_{d_1,d_2}\bigr)
\eeq
and we get
\beq
 Z_G(w) =  \;w( M_0\cap E^E_{\mathcal D}(G)) \;\sum_{k=1}^{2^{\tilde g}}
 \Lambda_k\;\Pf\bigl(A_k^{G,M_0}(w)\bigr) 
\eeq

\hfill\halmos

\bigskip\noindent 
{\bf Proof of Corollary \ref{cor:sum_o}.}\par

Let $G=(V,E)$ a graph of orientable genus
$g$. The surface in which it can be minimally embedded can be represented as a (fundamental) polygon with $4 g$ sides to be identified pairwise. Conventionally, the polygon can be represented 
as a succession of translations along the sides, round the polygon, as
\beq\label{poly_or}
P_{g} = A_1 B_1 A_1^{-1} B_1^{-1} A_2 B_2 A_2^{-1} B_2^{-1} \cdots 
A_{g} B_{g} A_{g}^{-1} B_{g}^{-1}
\eeq
where $A_1, B_1, \cdots, B_{g}^{-1}$ are the labels of the sides, and the index $^{-1}$ indicates 
an orientation opposite to the previous one. 

On the other hand, a non orientable surface with genus $\tilde g$ can be represented as a fundamental polygon with $2 \tilde g$ sides, in the form
\beq\label{poly_nor}
\tilde P_{\tilde g} = C_0 C_0  C_1 C_1  \cdots C_{\tilde g-1} C_{\tilde g-1} 
\eeq
where each side $C_k$ is followed twice in the same direction.

Now we consider \eqref{poly_or} as an element of the free group $\mathbb F_{2 g}$ generated by the  $2 g$ generators $A_1$, $B_1$,$\cdots$, $A_g$, $B_g$. and we show hereafter that it can be set in the form \eqref{poly_nor} with $\tilde g = 2 g+1$. 

We define $S_0= T_0=1_{\mathbb F_{2 g}}$ and recursively for all $1\le k\le g$,
\beq
S_k &=& S_{k-1} A_{k} B_{k} A_{k}^{-1} B_{k}^{-1}\nonumber\\
T_k &=& B_{k} A_{k}T_{k-1} 
\eeq
For every $1\le k\le g$,we write 
\beq
U_k&=& S_{k-1} A_k B_k T_{k-1}\nonumber\\
V_k&=& T_{k-1}^{-1} B_k^{-1} A_k^{-1} S_{k-1}^{-1} T_{k-1}^{-1} B_k T_{k-1}\\
W_k&=&T_{k-1}^{-1} B_k^{-1} T_{k-1} S_{k-1} T_{k-1}\nonumber
\eeq
From these definitions, it follows that ,
\beq
U_1 \bigl(V_1 V_1\bigr)\bigl( W_1 W_1 \bigr)=  A_{1}^{-1} B_{1}^{-1}
=  T_0 A_{1}^{-1} B_{1}^{-1}
\eeq
 and  for all $k\ge 2$,
 \beq
U_k \bigl( V_k V_k \bigr)\bigl(W_k W_k \bigr) U_{k-1}^{-1} = T_{k-1}^{-1}  \bigl(A_{k}^{-1} B_{k}^{-1}\bigr)
\bigl(B_{k-1} A_{k-1}\bigr) T_{k-2}
\eeq

Using these relations, we compute
\beq
 &&\bigl(U_g U_g\bigr)\bigl( V_g V_g \bigr)\bigl(W_g W_g \bigr)\cdots \bigl(V_k V_k\bigr)\bigl( W_k W_k\bigr) \cdots \bigl(V_1 V_1 \bigr)\bigl(W_1 W_1\bigr)\nonumber\\
 &&\qquad =\bigl(S_{g-1} A_g B_g T_{g-1} \bigr)\bigl( U_g V_g V_g W_g W_g U_{g-1}^{-1}\bigr)\cdots\nonumber\\
 &&\qquad\qquad\qquad\times\cdots
 \bigl( U_k V_k V_k W_k W_k U_{k-1}^{-1}\bigr)\cdots 
 \bigl( U_1 V_1 V_1 W_1 W_1 \bigr)\nonumber\\
 &&\qquad =\bigl(S_{g-1} A_g B_g T_{g-1} \bigr)\bigl( T_{g-1}^{-1}  A_{g}^{-1} B_{g}^{-1}
A_{g-1} B_{g-1} T_{g-2}\bigr)\cdots\nonumber\\
 &&\qquad\qquad\qquad\times\cdots\bigl(T_{k-1}^{-1}  A_{k}^{-1} B_{k}^{-1}
A_{k-1} B_{k-1} T_{k-2}\bigr)\cdots
 \bigl( A_{1}^{-1} B_{1}^{-1} \bigr)\nonumber\\
 &&\qquad = S_{g-1} A_g B_g A_{g}^{-1} B_{g}^{-1}
\eeq

Which gives the identity
\beq\label{og_nog_id}
 &&\bigl(U_g U_g\bigr)\bigl( V_g V_g \bigr)\bigl(W_g W_g \bigr)\cdots \bigl(V_k V_k\bigr)\bigl( W_k W_k\bigr) \cdots \bigl(V_1 V_1 \bigr)\bigl(W_1 W_1\bigr)\nonumber\\
 &&\qquad = \bigl(A_1 B_1 A_1^{-1} B_1^{-1} \bigr)\bigl(A_2 B_2 A_2^{-1} B_2^{-1}\bigr) \cdots 
\bigl(A_{g} B_{g} A_{g}^{-1} B_{g}^{-1}\bigr)
\eeq

\begin{figure}[htbp]
\begin{center}
\includegraphics[scale=1.,  angle = - 90, trim=  7cm 5.6cm 7cm 0cm, clip=true ]{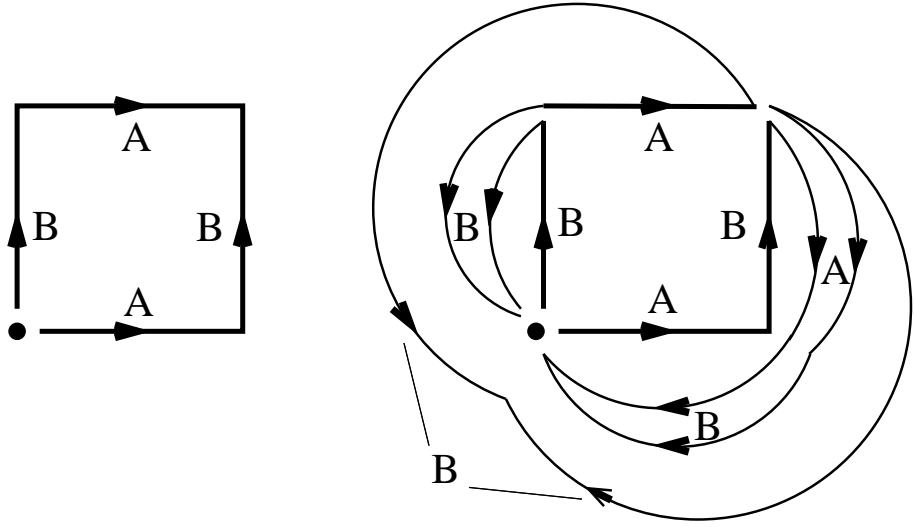}
\end{center}
\caption{Left: Fundamental polygon for a surface of genus one, described as a single curve closing at the point marked with a $\bullet$. Right: transformation of the same curve using decomposition \eqref{nogCk}.
The resulting curve now passes three times at the marked point; After 
identification of these three points, the curve splits into three lobes, each having the structure of a crosscap.  
}
\label{figure:g=1_nog=3}
\end{figure}

\begin{figure}[htbp]
\begin{center}
\includegraphics[scale=1.,  angle = - 90, trim=  5.5cm 5.6cm 5.5cm 0cm, clip=true ]{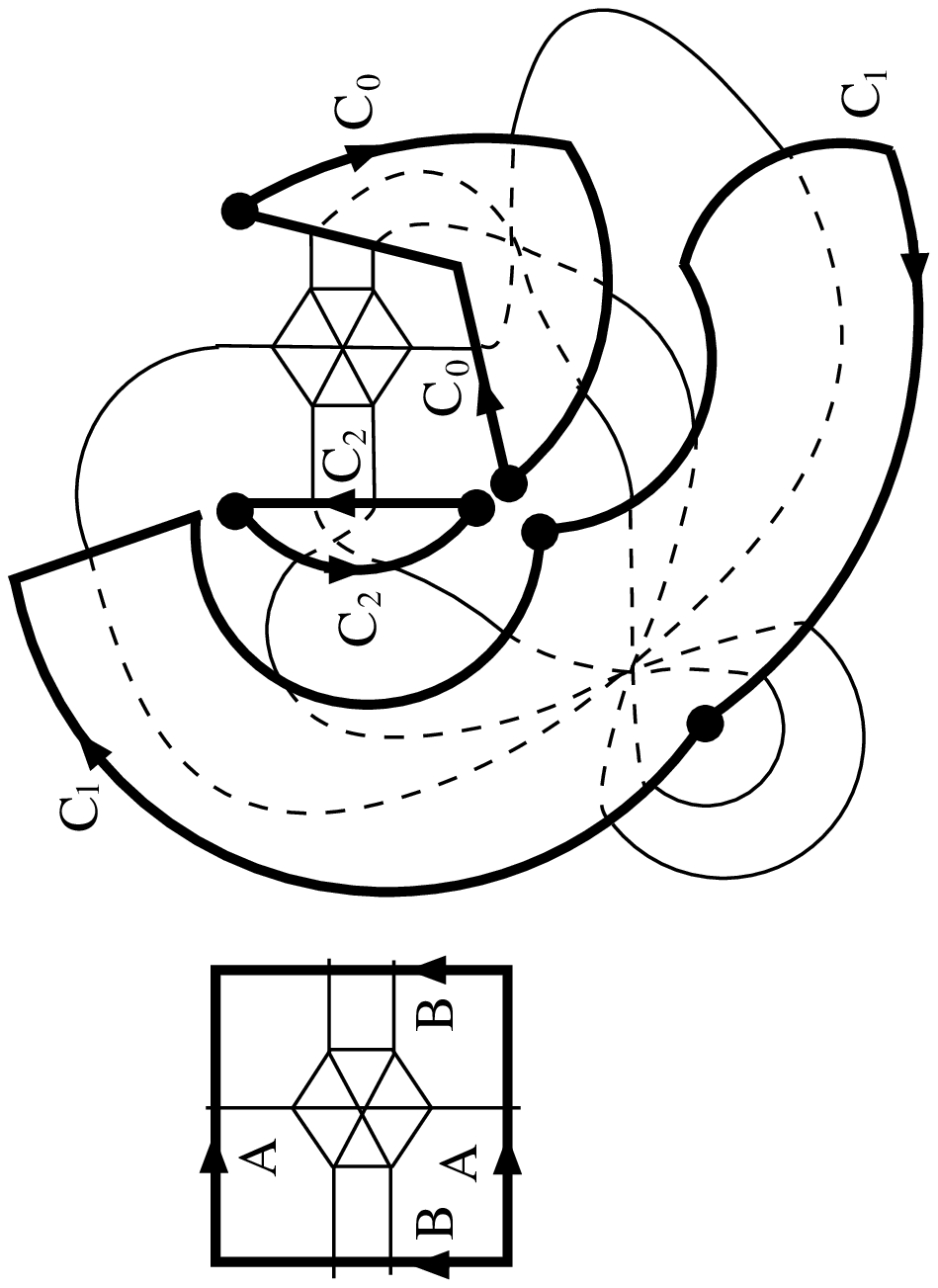}
\end{center}
\caption{Left: Embedding on the torus of a graph of orientable genus $1$. Right: Derived embedding in a surface with tree crosscaps, with $C_0 = A B$, $C_1= B^{-1} A^{-1} B$, $C_2= B^{-1}$. Multiple occurences of the same boundary are identified in two ways: inside each crosscap, opposite boundaries are identified (dashed lines); outside crosscaps, the paired 
new boundaries are joined by solid lines. By construction, edges do not cross outside crosscaps. Note that every edge goes through a crosscap
an even number of times.
}
\label{figure:g=1_nog=3b}
\end{figure}

In terms of elements of the free group $\mathbb F_{2 g}$, we have thus proven an identity between $P_g$ and $\tilde P_{\tilde g}$, with $\tilde g
= 2 g +1$, with the following choice for the factors $C_j$, $0\le j\le 2 g$,
\beq\label{nogC0}
C_0 = U_g
\eeq
and for all $0\le k\le g-1$,
\beq\label{nogCk}
C_{2 k +1} &=& V_{g-k}\\
C_{2 k +2} &=& W_{g-k}
\eeq
This decomposition proves directly that a graph of orientable genus
$g$ can be embedded in a nonorientable surface with $2 g +1$ crosscaps
and incenditally leads to a proof of inequality \eqref{nog_og},
longer but distinct from the original one \cite{St}.
In Figure \ref{figure:g=1_nog=3}, we give an example of transformation 
of a fundamental domain for a surface of orientable genus $1$
into that of a surface of nonorientable genus $3$, according to Identity 
\eqref{og_nog_id}. In Figure \ref{figure:g=1_nog=3b}, 
we show the embedding of a graph of orientable genus $1$ in a nonorientable surface with three crosscaps which results from decomposition \eqref{nogC0}--\eqref{nogCk}, starting from an orientable embedding.

Here we are interested in the explicit realization of such a nonorientable embedding. We introduce a multicomplex algebra $\C_{2 g +1}$ with generators $i_0$, $i_1$, $i_{2 g}$, so that $i_k^2= -1$ and $i_k i_{k'}=
 i_{k'}i_k$.
We now rely on the construction given in the proof of Theorem \ref{thm:mc}. We attach generator $i_k$  to the crosscap $[C_k C_k]$. Considering again a graph $\tilde G \succcurlyeq G$ such that an edge crosses at most one crosscap, we
recall that the edge entries in the incidence matrix $A^{\tilde G}_{(v,e),(w,e)}$ are linear in $i_k$ if and only if edge $e$ crosses the $k^{\hbox{th}}$ crosscap. By Theorem \ref{thm:mi_pf}, the induced incidence matrix for the original graph $G$ gets a similar property:
an edge on $G$ corresponds to a simple path in $\tilde G$ and
its edge entry in $A^G$ is proportional to the product of all edge entries in $A^{\tilde G}$ defined on this path. Accordingly, an edge entry in 
$A^G$ gets a factor $i_k$ each times the edge crosses the $k^{\hbox{th}}$ crosscap, and thus is linear in $i_k$ if and only if edge $e$ crosses the $k^{\hbox{th}}$ crosscap an odd number of times. 

Now starting from an orientable embedding of $G$, we have to read off from
expressions \eqref{nogC0}--\eqref{nogCk}, which crosscaps are crossed
oddwise by an edge in $G$ which was originally crossing (once )the boundary
of the fundamental domain through side $A_k$ (respectively $B_k$),
for all $1\le k\le g$.
For this purpose, we have thus to determine for each $C_j$, $0\le j\le 2 g$, the set of generators $\mathcal O (C_j)$ which occur an odd number of times. We have
\beq
\mathcal O (C_0)&=& \bigl\{ A_j,B_j;1\le j\le g\bigr\}\nonumber\\
\eeq
and for all $k\in\{0,\cdots, g-1\}$,
\beq
\mathcal O (C_{2k+1})&=& \bigl\{ A_j,B_j;1\le j\le g-k-1\bigr\}\cup
\bigl\{ A_{g-k}\bigr\} \nonumber\\
\mathcal O (C_{2k+2})&=& \bigl\{ A_j,B_j;1\le j\le g-k-1\bigr\}\cup
\bigl\{ B_{g-k}\bigr\} \nonumber
\eeq

It is easy to see that each generator appears in an even number of sets. Thus, for any edge crossing the boundary of the fundamental domain,
the corresponding edge entry gets proportional to the product of an even number of generators in $\C_{2 g +1}$. Each matrix entry can be thus
considered as an element of a subalgebra with generators $e_k = i_0 i_k$,
with $e_k e_{k'}= e_{k'} e_k$ and $e_k^2=1$. 

Simultaneously, the constant $\Lambda$ defined in Equation \eqref{Lambda} can be replaced by an element of the subalgebra without changing the real part of $\Lambda \Pf (A^{G,M_0}) (w)$. Namely, we set
\beq\label{Lambda_o}
\Lambda = \frac{1}{f_0} \bigl( \sum_{{S\subset \{1,\cdots,2 g\}}\atop {|S| even}} (-1)^{\frac{|S|}{2}}
\prod_{j\in S} e_j
+ \sum_{{S\subset \{1,\cdots,2 g\}}\atop {|S| odd}} (-1)^{\frac{|S|-1}{2}}
\prod_{j\in S} e_j \bigr)
\eeq

We follow the same way as in the proof of Corollary \ref{cor:sum_o}. We denote by $\R_{2 g}$ the 
real ring generated by $\{e_j\}_{1\le j\le 2g }$.
There are $2^{2 g}$  distincts algebra homomorphisms $H_k : \R_{2 g}  \rightarrow \R$, $k\in \{1,\cdots,2^{2 g}\}$
such that for all $j\in \{1,\cdots, 2 g\}$, $H_k (e_j) \in \{-1,+1\}$, and we have for every element  $w\in \C_{\mathcal g}$,
\beq
\operatorname{Re}(w) =\frac{1}{2^{2 g}}\; \sum_{k=1}^{2^{2 g}} H_k(w)
\eeq
Here we set
\beq
\Lambda_k = \frac{1}{2^{\tilde g}}\;  H_k(\Lambda)
\eeq
where $\Lambda$ is defined by Equation \eqref{Lambda_o}
and for all $(d_1,d_2)\in V_{\mathcal D}(G)\times V_{\mathcal D}(G)$
\beq
\bigl(A_k^G\bigr)_{d_1,d_2} =  H_k\bigl(A^G_{d_1,d_2}\bigr)
\eeq
Finally,  we get
\beq
 Z_G(w) =  \;w( M_0\cap E^E_{\mathcal D}(G)) \;\sum_{k=1}^{2^{2 g}}
 \Lambda_k\;\Pf\bigl(A_k^{G,M_0}(w)\bigr)
\eeq

\hfill\halmos

\bigskip\noindent 
{\bf Proof of Theorem \ref{thm:mi_pf}.}\par

We use the same notations as in the proof of Proposition \ref{prop:mi_pf}.
We first prove that if $G_1\preccurlyeq G_2$, there is an isomorphism between
$\mathcal C(G_1)$ and the subset
of closed curves on $\mathcal C(G_2)$ not containing deleted edges,
\be\label{cg1-cg2}
\check {\mathcal C}(G_2) = \{C\in \mathcal C(G_2) , C\cap E_2^d = \emptyset \}
\ee
 Under contraction, vertices connected through contracted edges are identified to a single vertex,
 which degree is thus equal modulo 2 to the sum of the degrees
 of the initial vertices. In particular, closed curves are sent to closed curves, and this allows to define a mapping $\pi_0$
 from $\check {\mathcal C}(G_2)$ to  $\mathcal C(G_1)$, which is surjective by hypothesis since $G_1\preccurlyeq G_2$. 
 The mapping is also injective: if two elements in  $ \check {\mathcal C}$ have the same image in $\mathcal C(G_1)$, 
 they have the same intersection with $E_2\setminus (E_2^c\cup E_2^d)$, so that their symmetric difference 
 has its support in $E_2^c$ and thus vanish, since $E_2^c$ contains no cycle. This defines a canonical injective mapping
 $\Pi : \mathcal C(G_1) \rightarrow \mathcal C(G_2)$, such that for all $C\in \mathcal C(G_1)$,
 \be
 \Pi(C) = \pi_0^{-1} (C)
 \ee
 
 Let $A^{G_2}$ be an incidence matrix on $\mathcal D(G_2)$. We first construct a modified incidence matrix $\check A^{G_2}$ as follows. For every pair of darts $d_1$, $d_2$ in  $V_{\mathcal D}(G_2)$,
 let $d_1=(v_1,e_1)$ and  $d_2=(v_2,e_2)$; we set
\be\label{checkA}
\check A^{G_2}_{d_1,d_2}=
\begin{cases}
0 &\hbox{ if } v_1=v_2  \hbox{ and } \{e_1,e_2\} \cap  E_2^d \not= \emptyset\cr
A^{G_2}_{d_1,d_2} &\hbox{ otherwise}
\end{cases}
\ee
Hereafter we take $M_2 = E_{\mathcal D}^E(G_2)\cong E_2$ as the reference perfect matching on $\mathcal D(G_2)$ and  consider a closed curve 
$C\in \check{\mathcal C}(G_2)$. Thus  $C\cap E_2^d = \emptyset$ and Equation \eqref{corresp} implies that  every perfect matching $M$ in $\phi_{M_2}^{-1}(C)$ contains all pairs of darts $\{(v,e),(v',e)\}$ in $E_{\mathcal D}(G_2)$ such that $e\in E_2^d$. 
In particular, $\check A^{G_2}$ and $A^{G_2}$ coincide on every element of $M$ and from Definition \eqref{FA}, we get
\be\label{FC2=F2}
 F_{\check A^{G_2}} (M_2,C) = F_{ A^{G_2}} (M_2,C)
\ee
for all $C \in \check {\mathcal C}(G_2)$.

Conversely, consider a closed curve $C\in\mathcal C(G_2)\setminus \check {\mathcal C}(G_2)$. By definition, there exists a pair of darts $d$, $d'$ in $V_{\mathcal D}(G_2)$ with $d=(v,e)$, $d'=(v',e)$ and $e\in C\cap E_2^d$.
 Since $\{d,d'\}\in M_2$,  Equation \eqref{corresp},
 implies that for every perfect matching $M$ in $\phi_{M_2}^{-1}(C)$,  $\{d,d'\}\not\in M$. Hence, for every such $M$,  there is  $d''=(v,e'')$ with $\{d,d''\}\in M$
 and by \eqref{checkA}, $\check A^{G_2}_{d,d''}=0$ . The contribution of every perfect matching $M$ in $\phi_{M_0}^{-1}(C)$ is thus zero and
 \be\label{FC2=0}
 F_{\check A^{G_2}} (M_2,C) =0
\ee
for all $C\in\mathcal C(G_2)\setminus \check {\mathcal C}(G_2)$.

 We now relate the terms in the Pfaffian expansion of  $\check A^{G_2}$ to the corresponding terms
 in the Pfaffian expansion of  the incidence matrix on $G_1$ constructed using Proposition \ref{prop:mi_pf}.
First, for every weight function $w: E_1\rightarrow \R^+$, we define its extension $\tilde w$ on $E_2$,
as
\be
\tilde w (e) = 
\begin{cases}
w(\tilde e) &\hbox{ if } e\in E_2\setminus (E_2^c\cup E_2^d)\cr
1 &\hbox{ if } e\in E_2^c\cup E_2^d
\end{cases}
\ee
where $\tilde e$ is the image in $E_1$ of edge $e$ in $E_2\setminus (E_2^c\cup E_2^d)$.

Let $\check G =(\check V,\check E)$ be the subgraph of $G_2$ with edge set $\check E = E_2^c\cup E_2^d$, and vertex set $\check V$ the set of vertices of $G_2$ incident with some edge in $\check E$.
The set of darts on $\check G$ is the set $K$ defined in Equation \eqref{K=EcEd}. By construction there
is no closed curve on $\check G$ except the null one and those with a non empty intersection with 
$E_E^d$. Since the arguments used to get equations \eqref{FC2=F2}--\eqref{FC2=0} apply equally to $\check G$, the nonzero terms in the expansion of  $\Pf\bigl([\check A^{G_2,M_0}(\tilde w)]_K\bigr)$
are associated to closed curves with no edge in $E^d$. Thus only the set of matchings associated to the null curve contributes, and it has only one element,  $M_2\vert_{\check E}$. More precisely, for every weight function $w$  on $E_1$, we have
\be\label{CG2}
\Pf\bigl([\check A^{G_2,M_0}(\tilde w)]_K\bigr) = \pm 1 \not= 0
\ee
which is thus independent on $w$.

The Pfaffian reduction formula \ref{hiera} reads here
\be\label{FA1=FC2}
Pf\bigl([\check A^{G_2,M_0}(\tilde w)]^{\overline K} \bigr)=
\Bigl(\Pf\bigl([\check A^{G_2,M_0}(\tilde w)]_K\bigr)\Bigr)^{(n-p-1)} 
\times 
\Pf(\check A^{G_2,M_0}(\tilde w))
\ee
where $2 n=|V_{\mathcal D}(G_2)|$ and  $p= |E_2^c\cup E_2^d|$.

By Proposition \ref{prop:mi_pf} , we already know that $[\check A^{G_2}]^{\overline K}$ is an incidence matrix on $\mathcal D(G_1)$.
We claim that the matrix $ [\check A^{G_2,M_0}(\tilde w)]^{\overline K}$ is that the weighted incidence matrix associated to it on $\mathcal D(G_1)$ for the weight function $w$ and reference perfect matching $M_1 = 
 E_{\mathcal D}^E(G_1)\cong E_1$. Therefore expanding both sides in equation \eqref{FA1=FC2}, for a generic weight function $w$ on $E_1$,
 and identifying terms on both sides using Equation \eqref{FA},
 one gets
 \beq
 \label{FA1=F2}
F_{[\check A^{G_2}]^{\overline K}} (M_1,C) &=& \Lambda_{1,2} F_{\check A^{G_2}} (M_2,\Pi(C))\nonumber\\
&=& \Lambda_{1,2} F_{A^{G_2}} (M_2,\Pi(C))
\eeq
for all $C \in {\mathcal C}(G_1)$, with $\Lambda_{1,2}=\pm 1$ by Equation 
\eqref{CG2} and $\Pi$ is the canonical mapping from ${\mathcal C}(G_1)$
onto $\check{\mathcal C}(G_2)$.

Suppose now that the partition function on $G_2$ admits a Pfaffian representation. Thus by Lemma \ref{lem:fa}, there exists an incidence 
matrix $A^{G_2}$ on $\mathcal D(G_2)$ such that $F_{\check A^{G_2}} (M_2,\cdot)$ is constant and different from zero on $\check{\mathcal C}(G_2)$.
By \eqref{FA1=F2} and \eqref{CG2}, $F_{[\check A^{G_2}]^{\overline K}} (M_1,\cdot)$
is also constant and different from zero on ${\mathcal C}(G_1)$. Using again Lemma \ref{lem:fa},
this implies that the partition function on $G_1$ also admits a Pfaffian representation.
 
In order to conclude the proof, we have to prove the claim. 

Given our choice of reference perfect matching, the characterization \eqref{def:wm} reduces to
the identity
\be\label{checkAKij=linear}
[\check A^{G_2,M_0}(\tilde w)]^{\overline K}_{d_1,d_2} = w^{-1} (\{d_1,d_2\}) [\check A^{G_2}]^{\overline K}_{d_1,d_2}
\ee
which has to be checked for every pair of darts  $d_1$, $d_2$ with a common edge in $E_2\setminus (E_2^c\cup\E_2^d)$. 

Let $d_1=(v_1,e)$, $d_2=(v_2,e)$ be such a pair of darts. The graph $\check G_e=(\check V\cup \{v_1,v_2\}, \check E\cup \{e\})$ is again a subgraph of $G_2$, and every closed curve 
on $\check G_e$ has either one edge in $E_2^c$ or has no edge.
We have
\be
[\check A^{G_2,M_0}(\tilde w)]^{\overline K}_{d_1,d_2} = \Pf ([\check A^{G_2,M_0}(\tilde w)]_{K\cup\{d_1,d_2\}})
\ee
and the expansion of the Pfaffian on the right hand side thus contains only one term, corresponding to 
the restriction of the reference perfect matching $M_0$ on $\mathcal D(\check G_e)$. In particular,we have
\beq
\Pf ([\check A^{G_2,M_0}(\tilde w)]_{K\cup\{d_1,d_2\}}) &=& 
\bigl(\prod_{e'\in \check E_e} \tilde w^{-1}(e') \bigr)\; \Pf ([\check A^{G_2}]_{K\cup\{d_1,d_2\}})\nonumber\\
&=&
w^{-1} (\tilde e)\; \Pf ([\check A^{G_2}]_{K\cup\{d_1,d_2\}})\nonumber\\
&=&
w^{-1} (\tilde e)\; [\check A^{G_2}]^{\overline K}_{d_1,d_2}\eeq
Thus Equation \eqref{checkAKij=linear} holds. The theorem is proven.

\hfill\halmos

\bigskip\noindent 
{\bf Proof of Proposition \ref{prop:K5K33}.}\par 

Consider first  the complete bipartite graph $K_{3,3}$. Since it is $3$-regular, its dart graph $\mathcal D(K_{3,3})$ is 
also $3$-regular and there is as many perfect matchings on $\mathcal D(K_{3,3})$
as there are closed curves on $K_{3,3}$. Thus given a reference perfect matching $M_0$, the mapping $\phi_{M_0}$
defined in Equation \eqref{corresp} is one to one.

\begin{figure}[htbp]
\begin{center}
\includegraphics[scale=.52, angle = - 90, trim=  4cm 1cm 4cm 1cm, clip=true ]{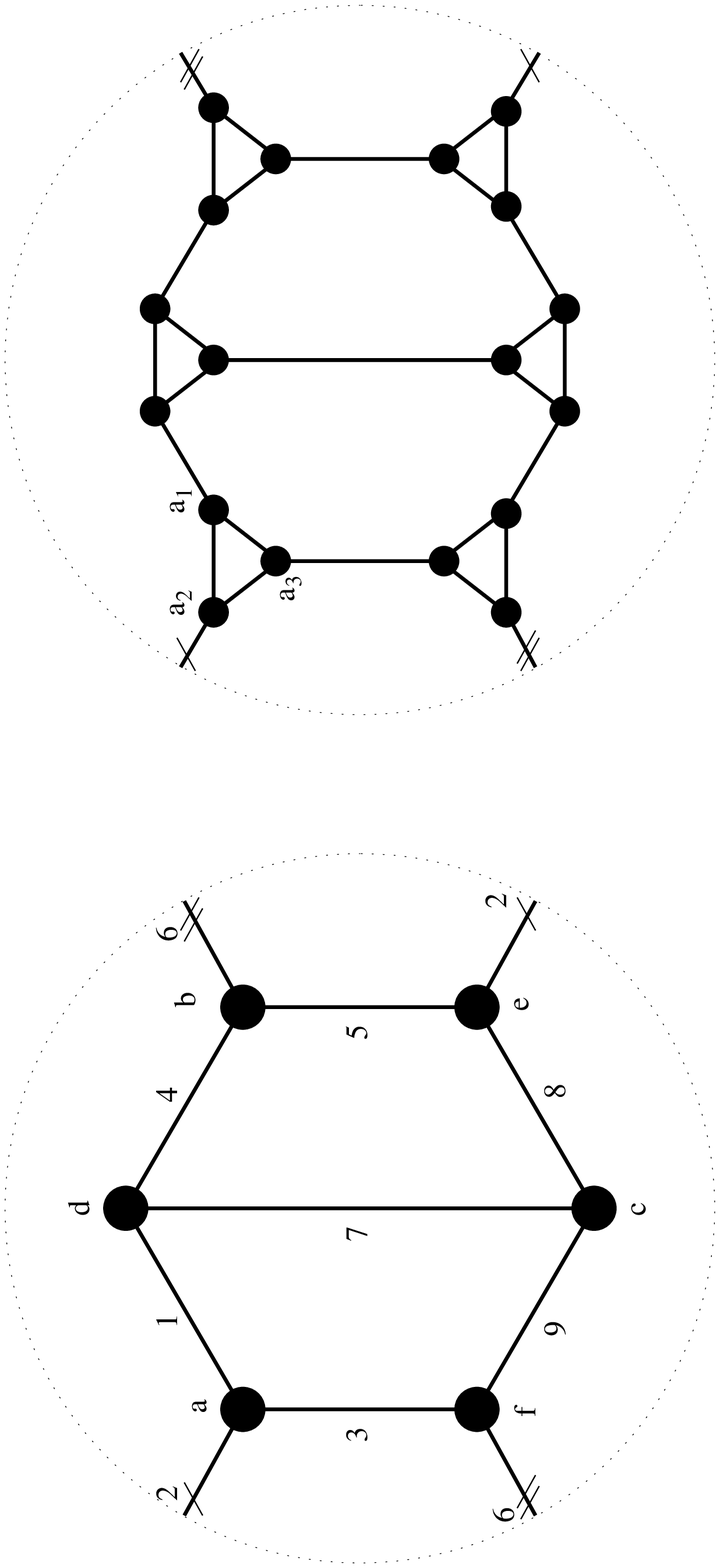}
\end{center}
\caption{ Representation of graph $K_{3,3}$ in the projective plane (Left) and its dart graph  (Right). Vertices and edges are labelled with letters and numbers, as used in text; darts are named accordingly.
}
\label{figure:3}
\end{figure}

We label the vertices and edges of $K_{3,3}$ by letters in $\{a,b,c,d,e,f\}$ and numbers in $\{1,\cdots,9\}$, respectively, as in Figure \ref{figure:3} 
and if vertex $v$ is incident with edge $k$, we use the short hand notation $v_k$ to denote the dart $(v,k)$.

We consider two sets of  cycles on $K_{3,3}$, $S=\{\gamma_1,\gamma_2,\gamma_3\}$ and 
$S'=\{\gamma_1',\gamma_2',\gamma_3'\}$, where
\beq\label{K33s}
\gamma_1&=&\{a,1,d,4,b,6,f,9,c,8,e,2\}\nonumber\\
\gamma_2&=&\{a,1,d,7,c,9,f,3\}\\
\gamma_3&=&\{b,4,d,7,c,8,e,5\}\nonumber
\eeq
and
\beq\label{K33sp}
\gamma_1'&=&\{a,1,d,4,b,5,e,8,c,9,f,3\}\nonumber\\
\gamma_2'&=&\{a,1,d,7,c,8,e,2\}\\
\gamma_3'&=&\{b,4,d,7,c,9,f,6\}\nonumber
\eeq

\begin{figure}[htbp]
\begin{center}
\includegraphics[scale=.52, angle = - 90, trim=  4cm 1cm 4cm 1cm, clip=true ]{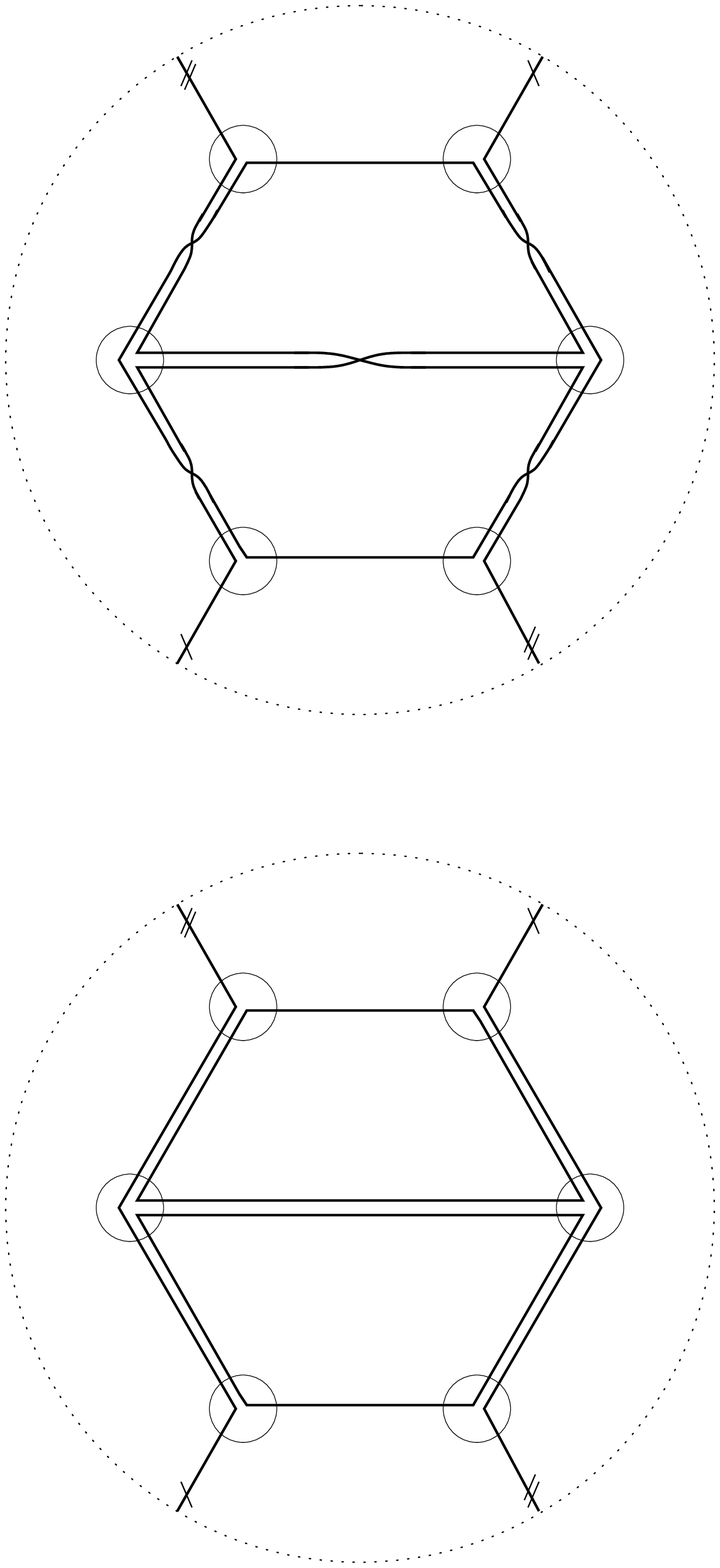}
\end{center}
\caption{The two sets of cycles $S$  \eqref{K33s},  and $S'$ \eqref{K33sp}  are drawn using the representation of $K_{3,3}$  shown in Figure \ref{figure:3}. Both sets have locally the same configurations but parity of the numbers of crossings differ. 
}
\label{figure:4}
\end{figure}

The main property of these two sets is that any subchain of length $4$ (a succession of two vertices and two edges in a cycle) appear
in both sets with the same multiplicity. For instance the subchain $(d,4,b,6)$ appears both in $\gamma_1$ and $\gamma_3'$.  
 Now, drawing simultaneously all cycles in a set on the same surface, slightly shifting each drawing, results in a figure as depicted 
 in Figure \ref{figure:4} with possibly some crossings between the lines of different cycles, and the number of crossings depends on 
 the surface and the way the cycles are drawn. However, when drawing the two sets $S$ and $S'$ on two copies of the same
 surface, the parity of the number of crossings always differs in both drawings (Figure \ref{figure:4}). 
 
 This property transfers to the perfects matchings on $\mathcal D(K_{3,3})$ in the following way:
 We take for instance  $E^E_{\mathcal D}(K_{3,3})$ as reference perfect matching, that is, we set
\beq 
M_0 &=& \{\{a_1,d_1\},\{a_2,e_2\},\{a_3,f_3\} ,\{b_4,d_4\},\nonumber\\
&&\;\;\{b_5,e_5\},\{b_6,f_6\} ,\{c_7,d_7\},\{c_8,e_8\},\{c_9,f_9\} \}
\eeq
so that the two sets of cycles $S$ and $S'$ are associated to two sets of perfect matchings, $S_M= \{M_1,M_2,M_3\}$ and $S_M'= \{M_1',M_2',M_3'\}$, respectively, where
\beq
M_1&=& \{\{a_3,f_3\} ,\{b_5,e_5\},\{c_7,d_7\},\{a_1,a_2\}, \{b_4,b_6\},\{c_8,c_9\},\{d_1,d_4\},\{e_2,e_8\},\{f_6,f_9\}\}\nonumber\\
M_2&=& \{\{a_2,e_2\},\{b_4,d_4\},\{b_5,e_5\},\{b_6,f_6\} ,\{c_8,e_8\},\{a_1,a_3\},\{c_7,c_8\},\{d_1,d_7\},\{f_3,f_9\}\}\nonumber\\
M_3&=& \{\{a_1,d_1\},\{a_2,e_2\},\{a_3,f_3\} ,\{b_6,f_6\} ,\{c_9,f_9\},\{b_4,b_5\},\{c_7,c_9\},\{d_4,d_7\},\{e_5,e_8\} \}\nonumber
\eeq
and
\beq
M_1'&=& \{\{a_2,e_2\},\{b_6,f_6\} ,\{c_7,d_7\},\{a_1,a_3\},\{b_4,b_5\},\{c_8,c_9\},\{d_1,d_4\},\{e_5,e_8\},\{f_3,f_9\} \}\nonumber\\
M_2'&=& \{\{a_3,f_3\} ,\{b_4,d_4\},\{b_5,e_5\},\{b_6,f_6\},\{c_9,f_9\},\{a_1,a_2\},\{c_7,c_8\},\{d_1,d_7\},\{e_2,e_8\}\}\nonumber\\
M_3'&=& \{\{a_1,d_1\},\{a_2,e_2\},\{a_3,f_3\} ,\{b_5,e_5\},\{c_8,e_8\}, \{b_4,b_6\},\{c_7,c_9\},\{d_4,d_7\},\{f_6,f_9\}\}\nonumber
\eeq
As a consequence of the properties of sets $S$ and $S'$, here each dimer appears in both sets $S_M$ and $S_M'$ with the same multiplicity.

Let $A$ be an arbitrary antisymmetric incidence matrix on  $\mathcal D(K_{3,3})$. The contributions of the cycles in $S$ and $S'$
to the Pfaffian expansion of $A$ (equation \eqref{FA} ) are related by
\be\label{K33-obstruction}
\prod_{i=1}^3 F_A(M_0,\gamma_i) = - \prod_{i=1}^3 F_A(M_0,\gamma_i')
\ee 
This relation derives from the properties of the two sets $S$ and $S'$. It can also be seen also as a property of the
perfect matchings in $S_M$ and $S_M'$ alone and any other choice of reference perfect matching would 
define two equivalent sets of cycles through equation \eqref{mapping}.   Thus for every antisymmetric incidence matrix on  $\mathcal D(K_{3,3})$
and every reference perfect matching there exist two sets of closed curves such that relation \eqref{K33-obstruction} holds. By Lemma
\ref{lem:fa}, the partition function on  $K_{3,3}$ has no Pfaffian representation.

The proof for the complete graph $K_{5}$ is similar. 

\begin{figure}[htbp]
\begin{center}
\includegraphics[scale=.52, angle = - 90, trim=  4cm 1cm 4cm 1cm, clip=true ]{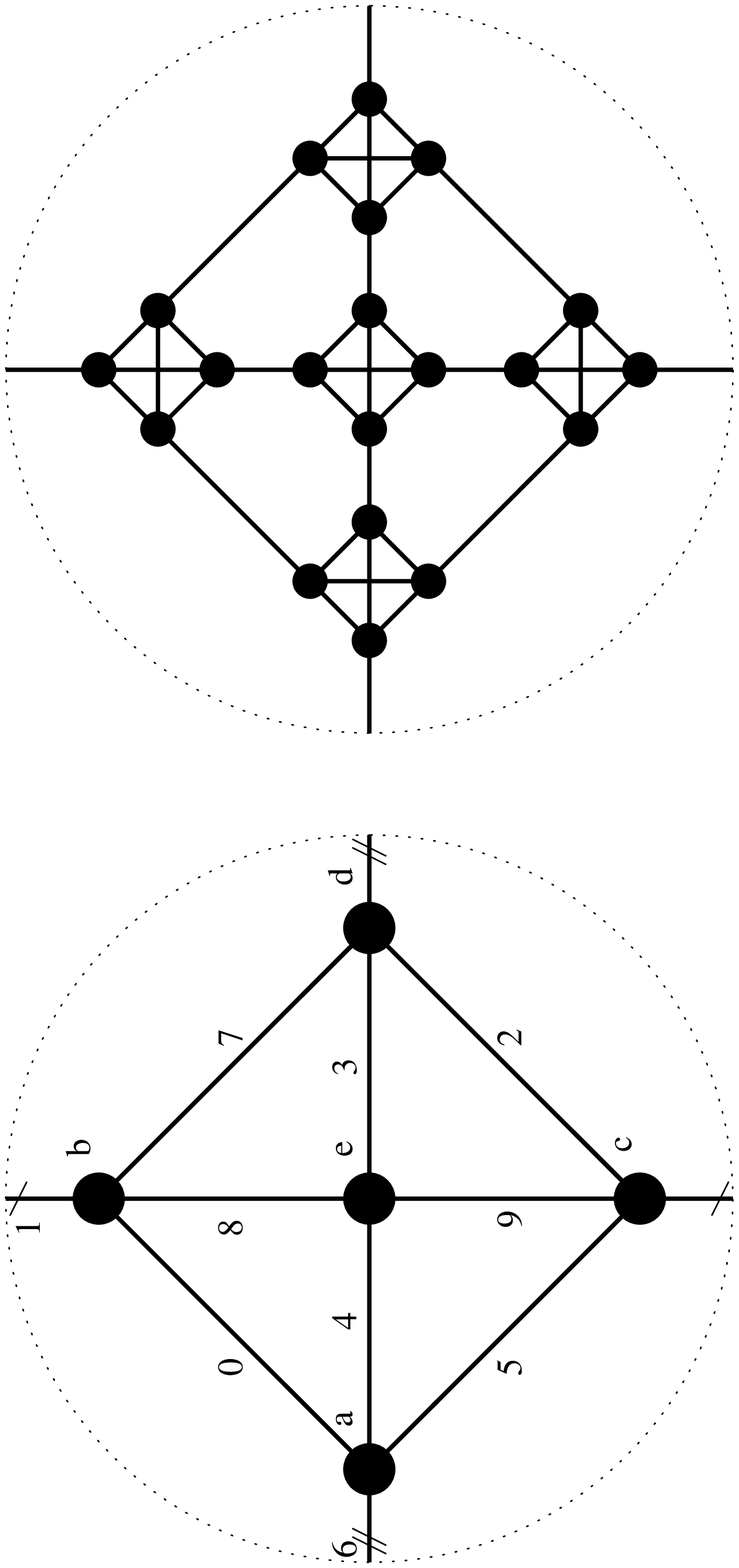}
\end{center}
\caption{ Representation of graph $K_{5}$ in the projective plane (Left) and its dart graph 
in the same representation (Right). Vertices and edges are labelled with letters and numbers, as used in text.
}
\label{figure:5}
\end{figure}

We label the vertices and edges of $K_{5}$ by letters in $\{a,b,c,d,e\}$ and numbers in $\{0,\cdots,9\}$, respectively, as in Figure \ref{figure:5} 
and we denote the dart $(v,k)$ by  $v_k$. $K_5$ is $4$-regular so that the mapping \eqref{corresp} is not one to one: in fact $K_5$ has $64$ closed curves and $\mathcal D(K_5)$ has $416$ dimer coverings.

We consider two sets of  cycles on $K_5$, $S=\{\gamma_1,\gamma_2,\gamma_3,\gamma_4\}$ and 
$S'=\{\gamma_1',\gamma_2',\gamma_3',\gamma_4'\}$, where
\beq\label{K5s}
\gamma_1&=&\{a,4,e,3,d,6\}\nonumber\\
\gamma_2&=&\{a,0,b,1,c,2,d,6\}\nonumber\\
\gamma_3&=&\{a,0,b,7,d,2,c,9,e,4\}\nonumber\\
\gamma_4&=&\{a,0,b,8,e,3,d,2,c,5\}
\eeq
and
\beq\label{K5sp}
\gamma_1'&=&\{a,4,e,9,c,2,d,6\}\nonumber\\
\gamma_2'&=&\{a,0,b,8,e,3,d,6\}\nonumber\\
\gamma_3'&=&\{a,0,b,7,d,2,c,5\}\nonumber\\
\gamma_4'&=&\{a,0,b,1,c,2,d,3,e,4\}
\eeq

\begin{figure}[htbp]
\begin{center}
\includegraphics[scale=.52, angle = - 90, trim=  4cm 1cm 4cm 1cm, clip=true ]{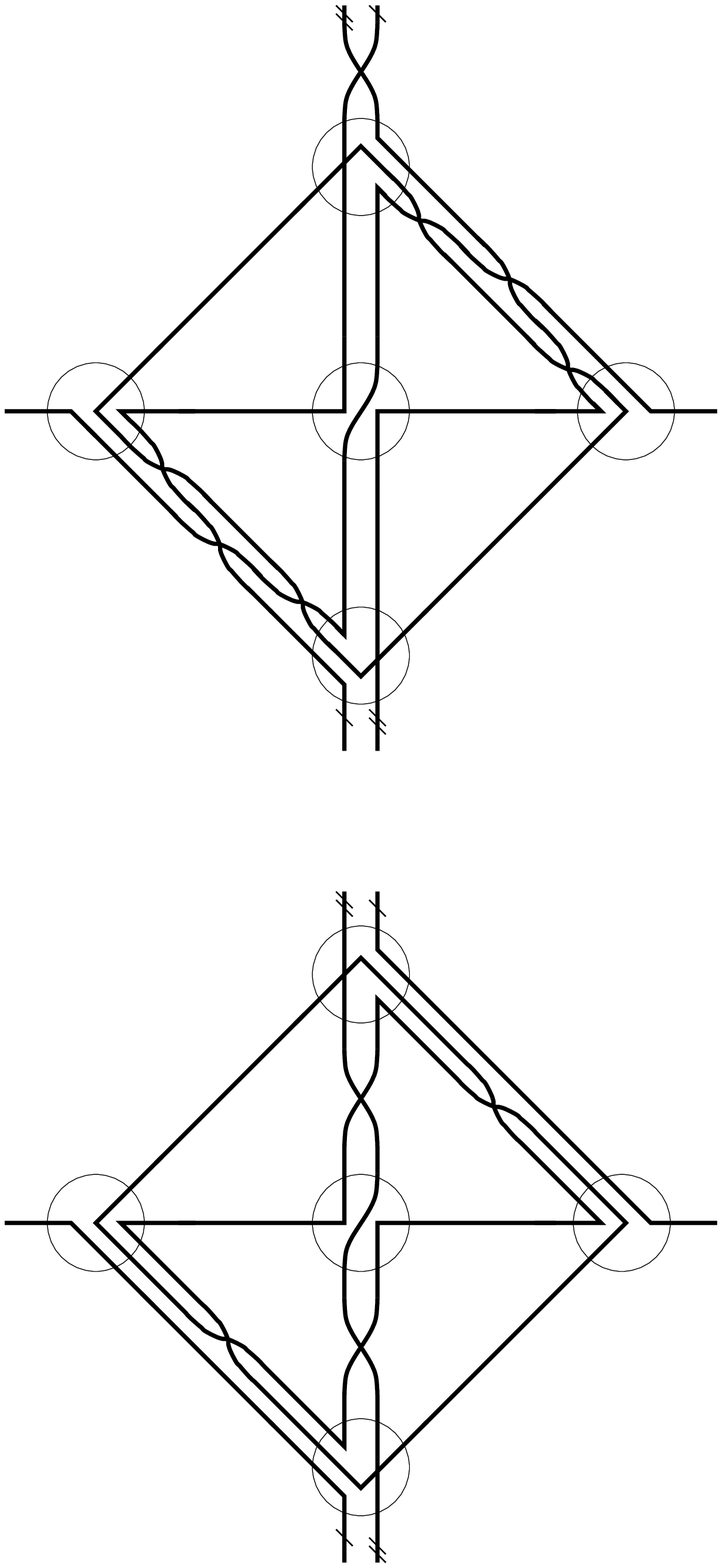}
\end{center}
\caption{The two sets of cycles $S$  \eqref{K5s},  and $S'$ \eqref{K5sp}  are drawn using the representation of $K_5$  shown in Figure \ref{figure:5}. Both sets have locally the same configurations but parity of the numbers of crossings differ. 
}
\label{figure:6}
\end{figure}

Here again, any subchain of length $4$ (two vertices and two edges)  appear
in both sets with the same multiplicity.
 Now, drawing simultaneously all cycles in a set on the same surface, slightly shifting each drawing, results in a figure  with some crossings between the lines of different cycles. Again, when drawing the two sets $S$ and $S'$ on two copies of the same
 surface, the parity of the number of crossings always differs in both drawings (Figure \ref{figure:6}). 
 
 We take  $E^E_{\mathcal D}(K_5)$ as reference perfect matching, that is
\beq 
M_0 &=& \{\{a_0,b_0\},\{b_1,c_1\},\{c_2,d_2\} ,\{d_3,e_3\}, \{a_4,e_4\},\nonumber\\
&&\;\;\{a_5,c_5\},\{a_6,d_6\} ,\{b_7,d_7\},\{b_8,e_8\},\{c_9,e_9\} \}
\eeq
so that each cycle in the two sets $S$ and $S'$ are associated to exactly one perfect matchings in $S_M= \{M_1,M_2,M_3,M_4\}$ and $S_M'= \{M_1',M_2',M_3',M_4'\}$, respectively, where
\beq
M_1&=& \{\{a_0,b_0\},\{b_1,c_1\},\{c_2,d_2\} ,\{a_5,c_5\},\{b_7,d_7\},\{b_8,e_8\},\{c_9,e_9\},
\{a_4,a_6\},\{d_3,d_6\},\{e_3,e_4\}
\}\nonumber\\
M_2&=& \{\{d_3,e_3\}, \{a_4,e_4\},\{a_5,c_5\},\{b_7,d_7\},\{b_8,e_8\},\{c_9,e_9\},
\{a_0,a_6\},\{b_0,b_1\},\{c_1,c_2\},\{d_2,d_6\}
\}\nonumber\\
M_3&=& \{\{b_1,c_1\},\{d_3,e_3\}, \{a_5,c_5\},\{a_6,d_6\} ,\{b_8,e_8\},
\{a_0,a_4\},\{b_0,b_7\},\{c_2,c_9\},\{d_2,d_7\},\{e_4,e_9\}
\}\nonumber\\
M_4&=& \{\{b_1,c_1\},\{a_4,e_4\},\{a_6,d_6\} ,\{b_7,d_7\},\{c_9,e_9\},
\{a_0,a_5\},\{b_0,b_8\},\{c_2,c_5\},\{d_2,d_3\},\{e_3,e_8\}
\}\nonumber
\eeq
and
\beq
M_1'&=&  \{\{a_0,b_0\},\{b_1,c_1\} ,\{d_3,e_3\},\{a_5,c_5\},\{b_7,d_7\},\{b_8,e_8\},
\{a_4,a_6\},\{c_2,c_9\},\{d_2,d_6\},\{e_4,e_9\}
\}\nonumber\\
M_2'&=&  \{\{b_1,c_1\},\{c_2,d_2\} ,\{a_4,e_4\},\{a_5,c_5\},\{b_7,d_7\},\{c_9,e_9\},
\{a_0,a_6\},\{b_0,b_8\},\{d_3,d_6\},\{e_3,e_8\}
\}\nonumber\\
M_3'&=& \{\{b_1,c_1\},\{d_3,e_3\}, \{a_4,e_4\},\{a_6,d_6\} ,\{b_8,e_8\},\{c_9,e_9\},
\{a_0,a_5\},\{b_0,b_7\},\{c_2,c_5\},\{d_2,d_7\}
\}\nonumber\\
M_4'&=& \{\{a_5,c_5\},\{a_6,d_6\} ,\{b_7,d_7\},\{b_8,e_8\},\{c_9,e_9\},
\{a_0,a_4\},\{b_0,b_1\},\{c_1,c_2\},\{d_2,d_3\},\{e_3,e_4\}
\}\nonumber
\eeq
As a consequence of the properties of sets $S$ and $S'$, each dimer appears in both sets $S_M$ and $S_M'$ with the same multiplicity.

Let $A$ be an arbitrary antisymmetric incidence matrix on  $\mathcal D(K_5)$. The contributions of the cycles in $S$ and $S'$
to the Pfaffian expansion of $A$ (equation \eqref{FA} ) are related by
\be\label{K5-obstruction}
\prod_{i=1}^4 F_A(M_0,\gamma_i) = - \prod_{i=1}^4 F_A(M_0,\gamma_i')
\ee 
This relation derives from the properties of the two sets $S$ and $S'$. It can also be seen also as a property of the
perfect matchings in $S_M$ and $S_M'$ alone and any other choice of reference perfect matching would 
define two equivalent sets of cycles through equation \eqref{mapping}.   Thus for every antisymmetric incidence matrix on  $\mathcal D(K_5)$
and every reference perfect matching there exist two sets of closed curves such that relation \eqref{K5-obstruction} holds. By Lemma
\ref{lem:fa}, the partition function on  $K_5$ has no Pfaffian representation.

\hfill\halmos

\bigskip\noindent 
{\bf Proof of Lemma \ref{lem:4r}.}\par

Let $G=(V,E)$ be a finite $2$-connected simple graph $2$-cell embedded in some smooth surface $\Sigma$. We construct  $4$-regular graph $\tilde G \succcurlyeq G$ by a succession of elementary transformations such as vertex splitting, edge addition and edge subdivision \cite{MT}, taking care that at each step, the resulting graph is still a $2$-connected simple graph embeddable in the same surface.

We first state the following three claims, which we prove for any $2$-connected simple graph $G$ $2$-cell embedded in some smooth surface $\Sigma$:

$\bullet$ Claim $1$: If some vertex in graph $G$ has odd valency, there exists a  $2$-connected simple graph $G'\succcurlyeq G$embeddable in the same surface, with a strictly smaller number of vertices of odd valency.

$\bullet$ Claim $2$: If all vertices in $G$ have even valency and some vertex in $G$ has valency larger than $4$, there exists a  $2$-connected simple graph $G'\succcurlyeq G$ embeddable in the same surface, with all vertices of even valency and a strictly smaller number of vertices of valency larger than $4$.

$\bullet$ Claim $3$:   If some vertex in $G$ has valency $2$ and all others of valency $4$,  there exists a  $2$-connected simple graph $G'\succcurlyeq G$embeddable in the same surface, with a strictly smaller number of vertices of valency $2$, and all others of valency $4$.

Clearly, the proof of Lemma \ref{lem:4r} follows easily from the above three claims and transitivity of $\cdot \succcurlyeq \cdot$ : a finite iteration of Claim $1$ leads to a graph $G'\succcurlyeq G$ with all vertices of even valency. Claim $2$ allows then for the construction of a graph $G''\succcurlyeq G'$ with all vertices of valency $2$ or $4$.
Finally a repeated use of Claim $3$ gives a $4$-regular graph $G'''\succcurlyeq G''$. 

We now prove the above three claims.

{\bf Proof of Claim 1.}  
Suppose  that $G$ is a $2$-connected simple graph $2$-cell embedded in some smooth surface $\Sigma$ and that some vertex has odd valency. Since the number of such vertices is necessarily even, one can pick a pair of them, says $v_1$, $v_2$. If they belong to the same face boundary,
we construct a graph $G'$ by adding an edge $\{v_1,v_2\}$ (or a new vertex $v$ and two edges  $\{v_1,v\}$, $\{v,v_2\}$ if $v_1$ and $v_2$ 
are adjacent on $G$). If  $v_1$, $v_2$ don't belong to the same face boundary, one considers a finite sequence of faces $\{F_j\}_{j=0,k}$ such that $v_1\in \overline  F_0$, $v_2\in \overline F_k$ and the boundaries of $F_{j-1}$ and $F_{j}$ share at least one edge, says $e_j$, for all $1< j\le k$. 
We chose this sequence to be of minimal length and construct  $G'$ by replacing each edge $e_j$ by a new vertex $w_j$ and two new edges, each with  endvertices $w_j$ and a distinct endvertex of $e_j$ (namely, subdividing each edge $e_j$), and adding $k+1$ edges $\{v_1,w_1\}$, $\{w_1,w_2\}$,$\cdots$,$\{w_k,v_2\}$.
In both cases, $v_1$ and $v_2$ have even valency in $G'$, while all new vertices have valency $2$ or $4$. In addition $G'$ is a $2$-connected
simple graph since $G$ is and the endvertices of new edges are not neighbors in $G$. Finally $G'$ can be embedded in the same surface $\Sigma$ since the transformation consists in splitting one or more faces of the embedding of $G$.

{\bf Proof of Claim 2.}  
Suppose now that all vertices of $G$ have even valency and let $v$ a vertex of valency $r>4$.
We construct a new graph $G'$ by replacing this vertex in $G$ and its $r$ edges  by a $4$-regular tree $T_v$ with $\frac{r-2}{2}$ points, $r$ external edges and $\frac{r-4}{2}$ internal edges. Since a sufficiently small neighborhood of $x_v$ is homeomorphic to  a disk and $T_v$ of genus zero, one can identify the external edges of this tree with the edges of $v$ in $G$ in such a way that $G'$ can still be embedded on $\Sigma$. 
By construction $G'$ is still a simple graph and it is $2$-connected since no new vertex is separating.

{\bf Proof of Claim 3.}  
Suppose that some vertices of $G$ have valency $2$  and all others valency $4$. Let $v$ be a vertex of valency $2$. If $G$ is a simple graph distinct from $K_3$, one of the faces of the embedding which contains $v$ has length at least $4$, so one can pick two distinct edges on the boundary of this face with endvertices distinct from $v$. Now one construct $G'$ from $G$ by adding one subdivision vertex on these two edges , says $w_1$
and $w_2$ and the three edges $\{v,w_1\}$, $\{v,w_2\}$ and $\{w_1,w_2\}$. the three points $v$, $w_1$ and $w_2$ have valency $4$ in $G'$
so $G'$ has one point of valency $2$ less than $G$. Furthermore the transformation consists in splitting a face so that $G'$ can still be embedded in $\Sigma$. Furthermore $G'$ is again a $2$-connected simple graph. If $G$ identifies to the graph $K_3$ , one may consider $K_{2,2} \succcurlyeq K_3$, and proceed as above.

\hfill\halmos

\bigskip\noindent 
{\bf Proof of Proposition \ref{prop:planar4r}.}\par 

Let $G=(V,E)$ be a planar $2$-connected $4$-regular graph and $\mathcal B_G$ a cycle basis on $G$ as in McLane's planarity criterion \ref{thm:ml}. An important property of such a basis is that none of its subsets can form a double cover of a proper subset of $E(v)$, for any vertex $v\in V$. For completeness,
we give here a short proof of this property \cite{BD}:

 Suppose that this property is false at some vertex $v$, and call a cluster the proper subset of $E(v)$ covered twice. Since $\mathcal B_G$ verifies McLane's criterion, every element of $\mathcal B_G$ 
either belong to the covering and has two edges in the cluster, or does not belong to it, and has thus  $0$ edges in it. The intersection with the cluster thus contains an even number of edges and this holds true also for all elements generated from $\mathcal B_G$ by finite difference.
However, by hypothesis, there is at least one edge at $v$ outside the cluster, and since $G$ is  $2$-connected, there exists a cycle in ${\mathcal C}(G)$ containing whichever pair of edges in $E(v)$; and in particular exactly that edge and one edge in the cluster. This is
in contradiction with the fact that $\mathcal B_G$ is a cycle basis for ${\mathcal C}(G)$, which proves the above property.

We now consider an incidence matrix $A^G$ and notations as in \eqref{6a}--\eqref{def:seg}. We prove that conditions of Proposition \ref{prop:cycle_eq} can hold simultaneously for all cycles in $\mathcal B_G$.

We first consider the site equations \eqref{seq}. Let $v\in V$ and $\gamma\in \mathcal B_G$ such that $\gamma\ni v$. When rewritten in the original variables (see Table \ref{table:1}), the related site equation may appear in three different forms, depending on which edges in $E(v)$ belong to $\gamma$:

\beq\label{seqov}
s_v {\bar s}_v +\; t_v {\bar t}_v =0&&\hbox{ if  }\, \gamma\cap E(v)=\{e_1,e_4\} \hbox{ or  }\{e_2,e_3\} \nonumber\\
t_v {\bar t}_v\, + u_v {\bar u}_v =0&&\hbox{ if  }\, \gamma\cap E(v)=\{e_1,e_2\} \hbox{ or  } \{e_3,e_4\} \\
u_v {\bar u}_v + \,s_v {\bar s}_v =0&&\hbox{ if  }\, \gamma\cap E(v)=\{e_1,e_3\} \hbox{ or  } \{e_2,e_4\} \nonumber
\eeq
\smallskip
Equations at different vertices are obviously independent.  Suppose that there are two cycles in $\mathcal B_G$ passing through $v$ and associated to a different equation in \eqref{seqov}. They share necessarily exactly one edge in $E(v)$. Now take any other cycle in $\mathcal B_G$ passing through $v$. By McLane's criterion, it can neither share the common edge (already appearing twice), nor share a distinct edge with each (which would form a cluster of $3$). Thus it is necessarily disjoint in $E(v)$
from one of the other cycle, and the associated equation is the same. Thus at every vertex $v$, the site equations associated to all elements of  $\mathcal B_G$ are at most two, and admit a nowhere zero solution. 

Note that all three equations together have no nowhere zero solution as it would give $s_v {\bar s}_v = t_v {\bar t}_v =u_v {\bar u}_v =0$.

We suppose from now on that site entries of the incidence matrix have been given non zero values so that
all site equations hold for all $\gamma\in \mathcal B_G$, and we consider the set of edge equations. 

We consider a cycle $\gamma=(V_\gamma, E_\gamma)$ in ${\mathcal B}_G$ with its elements ordered as in Equations \eqref{vgamma}--\eqref{egamma}. For every  
pair $(v,e)\in V_\gamma \times E_\gamma$ with $v$ incident on $e$, we define the following ratio
\beq\label{def:ratio}
R^\gamma_{v,e} = \begin{cases}
\displaystyle{
{{ U^{\scriptscriptstyle{\gamma}}_i} 
{\bar T^{\scriptscriptstyle{\gamma}}_i} \over 
S^{\scriptscriptstyle{\gamma}}_i} }
&\hbox{ if } \exists\,  i\in\{1,\cdots,r_\gamma\}  \hbox{ such that } e = e_i^\gamma \hbox{ and } v=v_i^\gamma \\
\displaystyle{
{{U^{\scriptscriptstyle{\gamma}}_{i+1}} T^{\scriptscriptstyle{\gamma}}_{i+1}} \over 
{\bar S^{\scriptscriptstyle{\Gamma}}_{i+1}} }
&\hbox{ if }  \exists\,  i\in\{1,\cdots,r_\gamma\}  \hbox{ such that } e = e_i^\gamma \hbox{ and } v=v_{i+1}^\gamma 
\end{cases}
\eeq

The ratio $R^\gamma_{v,e}$  depends on the intersection of $\gamma$ with $E(v)$, and is independent on its orientation. On each vertex, this ratio may take twelve different expressions in terms of the edge entries, which can be computed from Table \ref{table:1} and are reported here in Table \ref{table:2}. 

\begin{table}[htb]
\begin{center}
\hskip.1truecm\vbox{\offinterlineskip
\halign{
\vrule#&\strut\enskip\hfil#\hfil\enskip&&\vrule#&\strut\quad\hfil#\quad\cr
\noalign{\hrule}
            height5pt&                            \omit& &\omit& &\omit& &\omit& &\omit&\cr
                         &                         $\;\;e\;\;^{\displaystyle e'}$& &$^{\displaystyle e_v^1}$\;\;&    &$^{\displaystyle e_v^2}$\;\;&   &$^{\displaystyle e_v^3}$\;\;&     &$^{\displaystyle e_v^4}$\;\;&   \cr
           height3pt&                            \omit& &\omit& &\omit& &\omit& &\omit&\cr
\noalign{\hrule}
           height4pt&                            \omit& &\omit& &\omit& &\omit&  &\omit&\cr
                                                            &$e_v^1\;\;$& &
                                                            \omit&&   
                                                            $\displaystyle{ s_v \,u_v \over \bar t_v}$& &        
                                                              $\displaystyle{ t_v\, s_v \over \bar u_v}$&&
                                                              $\displaystyle{ u_v\, t_v \over \bar s_v}$&\cr
           height10pt&                            \omit& &\omit& &\omit& &\omit& &\omit&\cr
                                                            &$e_v^2\;\;$& &          
                                                            $\displaystyle{ s_v\, \bar u_v \over t_v}$& &
                                                            \omit&&
                                                            $\displaystyle{ \bar u_v \,\bar t_v \over \bar s_v}$& &
                                                             $\displaystyle{ \bar t_v \,s_v \over u_v}$&\cr
           height10pt&                            \omit& &\omit& &\omit& &\omit& &\omit&\cr
                                                            &$e_v^3\;\;$& &   
                                                            $\displaystyle{ t_v \,\bar s_v \over u_v}$ &&       
                                                             $\displaystyle{ \bar u_v \,t_v \over s_v}$& & 
                                                             \omit& &
                                                            $\displaystyle{\bar s_v\, \bar u_v\over \bar t_v}$&\cr
           height10pt&                            \omit& &\omit& &\omit& &\omit& &\omit& \cr
                                                            &$e_v^4\;\;$& &          
                                                            $\displaystyle{ u_v \,\bar t_v \over s_v}$& &  
                                                        $\displaystyle{ \bar t_v \,\bar s_v \over \bar u_v}$&&
                                                             $\displaystyle{ \bar s_v \, u_v\over t_v}$& &
                                                              \omit&\cr
           height5pt&                            \omit& &\omit& &\omit& &\omit& &\omit&\cr
\noalign{\hrule}}
\medskip
}
 \caption{Expression of the ratio $R^\gamma_{v,e}$ for a vertex $v$, an edge $
 e$ and a cycle $\gamma$ such that $\gamma\cap E(v)=\{e,e'\}$.
 This expression is written as a function of the location of  $e$, $e'$ in the ordered set  $E(v)=\{e_v^1,e_v^2,e_v^3,e_v^4\}$.
 }
 \label{table:2}
\end{center}
\end{table}

Let $e$ be an edge on graph $G$ with endvertices $v$ and $w$, and $\gamma$ be a cycle in $\mathcal B_G$
containing $e$. The associated edge equation reads
\beq\label{eeqo}
b_e^2 = - R^\gamma_{v,e} R^\gamma_{w,e}
\eeq

Now  consider two distinct cycles  $\gamma$ and $\gamma'$ in ${\mathcal B}_G$ passing through a same
edge $e$ with endvertex $v$. By McLane's criterion, they  cannot have
two successive edges in common, since it would form a cluster of two edges. Thus, up to a permutation, and possibly, exchanging the roles of $\gamma$ and $\gamma'$, one can suppose that $\gamma \cap E(v) = \{e_v^1,e_v^4\}$ and $\gamma' \cap E(v) = \{e_v^1,e_v^2\}$,
so that $e_v^1$ is their common edge at $v$. From Table \ref{table:2}, the associated ratios read
\beq
R^\gamma_{v,e_v^1} = \frac{u_v t_v}{\bar s_v}\nonumber\\
R^{\gamma'}_{v,e_v^1} = \frac{s_v u_v}{\bar t_v}\nonumber
\eeq
The site equation associated to $\gamma$ at $v$  reads
\beq
s_v {\bar s}_v +\; t_v {\bar t}_v =0
\nonumber
\eeq
which implies
\beq
R^\gamma_{v,e_v^1} = - R^{\gamma'}_{v,e_v^1} \nonumber
\eeq
This relation is invariant under any even permutation on $E(v)$, so we get the following result. Two distinct cycles  $\gamma$ and $\gamma'$ in ${\mathcal B}_G$ passing through a same
edge $e$ with endvertex $v$ have opposite ratio:
\beq
R^\gamma_{v,e} = - R^{\gamma'}_{v,e} 
\eeq
Thus, since the left hand side of Equation \eqref{eeqo} is quadratic in the ratios,it is invariant when considering different cycles in $\mathcal B_G$ passing through the same edge.

Thus edge equations are identical for different cycles at the same edge, and are also algebraically independent for different edges. Hence, the whole set of edge equation can be solved simultaneously for all edges and all cycles in $\mathcal B_G$.

We now assume that the coefficients of the incidence matrix have been chosen so that both site and edge equations are fulfilled, and consider the cycle equations \eqref{ceq}, for all cycles in $\mathcal B_G$. We first note that these equations are not fully independent from the edge and site equations.
Multiplying all edge equations associated to a given cycle $\gamma\in \mathcal B_G$ gives
\be
\prod_{i=1}^{r_\gamma} \bigl(B^{\scriptstyle{\gamma}}_i\bigr)^2 =  (-1)^{r_\gamma}
\prod_{i=1}^{r_\gamma} \frac{{T}^{\scriptstyle{\gamma}}_i {\bar T}^{\scriptstyle{\gamma}}_i\bigl({\bar U}^{\scriptstyle{\gamma}}_i\bigr)^2}{{S}^{\scriptstyle{\gamma}}_i {\bar S}^{\scriptstyle{\gamma}}_i}
=
\prod_{i=1}^{r_\gamma} \bigl({\bar U}^{\scriptstyle{\gamma}}_i\bigr)^2
\ee
where the last equality comes from the use of site equations \eqref{seq}.
Thus for every cycle $\gamma\in \mathcal B_G$, we have
\be\label{seq2}
\prod_{i=1}^{r_\gamma} B^{\scriptstyle{\gamma}}_i =  
- \epsilon_\gamma \prod_{i=1}^{r_\gamma}{\bar U}^{\scriptstyle{\gamma}}_i
\ee
with $\epsilon_\gamma \in\{-1,1\}$. 

We thus define a new set of variables $\{\epsilon_e \}_{e\in E}$ with value in $\{-1,1\}$,
and consider the set of equations

\be\label{rceq}
\prod_{e\in\gamma} \epsilon_e =  
\epsilon_\gamma
 \ee
for all $\gamma$ in ${\mathcal B}_G$.

Now the independence of the cycles as
elements of the (vector) basis ${\mathcal B}_G$  implies the independence 
of equations in \eqref{rceq} and thus insures the existence of a solution 
with $\{\epsilon_e \}_{e\in E}$ in $\{-1,1\}$. 
Now for every edge $e$ such that $\epsilon_e=-1$, we change the sign of the two associated edge entries in the incidence matrix. The new edge and site entries are all nonzero and such that 
Equations \eqref{seq}--\eqref{ceq} hold for all cycles in $\mathcal B_G$. 
Thus, by Proposition \ref{prop:cycle_eq}, there exists a reference perfect matching $M_0$ such that 
Equation \eqref{cycle_eq0} holds for all cycle $\gamma\in \mathcal B_G$. By Lemma \ref{lem:fa}, 
the partition function on $G$ admits a Pfaffian representation.

\hfill\halmos

\bigskip\noindent 
{\bf Proof of Lemma \ref{lem:perm}.}\par 

For every vertex $v$, the set of permutations on $E(v)$ is homeomorphic to $\mathcal S_4$. There are exactly two permutations 
$\sigma$ in $\mathcal S_4$, with a prescribed value of two images, which differ by a transposition. Thus they have a different signature
and only one has a positive one. The twelve permutations are listed explicitely in the first column of Table \ref{table:1}.
\hfill\halmos

 \bigskip\noindent 
{\bf Proof of  Lemma \ref{faces=cycles}.}\par 

Suppose $G_1$ is a $2$-regular simple graph embedded in some surface $\mathcal S$. Using Lemma \ref{lem:4r}, 
there exists a $4$-regular, $2$-connected simple graph $G_0' \succcurlyeq G_1$ embedded in the same surface.
Let $K$ be the number of  faces in this embedding which closure is not homeomorphic to a closed disk. If $k\not=0$, we construct recursively
a sequence of graphs $G_1'$, $\dots$, $G_K$ such that   $G_k' \succcurlyeq G_{k-1}'$ for all $k\in \{1,\cdots,K\}$ so that $G_K'$ has a strong embedding in $\mathcal S$.

For $k$ in $\{0,\cdots,K-1\}$, suppose that  $G_k' $ is embedded in $\mathcal S$ . We pick one face which closure is not homeomorphic to a closed disk, and denote by $\gamma$ the closed
path around this face, which can be described as a succession of vertices and edges as
\beq
\gamma=\{v_1,e_1,\cdots,v_i,e_i,\cdots,v_r,e_r\}
\eeq
where $r$ is the length of the path and $\{v_i\}_{i\in\{1,\cdots,r\}}$ (respectively   $\{e_i\}_{i\in\{1,\cdots,r\}}$) are vertices (respectively edges)
in $G_k' $ such that $v_i$ is incident with $e_{i-1}$ and $e_i$ for all $i\in \{1,\cdots,r\}$ (we assume that indices are defined modulo $r$).
Since $\gamma$ is a path round a face, every edge in $\gamma$ appears at most twice and successive edges have to be distinct (since $G_k'$ has no
vertex incident with only one edge).  Denote by $I_1\subset \{1,\cdots,r\}$ the indices of edges appearing only once in $\gamma$ and by $I_2$ the set of pairs of indices associated to edges appearing twice in $\gamma$
\beq
I_2 =\{(i,j), 1\le i< j\le r, e_i=e_j\}
\eeq
We construct $G_{k'+1}$ starting from $G_{k}'$ as follows: on each edge on the path, we add one (respectively two) subdivision points
according to its number of occurrences in $\gamma$. In other words, we add $r$ vertices $\{w_i\}_{i\in\{1,\cdots,r\}}$ to $G_{k}'$, delete all edges in $\gamma$ and replace $e_i$ by two edges $(v_i,w_i)$ and $(w_i,v_{i+1})$ if $i\in I_1$, or by three edges $(v_i,w_i)$ $(w_i,w_j)$ and $(w_j,v_{i+1})$ if $(i,j)\in I_2$.
Finally we add $r$ edges between subdivision points, $(w_1,w_2)$,  $(w_2,w_3)$, $\dots$, $(w_r,w_1)$. Note that each new edge has its endvertices on successive edges on the boundary of the face, they can be drawn on it without crossing, so that the resulting graph $G_{k+1}'$ can be embedded on the same surface $\mathcal S$. Furthermore, it is $4$-regular, since new vertices have four edges incident on it, and $2$-connected since $G_k'$ is. $G_{k'+1}$ has $r$ more vertices and $2 r$ more edges than $G_{k}$. On the embedding on $\mathcal S$, the $r+1$ new faces have length $3$, $4$, $5$ or $r$, and are all bounded by cycles since all their vertices are distinct. Other faces have not changed except by possibly adding one subdivision point on some edges, and there is one face less which boundary is not a cycle. Finally $G_{k'+1}\succcurlyeq G_k $
since it is constructed from it by adding subdivision points and edges.

Thus, starting from $G_0'$ and iterating $K$ times this construction, we
obtain a graph $G_K'$ embedded in $\mathcal S$ with all faces bounded by cycles. Graph $G_K'$ is $4$-regular and $2$-connected and by transitivity, we also have $G_K'\succcurlyeq G_1$

\hfill\halmos

 \bigskip\noindent 
{\bf Proof of  Lemma \ref{nonorientable=complex}.}\par

The proof of this lemma is partly similar to that of Proposition \ref{prop:planar4r}. In fact, as in the planar case, no subset of  $\mathcal F_0^*(G)$ can form a cluster, that is a double cover of a proper subset of E(v)) at some vertex $v\in V$. This implies that all results in the proof of  Proposition \ref{prop:planar4r}
hold true in  the non planar orientable case, for the restricted
set of cycles $\mathcal F_0^*$ which generates the face system and not for the whole basis. 

For every $v$ in $V$, the associated vertex equations consist in a system of two equations out of the three possible ones written in Equation \eqref{seqov}, which has clearly solutions in $\R$.
Therefore we suppose from now on that all vertex entries of the incidence matrix have been given a real non zero value so that the set of vertex equations for all vertices and all cycles in $\mathcal F_0^*(G)$, are fulfilled.

We now consider the ratios $R_{v,e}^\gamma$ defined in \ref{def:ratio}. We have already noted that their value does not depend on the orientation of the associated cycle $\gamma$, and that if two cycles pass trough the same edge, the ratios have opposite value. Here we show that the two ratios 
associated to the same cycle $\gamma$ at a given vertex, says $v_i^\gamma$, have values
with opposite signs.
Indeed, from the definition \eqref{def:ratio}, we have
\beq
R_{v_i^\gamma,e_i^\gamma}^\gamma= \frac{U_i^\gamma \bar T_i^\gamma}{S_i^\gamma}\\
R_{v_i^\gamma,e_{i-1}^\gamma}^\gamma= \frac{U_i^\gamma  T_i^\gamma}{\bar S_i^\gamma}
\eeq
while the vertex equation reads
\beq
 S_i^\gamma\bar S_i^\gamma+ T_i^\gamma\bar T_i^\gamma= 0
 \eeq
 and thus 
 \beq
 R_{v_i^\gamma,e_i^\gamma}^\gamma\;R_{v_i^\gamma,e_{i-1}^\gamma}^\gamma
 = - \bigl(U_i^\gamma \bigr)^2 < 0
\eeq

Therefore the ratios at every vertex are  alternating signs;
at a given vertex $v$,  either  $R_{v,e}^\gamma >0 $ and  $R_{v,\pi_v(e)}^\gamma <0$ for all edges $e\in E(v)$ and all cycles  $\gamma$ such that $\gamma\cap E(v) =\{e,\pi_v(e)\}$, or all signs are simultaneously reversed.
We also note that given a solution $(s_v,t_v,u_v,\bar s_v,\bar t_v, \bar u_v)$, one gets another solution by  reversing the signs of $u_v$ and $\bar u_v$, for which all ratio at $v$ change sign.
Thus given a rotation system $\Pi=(\pi_v)_{v\in V}$, one can choose the solutions of the site equations 
so that the ratios $R_{v,e}^\gamma >0$ simultaneously for all $v\in V$, all $e\in E(v)$ and all $\gamma
\in \mathcal F_0^*(G)$ such that $\gamma\cap E(v) =\{e,\pi_v(e)\}$. 

Now consider an edge $e\in E$ with endvertices $v$ and $w$. If its signature $\lambda(e) =+1$,
a cycle $\gamma$ passing through $e$ behaves  either according 
to $\pi_v$ on $v$ and  $\pi_w$ on $w$, or according 
to $\pi_v^{-1}$ on $v$ and  $\pi_w^{-1}$ on $w$. Thus either $\gamma\cap E(v) =\{\pi_v^{-1}(e),e \}$ and 
 $\gamma\cap E(w) =\{e, \pi_w(e) \}$, or $\gamma\cap E(w) =\{\pi_w^{-1}(e),e \}$ and  $\gamma\cap E(v) =\{e, \pi_v(e) \}$. Therefore the two ratios $R_{v,e}^\gamma$ and $R_{w,e}^\gamma$ have opposite sign and the associated edge equation
 \beq
 b_e^2= - R_{v,e}^\gamma R_{w,e}^\gamma >0
 \eeq
 has a  solution in $\R$.
 In the other case, $\lambda(e) =-1$, a cycle $\gamma$ passing through $e$ behaves  either according 
to $\pi_v$ on $v$ and  $\pi^{-1}_w$ on $w$, or according 
to $\pi_v^{-1}$ on $v$ and  $\pi_w$ on $w$. Thus either $\gamma\cap E(v) =\{\pi_v^{-1}(e),e \}$ and 
$\gamma\cap E(w) =\{\pi_w^{-1}(e),e \}$
, or  $\gamma\cap E(w) =\{e, \pi_w(e) \}$ and  $\gamma\cap E(v) =\{e, \pi_v(e) \}$. In that case, 
 the two ratios $R_{v,e}^\gamma$ and $R_{w,e}^\gamma$ have the same sign and the associated edge equation
 \beq
 b_e^2= - R_{v,e}^\gamma R_{w,e}^\gamma <0
 \eeq
 has only purely imaginary solutions.

If the surface is orientable, there exists an embedding scheme with $\lambda(\cdot)=+1$, which implies that there exists an incidence matrix $A^G$ with real coefficients such that Equation \eqref{cycle_eq0c} holds. 

Suppose now that there exists an incidence matrix with real coefficients
such that Equations \eqref{seq}--\eqref{ceq} hold for all  cycles in  $\mathcal F_0^*(G) $.
Starting from an embedding scheme $(\Pi,\lambda)$, we define a new rotation system $\tilde \Pi =(\tilde \pi_v)_{v\in V}$ such that on each vertex $v\in V$, 
\beq
\tilde \pi_v = 
\begin {cases} 
\pi_v  &\hbox{ if } \exists \gamma \in \mathcal F_0^*(G) \hbox{ such that }  \gamma\cap E(v) = \{e,\pi_v(e)\} \hbox{ and }  R_{v,e}^\gamma < 0\cr
 \pi_v ^{-1} &\hbox{otherwise} 
 \end{cases}
\eeq
Due to the alternating sign properties of the ratios studied previously, one gets that
for all $v\in V$ and all $\gamma \in \mathcal F_0^*(G)$ such that  $\gamma\cap E(v) = \{e,\tilde \pi_v(e)\}$, one has  $ R_{v,e}^\gamma < 0$. Since the right hand side of Equation \eqref{eeqo}
has to be positive for all edges $e\in E$,  all cycles in $\mathcal F_0^*(G)$ can be oriented so that
on each vertex the ratio on the incomming edge is negative, which in turn implies that they 
behave everywhere according  to the rotation system $\tilde \Pi$. The surface is thus orientable.
\hfill\halmos

 \bigskip\noindent

\section {Grassmann Algebra and representation of Pfaffians.}
\label{sec:5}

In this section, we introduce a Grassmann Algebra and use it to represent  Pfaffians
as Grassmann integrals. We refer to references \cite{B} and \cite{S} for a thorough introduction on this matter. We first recall the basic properties which are useful here, and use this representation to give a proof of Proposition \ref{prop:cycle_eq}. We end this section by a short proof of Proposition \ref{eq-hiera} using the same representation.

A Grassmann algebra over $\R$ (or $\C$) is an associative algebra 
with a set of generators $\{\xi_i\}_{i\in \{1,\cdots,n\}}$  obeying the relations
\be 
\xi_i \xi_j = - \xi_j \xi_i \qquad \forall  i,j \in \{1,\cdots,n\}
\ee
and in particular
\be
\xi_i^2 = 0 \qquad  \forall i \in \mathcal \{1,\cdots,n\}
\ee

Integration on the Grassmann algebra \cite{B} is defined through the following formulas, 

\be
\int d\xi_i =0 \qquad\qquad \int d\xi_i \;\xi_i = 1 
\ee

and multiple integrals are defined as iterates 
\be
\int d\xi_{i_r} d\xi_{i_{r-1}} \cdots d\xi_{i_1}\; \xi_{i_1} \cdots \xi_{i_{r-1}} \xi_{i_r} =1 \qquad\hbox{ for all } 1\le i_1 < \cdots< i_{r-1}< i_r\le n
\ee
With this prescription, integrals of products of factors depending on disjoints subsets of Grassmann variables is just the product of the integrals,

\beq\label{integralproduct}
&&\int d\xi_{i_{r+s}} \cdots d\xi_{i_{r+1}}  d\xi_{i_r} \cdots d\xi_{i_1} \phi_1\bigl(\xi_{i_{1}}, \cdots ,\xi_{i_{r}}\bigr) \phi_2\bigl(\xi_{i_{r+1}}, \cdots ,\xi_{i_{r+s}}\bigr) =\nonumber\\
&&\qquad\qquad
\Bigl[\int d\xi_{i_r} \cdots d\xi_{i_1} \phi_1\bigl(\xi_{i_{1}}, \cdots ,\xi_{i_{r}}\bigr)\Bigr]
\Bigl[\int d\xi_{i_{r+s}} \cdots d\xi_{i_{r+1}} \phi_2\bigl(\xi_{i_{r+1}}, \cdots ,\xi_{i_{r+s}}\bigr)\Bigr]
\eeq

Here, our main interest in the introduction of a Grassmann algebra is the following. Let $A$ be an antisymmetric matrix of even order $n$,
then its Pfaffian can be expressed as the following Grassmann integral \cite{S}:
\be\label{pfaffgrassmann}
Pf(A) = \int d\xi_{n} \cdots d\xi_{1} \exp\bigl\{ \frac{1}{2} \sum_{i,j=1}^n \xi_i A_{i,j} \xi_j\bigr\}
\ee

\bigskip\noindent
{\bf Proof of Proposition \ref{prop:cycle_eq}}

For the purpose of using identity \eqref{pfaffgrassmann} in the context of Proposition \ref{prop:cycle_eq}, we introduce a Grassmann algebra of order $n= |V_{\mathcal D}(G)|$, with generators
indexed by the darts on $G$, $\{\xi_{d}\}_{\{d\in V_{\mathcal D}(G) \}}$ and write the Pfaffian of the antisymmetric incidence matrix $A$ defined in \eqref{def:A} as
\be\label{def:pfaffAgrassmann}
Pf(A) = \int \overleftarrow{\phantom{\prod}}\hskip-.9cm \prod_{d\in V_{\mathcal D}(G)} \hskip-.3cm d\xi_d \;\;
\exp\bigl\{\mathcal L_G(\underline \xi)\bigr\}
\ee
with
\be\label{qf}
\mathcal L_G(\underline \xi) =\sum_{\{d,d'\}\in E_{\mathcal D}(G)} \xi_{d} A_{d,d'} \xi_{d'} 
\ee 
In Equation \eqref{def:pfaffAgrassmann}, we have also used the shorthand notation $\overleftarrow \prod$ to indicate that the differential elements in the product have to be ordered from left to right in decreasing order with respect to the (lexicographic) order defined on $V_{\mathcal D}(G)$. Note also that in \eqref{qf}, the sum is over 
all edges in $E_{\mathcal D}(G)$, which are defined as unordered pairs of elements in $V_{\mathcal D}(G)$; 
since $A$ is antisymmetric and Grassmann generators anticommute,  $\xi_{d} A_{d,d'} \xi_{d'} =\xi_{d'} A_{d',d} \xi_{d}$, that is,  the value of each summand is independent
on the choice of which representation, $(d,d')$ or $(d',d)$ is chosen for the unordered pair  $\{d,d'\}$, the right hand side of \eqref{qf} is well defined  and Equation \eqref{def:pfaffAgrassmann} identifies with \eqref{pfaffgrassmann}.
Up to a permutation of positive signature, we may write
\beq
\overleftarrow{\phantom{\prod}}\hskip-.9cm \prod_{d\in V_{\mathcal D}(G)} \hskip-.3cm d\xi_d &=& 
\prod_{v\in V} d\xi_{(v,e_v^4)}d\xi_{(v,e_v^3)} d\xi_{(v,e_v^2)} d\xi_{(v,e_v^1)} 
\eeq
where for all $v\in V$,
$e_v^1<e_v^2<e_v^3<e_v^4$ denote the four ordered edges at $v$. 

Given a weight function $w$ on $E$, the Pfaffian of the related weighted incidence matrix
can be expanded as a sum over the set of closed curves on $G$
\be
\Pf(A(w)) = \sum_{C\in \mathcal C(G)} \prod_{e\in C} w(e) F_A(C)
\ee

For any closed curve $C\in \mathcal C(G)$, the expression of $F_A(C)$ in terms of a Grassmann intergral can be obtained through an
expansion of the exponential in \eqref{def:pfaffAgrassmann} over all subsets of $E_{\mathcal D}^E(G)$. Non zero terms necessarily involve 
an even number of generator at each vertex $v\in V$, and thus correspond to closed curves on $G$. The contribution to $F_A(C)$ can
then be readily identified as,
\beq\label{def:fagrassmann}
F_A(C) = && \int  \overleftarrow{\phantom{\prod}}\hskip-.9cm \prod_{d\in V_{\mathcal D}(G)}\hskip-.3cm d\xi_d 
\; \Bigl[\prod_{{\{d,d'\}\in E_{\mathcal D}^E(C)}}  A_{d,d'}  \xi_{d} \xi_{d'}\Bigr]
\exp\bigl\{\sum_{\{d,d'\}\in E_{\mathcal D}^V(G)} \xi_{d} A_{d,d'} \xi_{d'}  \bigr\}
\eeq
where for clarity, we have denoted by $E_{\mathcal D}^E(C)$ the (canonical) image of the closed curve $C$ on the dart graph,
\be
E_{\mathcal D}^E(C) =\bigl\{ \{(v,e),(v',e)\}\in E_{\mathcal D}^E(G) : e\in C \bigr\}
\ee 

Now, let $\Gamma=(V_\Gamma,E_\Gamma)$ be a cycle of length $r$ on $G$. We note $V_{\mathcal D}(\Gamma)$ the set of darts with vertex in $V_\Gamma$ 
\be\label{dcg}
 V_{\mathcal D}(\Gamma) = \bigl\{ (v,e)\in V_{\mathcal D}(G), v\in V_\Gamma, e\in E(v) \bigr\}
 \ee
 and we partition the elements of $E_{\mathcal D}^E(C)$ and $E_{\mathcal D}^V(G)$ according to their
 intersection with $V_{\mathcal D}(\Gamma)$,
 \beq
 &&E_k^E(C) =\bigl\{ \{(v,e),(v',e)\}\in E_{\mathcal D}^E(C): |V_{\mathcal D}(\Gamma)\cap \{(v,e),(v',e)\}| = k
\bigr\}, \quad k= 0,1,2\\
 &&E_k^V =\bigl\{ \{(v,e),(v,e')\}\in E_{\mathcal D}^V(G): |V_{\mathcal D}(\Gamma)\cap \{(v,e),(v,e')\}| = k
\bigr\},  \quad k= 0,2
 \eeq
 These sets are pairwise disjoints and we have
 \beq
 && E_{\mathcal D}^E(C) = E_0^E(C) \cup E_1^E(C) \cup E_2^E(C)\\
 && E_{\mathcal D}^V(G) = E_0^V \cup E_2^V
 \eeq
 We thus rewrite the expression of $F_A(C)$ using this decomposition,
 \beq\label{def:fagrassmann2}
F_A(C) = && \int  \overleftarrow{\phantom{\prod}}\hskip-.9cm \prod_{d\in V_{\mathcal D}(G)}\hskip-.3cm d\xi_d 
\; \Bigl[\prod_{{\{d,d'\}\in E_0^E(C)}}  A_{d,d'}  \xi_{d} \xi_{d'}\Bigr] exp\bigl\{\sum_{\{d,d'\}\in E_0^V} \xi_{d} A_{d,d'} \xi_{d'}  \bigr\}\nonumber\\
&&\qquad\times\Bigl[\prod_{{\{d,d'\}\in E_1^E(C)}}  A_{d,d'}  \xi_{d} \xi_{d'}\Bigr]\nonumber\\
&&\qquad\times\Bigl[\prod_{{\{d,d'\}\in E_2^E(C)}}  A_{d,d'}  \xi_{d} \xi_{d'}\Bigr]
exp\bigl\{\sum_{\{d,d'\}\in E_2^V} \xi_{d} A_{d,d'} \xi_{d'}  \bigr\}\nonumber
\eeq
In order to factorize this expression, we rewrite the product over  $ E_1^E$ 
 as
 \beq
 \prod_{{\{d,d'\}\in E_1^E(C)}}  A_{d,d'}  \xi_{d} \xi_{d'} = 
 \Bigl[\hskip.5cm\overleftarrow{\phantom{\prod}}\hskip-1.2cm\prod^{{(\Gamma)}\atop\phantom{-}}_{{\{d,d'\}\in E_1^E(C)}\atop {d'\in V_{\mathcal D}(\Gamma)}}  A_{d,d'}  \xi_{d}\Bigr]
 \Bigl[\hskip.5cm\overrightarrow{\phantom{\prod}}\hskip-1.2cm\prod^{{(\Gamma)}\atop\phantom{-}}_{{\{d,d'\}\in E_1^E(C)}\atop {d'\in V_{\mathcal D}(\Gamma)}}  \xi_{d'}\Bigr]
 \eeq
 where the notation $\overleftarrow\prod^{{(\Gamma)}\atop\phantom{-}}$ (respectively $\overrightarrow\prod^{{(\Gamma)}\atop\phantom{-}}$) indicates that the factors have to be written in decreasing (respectively increasing) order 
 with respect to an order on $V_{\mathcal D}(\Gamma)$ to be precised later. The above identity just states that 
the reordering involves an even number of transpositions, whatever the order chosen on $V_{\mathcal D}(\Gamma)$.

Using identity \eqref{integralproduct}, we can now write  $F_A(C) $ as a product of two factors
\be\label{facsplit}
F_A(C)  = \overline F_A(C,\Gamma)  F_A(C,\Gamma)
\ee
where
  \beq\label{def:fabgrassmann2}
\overline F_A(C,\Gamma) =  \int \hskip.4cm \overleftarrow{\phantom{\prod}}\hskip-1.4cm \prod_{d\in V_{\mathcal D}(G)\setminus V_{\mathcal D}(\Gamma)}\hskip-.6cm d\xi_d &&
\; \Bigl[\prod_{{\{d,d'\}\in E_0^E(C)}}  A_{d,d'}  \xi_{d} \xi_{d'}\Bigr] 
\Bigl[\hskip.5cm\overleftarrow{\phantom{\prod}}\hskip-1.2cm\prod^{{(\Gamma)}\atop\phantom{-}}_{{\{d,d'\}\in E_1^E(C)}\atop {d'\in V_{\mathcal D}(\Gamma)}}  A_{d,d'}  \xi_{d}\Bigr]
\nonumber\\&&\times
 exp\bigl\{\sum_{\{d,d'\}\in E_0^V} \xi_{d} A_{d,d'} \xi_{d'}  \bigr\}
\eeq

and 
\beq
F_A(C,\Gamma) =  \int  \overleftarrow{\phantom{\prod}}\hskip-.9cm \prod_{d\in  V_{\mathcal D}(\Gamma)}\hskip-.3cm d\xi_d &&
\Bigl[\prod_{{\{d,d'\}\in E_2^E(C)}}  A_{d,d'}  \xi_{d} \xi_{d'}\Bigr]
 \Bigl[\hskip.5cm\overrightarrow{\phantom{\prod}}\hskip-1.2cm\prod^{{(\Gamma)}\atop\phantom{-}}_{{\{d,d'\}\in E_1^E(C)}\atop {d'\in V_{\mathcal D}(\Gamma)}}  \xi_{d'}\Bigr]\nonumber\\
&&\times \exp\bigl\{\sum_{\{d,d'\}\in E_2^V} \xi_{d} A_{d,d'} \xi_{d'}  \bigr\}
\eeq

Since by construction, $C$ and $C\triangle\Gamma$ coincide on $E\setminus \Gamma$, one has necessarily
\beq
&&E_0^E(C\triangle\Gamma)= E_0^E(C)\nonumber\\ 
&&E_0^V(C\triangle\Gamma)= E_0^V(C)
\eeq
Hence
\be\label{bfacg}
\overline F_A(C\triangle \Gamma,\Gamma) = \overline F_A(C,\Gamma)
\ee
We now prove that $F_A(C\triangle\Gamma,\Gamma)$ and $F_A(C,\Gamma)$ are equal.
We first define the following notation for the generators indexed in $V_{\mathcal D}(\Gamma)$:
for all $i\in \{1,\cdots,r\}$ and all $k\in \{1,2,3,4\}$, we write
\be
\eta_i^k = \xi_{(v_i,\sigma_i(e_{v_i}^k))}
\ee
where $\sigma_i$ is the permutation defined in Lemma \ref{lem:perm}. We also define
\be
\psi_i^k(C)= 
\begin{cases}
\eta_i^k &\hbox{ if } \sigma_i(e_{v_i}^ k)\in C\cr
1 & \hbox{ otherwise.}
\end{cases}
\ee
and given $\underline \eta_i=\{ \eta_i^1, \eta_i^2, \eta_i^3, \eta_i^4\}$,
\be
\mathcal L_i(\underline \eta_i) = S_i \eta_i^1 \eta_i^2 + \overline S_i \eta_i^3 \eta_i^4
+ T_i \eta_i^1 \eta_i^3 + \overline T_i \eta_i^4 \eta_i^2
+ U_i \eta_i^1 \eta_i^4 + \overline U_i \eta_i^2 \eta_i^3
\ee
Using these and notations \eqref{def:seg}, \eqref{def:eeg}, we rewrite the expression of
 $F_A(C,\Gamma)$ as
\beq
F_A(C,\Gamma) =  \int  
\hskip.5cm\overleftarrow{\phantom{\prod}}\hskip-1.cm\prod^{{\scriptstyle{(r,4)}}\atop\phantom{-}}_{(i,k)=(1,1)}
\hskip-.3cm d\eta_i^k &&
\Bigl[\prod_{i : e_i \in C}  B_i \; \eta_{i-1}^4 \eta_i^1\Bigr]\nonumber\\
&&\times \Bigl[ 
\hskip.5cm\overrightarrow{\phantom{\prod}}\hskip-.6cm \prod_{i=1}^{\scriptstyle{r}\atop\phantom{-}}
\bigl(\psi_i^2(C) \psi_i^3(C)\bigr) \Bigr]
\exp\bigl\{\sum_{i} \mathcal L_i(\underline \eta_i)  \bigr\}\nonumber\\
= \int  
\hskip.5cm\overleftarrow{\phantom{\prod}}\hskip-1.cm\prod^{{\scriptstyle{(r,4)}}\atop\phantom{-}}_{(i,k)=(1,1)}
\hskip-.3cm d\eta_i^k &&
\Bigl[\prod_{i : e_i \in C}  B_i \Bigr]\; \Bigl[
\prod_{i=1}^r
\psi_{i-1}^4(C) \psi_i^1(C) \Bigr]\nonumber\\
&&\times \Bigl[ 
\hskip.5cm\overrightarrow{\phantom{\prod}}\hskip-.6cm \prod_{i=1}^{\scriptstyle{r}\atop\phantom{-}}
\bigl(\psi_i^2(C) \psi_i^3(C)\bigr) \Bigr]
\exp\bigl\{\sum_{i} \mathcal L_i(\underline \eta_i)  \bigr\}\nonumber\\
\eeq
where the order of the ``differentials''  is induced by the lexicographic order 
on $\{1,\cdots,r\}\times \{1,\cdots,4\}$. Each factor in the first product contains an even number of generators ($2$ or $0$
depending on whether $e_i\in C$ and thus commute with any other term. Thus we can write

\beq
F_A(C,\Gamma) &=& \int  
\hskip.5cm\overleftarrow{\phantom{\prod}}\hskip-1.cm\prod^{{\scriptstyle{(r,4)}}\atop\phantom{-}}_{(i,k)=(1,1)}
\hskip-.3cm d\eta_i^k\; 
\Bigl[\prod_{i : e_i \in C}  B_i \Bigr]\nonumber\\
&&\qquad\times \psi_0^4(C)
 \Bigl[
 \hskip.5cm\overrightarrow{\phantom{\prod}}\hskip-.6cm \prod_{i=1}^{\scriptstyle{r-1}\atop\phantom{-}}
 \bigl(\psi_i^1(C) \psi_i^2(C) \psi_i^3(C) \psi_i^4(C)\bigr)\Bigr]
\Bigl[ \bigl(\psi_r^1(C) \psi_r^2(C) \psi_r^3(C) \Bigr]
\exp\bigl\{\sum_{i} \mathcal L_i(\underline \eta_i)  \bigr\}\nonumber\\
&=&   (-1)^{|\{e_0\}\cap C|} \; \int  
\hskip.5cm\overleftarrow{\phantom{\prod}}\hskip-1.cm\prod^{{\scriptstyle{(r,4)}}\atop\phantom{-}}_{(i,k)=(1,1)}
\hskip-.3cm d\eta_i^k\; 
\Bigl[\prod_{i : e_i \in C}  B_i \Bigr]\; 
 \Bigl[ \prod_{i=1}^{r}\bigl(\psi_i^1(C) \psi_i^2(C) \psi_i^3(C) \psi_i^4(C)\bigr)
\exp\bigl\{\mathcal L_i(\underline \eta_i)  \bigr\}\Bigr]\nonumber
\eeq
where in the last line, the term $\psi_0^4(C)=\psi_r^4(C)$ has been shifted on the right of all other terms, taking into
account that the total number of generators is even, leading to a possible minus sign by anticommutation. 
Factorization and shifts of the exponential terms takes
into account that the arguments are quadratic in the generators and thus commute with other terms. 

$F_A(C,\Gamma)$ can then be expressed as a product of $r$ Grassmann integrals,
\be\label{fawa}
F_A(C,\Gamma) =  (-1)^{|\{e_0\}\cap C|} \;  \bigl[\prod_{i : e_i \in C}  B_i \bigr]  \prod_{1}^r W_i(C,\Gamma)
\ee
with
\beq\label{wic} 
W_i(C,\Gamma) =  \int   d\eta_i^4 \, d\eta_i^3 \, d\eta_i^2 \, d\eta_i^1 \; 
\bigl[\psi_i^1(C) \psi_i^2(C) \psi_i^3(C) \psi_i^4(C) \bigr]\exp\bigl\{\mathcal L_i(\underline \eta_i)  \bigr\}
\eeq

The expression of $W_i(C,\Gamma)$ can be computed for all possible values of the $\{ \psi_i^k(C)\}_{1\le k\le 4}$. Since $C$
is a closed curve, $v_i$ is incident with an even number of edges in $C$, and there are thus eight possible configurations,
giving rise to eight possible expressions for $W_i(C,\Gamma)$, listed in the table below as a function of which edges (indexed relatively to
$\Gamma$) are in $C$

\begin{table}[htb] 
\begin{center}
\hskip.0truecm\vbox
{
\offinterlineskip
\tabskip=0pt
\halign{ 
\vrule  height2.75ex depth1.25ex width 0.6pt #\tabskip=1em &
\hfil $#$\hfil &\vrule # &
\hfil $#$\hfil &\vrule width 0.6pt #\tabskip=1pt &\vrule width 0.6pt # \tabskip=1em &
\hfil $#$\hfil &\vrule # &
\hfil $#$\hfil &\vrule width 0.6pt #\tabskip=0pt\cr
\noalign{\hrule height 0.6pt}
&\{k|\sigma_i(e_i^k)\in C\} && W_i(C) &&
& \{k|\sigma_i(e_i^k)\in C\} & &W_i(C) &\cr
\noalign{\hrule}
& \emptyset && S_i \overline S_i + T_i  \overline T_i +U_i\overline U_i \ &&
&  \{1,4\}  & &\overline U_i& \cr
& \{1,2,3,4\} && 1 &&
&  \{2,3\}  && U_i &\cr
&  \{1,2\} && \overline S_i  &&
&  \{2,4\}  && -  T_i  &\cr
&  \{3,4\} && S_i  &&
&  \{1,3\}  &&  \overline T_i  &\cr
\noalign{\hrule height 0.6pt}
}
}
\medskip
 \caption{Expression of the Grassmann integral $W_i(C)$ defined in Equation \eqref{wic} for the eight possible configurations for $C$ on $E(v_i)$, as a function of their ordering with respect to cycle $\Gamma$ (Lemma \ref{lem:perm}). Note that first and third columns are exchanged under the transformation $C \rightarrow C\triangle \Gamma$.}
 \label{table:wic}
\end{center}
\end{table}  
Now define the following two functions
\beq\label{phic}
\varphi _i(C) &=& 
\begin{cases} 
\displaystyle{
\frac{ U_i \overline T_i}{ S_i} }&\hbox{ if } \sigma_i(e_i^4) \in C\cr
1 & \hbox{ otherwise.}
\end{cases}
\\\label{bphic}
\overline \varphi _i(C) &=& 
\begin{cases} 
\displaystyle{
\frac{ U_i S_i}{  \overline  T_i} }&\hbox{ if } \sigma_i(e_i^1) \in C\cr
1 & \hbox{ otherwise.}
\end{cases}
\eeq

Under the hypothesis of Proposition \ref{prop:cycle_eq},  Equations  \eqref{seq}
are fulfilled. In Table \ref{table:wic2}, we check by direct inspection that for all $i\in \{1,\cdots,r\}$,
the following relation holds between $W_i(C)$ and $W_i(C\triangle \Gamma)$:
\be\label{wicwicg}
\displaystyle{\frac{1}{U_i}}\overline \varphi _i(C) W_i(C) \varphi _i(C) =W_i(C\triangle \Gamma)
\ee

\begin{table}[htb] 
\begin{center}
\hskip.0truecm\vbox
{
\offinterlineskip
\tabskip=0pt
\halign{ 
\vrule  height3.6ex depth2.9ex width 0.6pt #\tabskip=1em &
\hfil $#$\hfil &\vrule # &
\hfil $#$\hfil &\vrule # &
\hfil $#$\hfil &\vrule # &
\hfil $#$\hfil &\vrule # &
\hfil $#$\hfil &\vrule width 0.6pt #\tabskip=1pt &\vrule width 0.6pt # \tabskip=1em &
\hfil $#$\hfil &\vrule width 0.6pt #\tabskip=0pt\cr
\noalign{\hrule height 0.6pt}
&\{k|\sigma_i(e_i^k)\in C\} && \overline \varphi _i(C) &
& W_i(C)& &\varphi _i(C)  &
&\displaystyle{\frac{\overline \varphi _i(C) W_i(C) \varphi _i(C)}{U_i}}&&
& W_i(C\triangle\Gamma))&\cr
\noalign{\hrule}
& \emptyset && 1&& U_i\overline U_i &&1&&\overline U_i&&&\overline U_i&\cr
&  \{1,4\}  & &\displaystyle{\frac{ U_i  S_i}{\overline T_i} }&&\overline U_i&
 &\displaystyle{\frac{ U_i  \overline T_i}{ S_i} } &&U_i\overline U_i&&&U_i\overline U_i&\cr
& \{1,2,3,4\} &&\displaystyle{\frac{ U_i  S_i}{\overline T_i} }&& 1 &&\displaystyle{
\frac{ U_i \overline T_i}{S_i} }&&U_i &&&U_i &\cr
&  \{2,3\}  && 1&& U_i &&1&&1&&&1&\cr
&  \{1,2\} &&\displaystyle{\frac{ U_i  S_i}{ \overline T_i} }&& \overline S_i  &&1&& -  T_i &&& -  T_i &\cr
&  \{2,4\}  && 1&& -  T_i  &&\displaystyle{
\frac{ U_i \overline T_i}{ S_i} }&&\overline S_i &&&\overline S_i &\cr
&  \{3,4\} && 1&& S_i  &&\displaystyle{
\frac{ U_i \overline T_i}{ S_i} }&&\overline T_i &&&\overline T_i &\cr
&  \{1,3\}  &&\displaystyle{\frac{ U_i  S_i}{\overline T_i} }&&  \overline T_i  &&1&&S_i &&&S_i &\cr
\noalign{\hrule height 0.6pt}
}
}
\medskip
 \caption{ Relation between $ W_i(C) $ and $W_i(C\triangle\Gamma))$ under the hypothesis that 
  $S_i \overline S_i + T_i  \overline T_i =0$ (Equation \eqref{seq}). The table shows that relation 
  \eqref{wicwicg} holds for the eight possible local configurations}
 \label{table:wic2}
\end{center}
\end{table}  

We suppose now that Equations \eqref{seq}, \eqref{seq} hold for all $i\in \{1,\cdots,r\}$ together with 
\eqref{ceq}. We have
\beq\label{facgfacgg}
F_A(C,\Gamma) &=&  (-1)^{|\{e_0\}\cap C|} \;  \bigl[\prod_{i : e_i \in C}  B_i \bigr] 
\bigl[ \prod_{i=1}^r W_i(C,\Gamma) \bigr] \nonumber\\
&=& - (-1)^{|\{e_0\}\cap C|} \;  \bigl[\prod_{i : e_i \in C}  B_i \bigr]  
\bigl[\prod_{i =1}^r  B_i \bigr] \bigl[\prod_{i =1}^r  \displaystyle{\frac{1}{U_i}} \bigr] 
\bigl[ \prod_{i=1}^r W_i(C,\Gamma) \bigr]\nonumber\\
&=& - (-1)^{|\{e_0\}\cap C|} \;  \bigl[\prod_{i : e_i \in C}  B_i^2 \bigr]  
\bigl[\prod_{i: e_i\not\in C} B_i \bigr] \bigl[\prod_{i =1}^r  \displaystyle{\frac{1}{U_i}} \bigr]  
\bigl[ \prod_{i=1}^r W_i(C,\Gamma) \bigr]\nonumber\\
&=& - (-1)^{|\{e_0\}\cap C|} \;  
\bigl[\prod_{i: e_i\not\in C} B_i \bigr] 
 \bigl[\prod_{i : e_i \in C}  \displaystyle{\frac{U_i \overline T_i}{S_i}}  \displaystyle{\frac{U_{i+1}S_{i+1}}{ \overline  T_{i+1}}}\bigr] 
\bigl[\prod_{i =1}^r  \displaystyle{\frac{1}{U_i}} \bigr]  
\bigl[ \prod_{i=1}^r W_i(C,\Gamma) \bigr]\nonumber\\
&=& - (-1)^{|\{e_0\}\cap C|} \;  
\bigl[\prod_{i: e_i\not\in C} B_i \bigr] 
 \bigl[\prod_{i =1}^r  \varphi_i(C)  \overline \varphi_{i+1}(C)\bigr] 
\bigl[\prod_{i =1}^r  \displaystyle{\frac{1}{U_i}} \bigr] 
\bigl[ \prod_{i=1}^r W_i(C,\Gamma) \bigr]\nonumber\\
&=& - (-1)^{|\{e_0\}\cap C|} \;  
\bigl[\prod_{i: e_i\not\in C} B_i \bigr] 
\bigl[\prod_{i =1}^r  \displaystyle{\frac{ \overline \varphi_{i}(C) W_i(C,\Gamma) \varphi_i(C) }{U_i}} \bigr] 
\nonumber\\
&=&  (-1)^{|\{e_0\}\cap (C\triangle\Gamma)|} \;  \bigl[\prod_{i : e_i \in C\triangle\Gamma}  B_i \bigr] 
\bigl[ \prod_{i=1}^r W_i(C\triangle\Gamma,\Gamma) \bigr] \nonumber\\
&=& F_A(C\triangle\Gamma,\Gamma)
 \eeq
To get the second line we have introduced \eqref{ceq} as an identity
\be
\bigl[\prod_{i =1}^r  B_i \bigr] \bigl[\prod_{i =1}^r  \displaystyle{\frac{1}{U_i}} \bigr]= -1
\ee
For the fourth line of \eqref{facgfacgg}, we have used Equations \eqref{eeq} for all $i$ such that $e_i\in C$;
To get the fifth line, we used the definitions \eqref{phic}--\eqref{bphic}; In the sixth line appears the left hand side of \eqref{wicwicg} and the seventh is just expression \eqref{fawa} for $C\triangle\Gamma$.

Now inserting Equation \eqref{bfacg} and the result of Equation \eqref{facgfacgg} in Equation \eqref{facsplit},
one gets
\beq
 F_A(C)  &=& \overline F_A(C,\Gamma)  F_A(C,\Gamma)\nonumber\\
 &=&  \overline F_A(C\triangle\Gamma,\Gamma)  F_A(C\triangle\Gamma,\Gamma)\nonumber\\
&=&  F_A(C\triangle\Gamma)
\eeq
Proposition \ref{prop:cycle_eq} Êis proven.
\hfill\halmos

\bigskip\noindent 
{\bf Proof of Proposition \ref{eq-hiera}.}\par 

We conclude this section by a proof of Proposition \ref{eq-hiera}. The Pfaffian reduction formula is obviously not new but we found no clear recent textbook reference. In addition, 
one of the few available references \cite{GH} appears to have some misprints. We report here an easy but rather lengthy proof.

We prove Proposition \ref{hiera} by induction  on $p$.
Let us first prove Equation \eqref{eq-hiera} for $p=1$. We set $K=\{\alpha, \beta\}$, where $\alpha < \beta$ is  a pair of indices  in $\{1,\cdots,2n\}$ such that $A_{\alpha,\beta}\not=0$ and write the Pfaffian of $A$ as an integral over a Grassmann algebra of  dimension $2 n$ (Equation \eqref{pfaffgrassmann}).
Expand the terms in the exponential 
containing elements indexed by $K$ gives,
\beq
Pf (A)&& = \int 
\hskip.2cm\overleftarrow{\phantom{\prod}}\hskip-.6cm\prod^{{\scriptstyle{2 n}}\atop\phantom{-}}_{k=1} d\xi_k \;\;
 \exp \bigl\{ {1\over 2} \sum_{i,j=1}^{2 n}  A_{i,j} \xi_i \xi_j \bigr\}\nonumber\\
&&= \int 
\hskip.2cm\overleftarrow{\phantom{\prod}}\hskip-.6cm\prod^{{\scriptstyle{2 n}}\atop\phantom{-}}_{k=1} d\xi_k \;\;
 \exp \bigl\{ {1\over 2}\sum_{\scriptstyle{{i,j=1}\atop |\{i,j\}\cap \{\alpha,\beta\}|=\emptyset}}^{2 n} A_{i,j} \xi_i \xi_j +
    \sum_{i\not=\alpha} A_{i,\beta} \xi_i \xi_\beta +\sum_{j\ne\beta} A_{\alpha} \xi_q \xi_\beta +
    A_{\alpha,\beta} \xi_\alpha \xi_\beta \bigr\}\nonumber\\
&&= \int 
\hskip.2cm\overleftarrow{\phantom{\prod}}\hskip-.6cm\prod^{{\scriptstyle{2 n}}\atop\phantom{-}}_{k=1} d\xi_k \;\;
 \exp \bigl\{ {1\over 2}\sum_{\scriptstyle{{i,j=1}\atop |\{i,j\}\cap \{\alpha,\beta\}|=\emptyset}}^{2 n} A_{i,j} \xi_i \xi_j\bigr\}\nonumber\\
 &&\qquad\qquad\times \bigl( 1 +  \sum_{i\ne\alpha} A_{i,\beta} \xi_i \xi_\beta \bigr)
 \bigl( 1 +  \sum_{j\not=\beta} A_{\alpha} \xi_\alpha \xi_j\bigr)
 \bigl( 1 +  A_{\alpha,\beta} \xi_\alpha \xi_\beta \bigr) \nonumber\\
&&=\int 
\hskip.2cm\overleftarrow{\phantom{\prod}}\hskip-.6cm\prod^{{\scriptstyle{2 n}}\atop\phantom{-}}_{k=1} d\xi_k \;\;
  \exp \bigl\{ {1\over 2}\sum_{\scriptstyle{{i,j=1}\atop |\{i,j\}\cap \{\alpha,\beta\}|=\emptyset}}^{2 n} A_{i,j} \xi_i \xi_j\bigr\}
 \bigl( A_{\alpha,\beta} -  \sum_{\scriptstyle{{i\not=\alpha}\atop{j\not=\beta}}} A_{i,\beta} A_{\alpha,j} \eta_i \eta_j\bigr) \eta_\alpha \eta_\beta 
\eeq
After integration over $(\eta_{\alpha},\eta_{\beta})$, the expanded part reduces to the sum of a non zero constant ($A_{\alpha,\beta}$), and a product of two linear combinations of Grassmann generators.
Thus successive powers of this second term are zero and one can re-exponentiate this part. We have,
\beq
Pf (A)&&= (-1)^{\beta - \alpha + 1} \int 
\hskip.2cm\overleftarrow{\phantom{\prod}}\hskip-.7cm\prod^{{\scriptstyle{2 n}}\atop\phantom{-}}
_{{\scriptstyle{{k=1}\atop{k\not=\alpha,\beta}}}} d\xi_k \;\;
\exp \bigl\{ {1\over 2}\sum_{i,j\ne \alpha,\beta} A_{i,j} \eta_i \eta_j\bigr\} \times 
\bigl( A_{\alpha,\beta} -  \sum_{p,q\ne \alpha,\beta} A_{p,\alpha} A_{q,\beta} \eta_p \eta_q\bigr)\nonumber\\
&&=(-1)^{\beta - \alpha + 1} \int 
\hskip.2cm\overleftarrow{\phantom{\prod}}\hskip-.6cm\prod^{{\scriptstyle{2 n}}\atop\phantom{-}}_{k=1} d\xi_k \;\;
\exp \bigl\{ {1\over 2}\sum_{i,j\ne \alpha,\beta} A_{i,j} \eta_i \eta_j -  
A_{\alpha,\beta}^{-1}\sum_{p,q\ne \alpha,\beta} A_{p,\alpha} A_{q,\beta} \eta_p \eta_q \bigr\}\nonumber\\
&&=(-1)^{\beta - \alpha + 1} \int 
\hskip.2cm\overleftarrow{\phantom{\prod}}\hskip-.6cm\prod^{{\scriptstyle{2 n}}\atop\phantom{-}}_{k=1} d\xi_k \;\;
\exp \bigl\{ {1\over 2 A_{\alpha,\beta}}\sum_{i,j\ne \alpha,\beta} 
 (A_{i,j} A_{\alpha,\beta} - A_{i,\alpha} A_{j,\beta} + A_{j,\alpha} A_{i,\beta} )\eta_i \eta_j \bigr\}
\eeq
In order to get the correct signs in all terms of the sum, we introduce the following change of variables
\beq
\tilde \eta_i=
\begin{cases}
- \eta_i &\hbox{ if } \alpha< i <\beta \cr
\eta_i & \hbox{ otherwise}
\end{cases}
\eeq
The Jacobian of the transformation is $(-1)^{\alpha +\beta -1}$ and we get, using homogeneity
\beq
\Pf (A)&&=\bigl(\frac{1}{A_{\alpha,\beta}}\bigr)^{n-2} \int \bigl(\prod_{k\ne \alpha,\beta}  d\tilde \eta_k \bigr)  
\exp \bigl\{ \sum_{\scriptstyle{{i<j}\atop{i,j\ne \alpha,\beta}}}
\Pf( A_{\{i,j,\alpha,\beta\}})\tilde \eta_i \tilde \eta_j \bigr\}\nonumber\\
&&=\bigl(\frac{1}{A_{\alpha,\beta}}\bigr)^{n-2} \int \bigl(\prod_{k\ne \alpha,\beta}  d\eta_k \bigr) 
\exp \bigl\{ {1\over 2 }\sum_{i,j\ne \alpha,\beta} 
A^{\overline{\{\alpha,\beta\}}}_{i,j}\;\eta_i \eta_j \bigr\}\nonumber\\
&&=\Pf(A_{\{\alpha,\beta\}})^{-(n-2)} \Pf( A^{\overline{\{\alpha,\beta\}}} )
\eeq
Equation \eqref{eq-hiera} is proven for $p=1$. We now fix  $p$ with $1< p <n$ and suppose that equation \eqref{eq-hiera} holds for all $p' < p$.
We consider $K$ a subset of indices of order $2 p $ in $\{1,\cdots, 2 n\}$ such that $\Pf(A_K)\not=0$. Thus there exists at least one proper submatrix of $A_K$, says $A_{K_1}$  with $|K_1|= 2 p_1$, $0<p_1<p$, such that $\Pf(A_{K_1})\not=0$. Let $K_2 = K\setminus K_1$ and $p_2=p-p_1$. 
We first have the two following identities which are a direct consequence of the definitions \eqref{submatrix} and \eqref{supmatrix}:

\be\label{id1}
\bigl[ A_{K_1\cup K_2}\bigr]_{K_1} = A_{K_1}
\ee
\be\label{id2}
\bigl[ A^{\overline{K_1}}\bigr]_{K_2} = \bigl[A_{K_1\cup K_2}\bigr]^{\overline{K_1}}
\ee

Now applying the recurrence hypothesis for $p_1$ for the matrix $A_{K_1\cup K_2}$ and both \eqref{id1} and \eqref{id2}, we first get,
\beq\label{id3}
\Pf(A_{K_1\cup K_2})  &=&   \Pf \bigl( \bigl[A_{K_1\cup K_2}\bigr]_{K_1} )^{-(p_2-1)} 
\Pf \bigl(\bigl[A_{K_1\cup K_2}\bigr]^{\overline{K_1}} \bigr)
\nonumber\\
&=&  \Pf ( A_{K_1} )^{-(p_2-1)} \Pf \bigl(\bigl[A^{\overline{K_1}}\bigr]_{K_2} \bigr)
\eeq
We also need to prove that both matrices $A^{\overline{{K_1\cup K_2}}}$ and $\bigl[A^{\overline{K_1}}\bigr]^{\overline{K_2}}$
are proportional. For every pair of indices $i,j \in K\setminus (K_1\cup K_2)$, we have
\beq
\bigl( A^{\overline{K_1\cup K_2}}\bigr)_{i,j}&=& \Pf(A_{K_1\cup (K_2 \cup \{i,j\})})\nonumber\\
 &=& \Pf \bigl(A_{K_1} \bigr)^{-p_2} \Pf \bigl(\bigl[A^{\overline{K_1}}\bigr]_{(K_2 \cup \{i,j\})} \bigr)\nonumber\\
 &=& \Pf \bigl(  A_{K_1}  \bigr)^{-p_2}  \bigl(\bigl[ A^{\overline{K_1}}\bigr]^{\overline{K_2}}\bigr)_{i,j}
 \eeq
Since this relation holds for every entry of both matrices, we have
\be\label{id4}
A^{\overline{K_1\cup K_2}} =\Pf(A_{K_1})^{-p2} \bigl[ A^{\overline{K_1}}\bigr]^{\overline{K_2}}
\ee
Now we can write
\beq
\Pf(A) &=&  \Pf ( A_{K_1} )^{-(n-p_1-1)} \Pf (A^{K_1})\nonumber\\
&=& \Pf ( A_{K_1} )^{-(n-p_1-1)} \Pf ([A^{\overline{K_1}}]_{K_2})^{-(n-p_1-p_2-1)} \Pf ((A^{\overline{K_1}})^{\overline{K_2}})\nonumber\\
&=&  \Pf ( A_{K_1} )^{-(n-p_1-1)+p_2 (n-p_1-p_2)} \Pf ([A^{\overline{K_1}}]_{K_2})^{-(n-p_1-p_2-1)}  \Pf (A^{\overline{K_1\cup K_2}})\nonumber\\
&=& \Pf ( A_{K_1} ){-(n-p_1-1)+p_2 (n-p_1-p_2)-(p_2-1)(n-p_1-p_2-1)} 
\Pf (A_{K_1\cup K_2})^{-(n-p_1-p_2-1)}  \Pf (A^{\overline{K_1\cup K_2}})\nonumber
\eeq
we have used the hypothesis in the first two lines and equations \eqref{id2}, \eqref{id3},\eqref{id4}
in the last two lines, respectively. The exponent of $Pf ( A_{K_1})$ being zero in the last line, we finally get
\be
\Pf(A) = \Pf (A_{K_1\cup K_2})^{-(n-(p_1+p_2)-1)} \Pf (A^{\overline{K_1\cup K_2}})
\ee
which is \eqref{eq-hiera} for $p=p_1+p_2$.

Proposition \ref{hiera} is proven.
\hfill\halmos

\bigskip\goodbreak\noindent
\section{Acknowledgments}
Dedicated to the memory of Jean Lascoux: this work is a far-off answer to one of his illuminating questions. I also acknowledge the kind hospitality of Institut Henry Poincar\'e, Paris, where part of this work has been done.

\goodbreak
\bigskip
\bibliographystyle{amsalpha}

\providecommand{\bysame}{\leavevmode\hbox to3em{\hrulefill}\thinspace}
\providecommand{\MR}{\relax\ifhmode\unskip\space\fi MR }
\providecommand{\MRhref}[2]{%
  \href{http://www.ams.org/mathscinet-getitem?mr=#1}{#2}
}
\providecommand{\href}[2]{#2}
\begin{thebibliography}{}

\end{thebibliography}


\begin{thebibliography}{A}
\medskip

\bibitem{O}  Onsager, L. ,{\sl  Crystal Statistics. I. A Two-Dimensional Model with an Order-Disorder Transition.}
 Phys. Rev.  {\bf  65}, 117�149 (1944).
\bibitem{KW}  Kac, M. , Ward, J. C.  {\sl A Combinatorial Solution of the Two-Dimensional Ising Model}, 
Phys. Rev.  {\bf 88}, 1332--1337 (1952).
\bibitem{PW}  Potts, R. B. , Ward,  J. C.   {\sl The Combinatorial Method and the Two-Dimensional Ising Model}, 
Prog. Theor. Phys. {\bf  13} 38-46 (1955). 
\bibitem{HG}  Hurst, C. A. , Green, H. S.  {\sl New Solution of the Ising Problem for a Rectangular Lattice}, 
J. Chem. Phys. {\bf 33}, 1059 (1960).
\bibitem{F2}  Fisher, M. E.  {\sl Statistical mechanics of dimers on a plane lattice}, 
Phys. Rev. {\bf 124}, 1664-1672 (1961).
\bibitem{K1}  Kasteleyn, P. W. {\sl  The statistics of dimers on a lattice.}
 Physica  {\bf  27}, 1209--1225 (1961).
\bibitem{K2}   Kasteleyn, P. W.  {\sl Dimer statistics and phase transitions.} 
J. Math. Phys. {\bf 4}, 287 (1963).
\bibitem{K3}  Kasteleyn, P.W. {\sl Graph theory and crystal Physics.} 
in Graph Theory and Theoretical Physics, pp. 43--110. 
F. Harary ed.,  Academic Press, London (1967).
\bibitem{GL}     Galluccio, A., Loebl, M. {\sl On the theory of Pfaffian orientations. I. Perfect matchings
and permanents.} Electron. J. Comb.  {\bf  6}, 18 (1999)
\bibitem{CR}    Cimasoni, D., Reshetikhin, N. {\sl Dimers on surface graphs and spin structures. I.} Commun.
Math. Phys.  {\bf  275}, 187--208 (2007)
\bibitem{Te}     Tesler, G.  {\sl Matchings in graphs on non-orientable surfaces.} J. Combin. Theory Ser. B
 {\bf  78}, 198--231 (2000).
\bibitem{Po}    P\'olya, G.  {\sl Aufgabe 424.} Arch. Math. Phys. Ser. {\bf 20},  271 (1913).
\bibitem{V}     Valiant, L.G. {\sl The complexity of computing the permanent.} Theor. Comput. Sci.
 {\bf  8}, 189--201 (1979).
\bibitem{T}  Thomas, R.   {\sl A Survey of Pfaffian Orientations of Graphs.} in  Proceedings of the International Congress of Mathematicians 2006, Vol III, pp 963--984, M.Sanz-Sole, J. Soria, J. L. Varona, J. Verdera Ed.,  EMS Publishing House, Z\"urich (2007).
\bibitem{Tu}  W. T. Tutte {\sl The factorization of linear graphs}, 
J. London Math. Soc. {\bf 22}, 107 (1947).
\bibitem{LP}      Lov\'asz, L., Plummer, M.D.{\sl Matching theory} , North-Holland, Amsterdam (1986)
\bibitem{LW1}      Lu, W.T., Wu, F.Y. {\sl Dimer statistics on the M\"obius strip and the Klein bottle.} Phys.
Lett. . {\bf A 259}, 108--114 (1999)
\bibitem{LW2}      Lu, W.T., Wu, F.Y. {\sl Close-packed dimers on nonorientable surfaces.} Phys. Lett. A
 {\bf 293}, 235--246 (2002)
\bibitem{ML}    MacLane, S. {\sl A combinatorial condition for planar graphs}, Fund. math. {\bf 28}, 22 (1937).
\bibitem{BD}    Bruhn, H.,  Diestel, R. {\sl MacLane's theorem for arbitrary surfaces}, J. Comb. Theory, {\bf B 99}, 275--286 (2009) .
\bibitem{H}   Harary, F. {\sl Graph Theory}, Addison-Wesley, Reading (1969).
\bibitem{D}  Diestel, R. {\sl Graph theory}, Graduate Texts in Mathematics, Volume 173,
4th edition, Springer-Verlag ed., Heidelberg  (2010).
\bibitem{GH}   Green,H. S., Hurst, C. A.  {\sl Order-disorder phenomena}, Interscience Publishers (London 1964).
\bibitem{B}   Berezin, F.A.  {\sl The method of second quantification},
Academic Press, New York (1966).
\bibitem{S}  Samuel, S. {\sl The use of anti-commuting variable integrals
in statistical mechanics, I}, J. Math. Phys. {\bf 21}, 2806 (1980).
\bibitem{F}   Fisher, M. E. {\sl On the dimer solution of the planar Ising Model}, 
J. Math. Phys. {\bf 7}, 1776 (1966).
\bibitem{K} Kuratowski, C. {\sl Sur le probl\`eme des courbes gauches en Topologie},
Fund. Math. {\bf 15} 271--283 (1930).
\bibitem{HN} Harari, F. Norman, R.Z.  {\sl Some properties of line digraphs}, 
Rendiconti del Circolo Matematico di Palermo {\bf 9}, 161 (1960).
\bibitem{BP} Baley Price, G. {\sl An introduction to multicomplex spaces and functions}, Marcel Dekker,  New York (1991).
\bibitem{MT} Mohar,B., Thomassen, C. {\sl Graphs on Surfaces}
Johns Hopkins University Press, Baltimore (2001).
\bibitem{St}  Stahl, S. {\sl Generalized embedding schemes}, J. Graph Theory {\bf 2}, 41 (1978).


\end{thebibliography}

\end{document}